\documentclass{tep}
\pdfoutput=1 
\usepackage{dynonfact}

\begin{document}

\thispagestyle{empty}
\def\thefootnote{\fnsymbol{footnote}}
\setcounter{footnote}{1}
\null
\mbox{}\hfill  
FR-PHENO-2013-014  \\
\vskip 0cm
\vfill
\begin{center}
  {\Large \boldmath{\bf 
      Mixed QCD--electroweak \order{\alphas\alpha} corrections to \\
      Drell--Yan processes in the resonance region:\\
      pole approximation and non-factorizable corrections
    }
    \par} \vskip 2.5em
  {\large
    {\sc Stefan Dittmaier, Alexander Huss \\[.3em]
      and Christian Schwinn
    }\\[1ex]
    {\normalsize \it 
      Albert-Ludwigs-Universit\"at Freiburg, Physikalisches Institut, \\
      D-79104 Freiburg, Germany
    }
    \\[2ex]
  }
  \par \vskip 1em
\end{center}\par
\vskip .0cm \vfill {\bf Abstract:} \par
Drell--Yan-like \PW-boson and \PZ-boson production in the resonance region
allows for high-precision measurements that are crucial to carry
experimental tests of the Standard Model to the extremes, such as the
determination of the \PW-boson mass and the effective weak mixing angle.
In this article, we establish a framework for the calculation of the mixed
QCD--electroweak \order{\alphas\alpha} corrections to Drell--Yan
processes in the resonance region, which are one of the main remaining
theoretical uncertainties.  We describe how the Standard Model
prediction can be successfully performed in terms of a consistent
expansion about the resonance poles, which classifies the corrections
in terms of factorizable and non-factorizable contributions.  The
former can be attributed to the \PW/\PZ production and decay subprocesses
individually, while the latter link production and decay by
soft-photon exchange.  At next-to-leading order we compare the full
electroweak corrections with the pole-expanded approximations,
confirming the validity of the approximation.  At
\order{\alphas\alpha}, we describe the concept of the expansion and
explicitly give results on the non-factorizable contributions, which
turn out to be phenomenologically negligible. 
Our results, thus,
demonstrate that for phenomenological purposes the
\order{\alphas\alpha} corrections can be factorized into terms
associated with initial-state and/or final-state corrections.
Moreover, we argue that the factorization properties of the non-factorizable 
corrections at \order{\alphas\alpha} from lower-order \order{\alphas} 
graphs generalize to any order in \order{\alphas^n\alpha}.
\par
\vskip 1cm
\noindent
{\bf March 2014}
\par
\null
\setcounter{page}{0}
\clearpage
\def\thefootnote{\arabic{footnote}}
\setcounter{footnote}{0}

%%%%%%%%%%%%%%%%%%%%%%%%%%%%%%%%%%%%%%%%%%%%%%%%%%%%%%%%%%%%%%%%%%%%%%
\section{Introduction}
\label{sec:intro}

Drell--Yan-like \PW- or \PZ-boson production is among the most important classes of standard-candle processes at the LHC (see, e.g.\ Refs.~\cite{Gerber:2007xk,Abdullin:2006aa}).
Apart from delivering important information on parton distributions and allowing for the search for new gauge bosons in the high-mass range, these processes allow for high-precision measurements in the resonance regions.
The weak mixing angle might be extracted from data with LEP precision~\cite{Haywood:1999qg}, and the \PW-boson mass \MW, whose world average is dominated by measurements via Drell--Yan-like \PW-boson production at the Tevatron~\cite{Aaltonen:2013iut}, might be measured with a sensitivity of about $7~\MeV$~\cite{Besson:2008zs}.

In the past two decades, great effort was made in the theory community to deliver precise predictions matching the required accuracy.
QCD corrections are known up to next-to-next-to-leading order~\cite{Hamberg:1990np,Harlander:2002wh,Anastasiou:2003ds,Melnikov:2006di,Melnikov:2006kv,Catani:2009sm,Gavin:2010az,Gavin:2012sy}, electroweak (EW) corrections up to next-to-leading order (NLO)~\cite{Baur:1997wa,Zykunov:2001mn,Baur:2001ze,Dittmaier:2001ay,Baur:2004ig,Arbuzov:2005dd,CarloniCalame:2006zq,Zykunov:2005tc,CarloniCalame:2007cd,Arbuzov:2007db,Brensing:2007qm,Dittmaier:2009cr}.%
\footnote{A recent update with QED-corrected PDFs has been presented in Ref.~\cite{Boughezal:2013cwa}.}
Both on the QCD and on the EW sides, there are further refinements such as leading higher-order effects~\cite{Baur:1999hm,CarloniCalame:2003ux,Placzek:2003zg,Moch:2005ba,Laenen:2005uz,Idilbi:2005ky,Ravindran:2007sv,Brensing:2007qm,Dittmaier:2009cr} and generalizations to the supersymmetric extension of the Standard Model~\cite{Brensing:2007qm,Dittmaier:2009cr}.
 The treatment of small transverse momenta requires a resummation of large logarithms through matched parton showers~\cite{Frixione:2006gn,Alioli:2008gx,Hamilton:2008pd} or dedicated calculations supplemented with fits to non-perturbative functions~\cite{Balazs:1997xd,Landry:2002ix} that are available at next-to-next-to-leading logarithmic accuracy~\cite{Bozzi:2010xn,Mantry:2010mk,Becher:2011xn}.
First approaches to the combination of QCD and EW corrections can be found in Refs.~\cite{Cao:2004yy,Richardson:2010gz,Bernaciak:2012hj,Barze:2012tt,Li:2012wna,Barze:2013yca}.
In view of fixed-order calculations, the largest missing piece seems to be the mixed QCD--EW corrections of \order{\alphas\alpha}.
Knowing the contribution of this order will also answer the question how to properly combine QCD and EW corrections in predictions.
In Ref.~\cite{Balossini:2009sa} this issue is quantitatively discussed with special emphasis on observables that are relevant for the \MW determination, revealing percent corrections of \order{\alphas\alpha} that should be calculated.
First steps towards this direction have been taken by calculating two-loop contributions~\cite{Kotikov:2007vr,Kilgore:2011pa,Bonciani:2011zz}, the full \order{\alphas\alpha} corrections to the \PW and \PZ decay widths~\cite{Czarnecki:1996ei,Kara:2013dua}, the \order{\alpha} EW corrections to $\PW/\PZ+\jet$ production~\cite{Kuhn:2005az,Kuhn:2007cv,Hollik:2007sq,Denner:2009gj,Denner:2011vu,Denner:2012ts}, and the \order{\alphas} QCD corrections to $\PW/\PZ+\Pgg$ production~\cite{Smith:1989xz,Ohnemus:1992jn,Ohnemus:1994qp,Dixon:1998py,Campbell:1999ah,DeFlorian:2000sg,Hollik:2007sq,Campbell:2011bn}.

In this article, we report on the calculation of the \order{\alphas\alpha} corrections to Drell--Yan processes in the resonance region via the so-called \emph{pole approximation}~(PA), which is based on a systematic expansion about the resonance pole.
Specifically, we describe the concept of the approximation in detail, define all ingredients in an explicit manner, and discuss all its contributions and its success at NLO.
It turns out that at \order{\alpha}, the corrections near the resonance are almost entirely due to \emph{factorizable} corrections to the \PW/\PZ decay subprocesses, which show the well-known enhancements due to photonic final-state radiation off charged leptons, while the factorizable corrections to the production process are suppressed below the percent level.
The \emph{non-factorizable} corrections, which are due to soft-photon exchange between production, decay, and intermediate nearly resonant gauge bosons (for \PW production), are also suppressed way below the percent level.
In summary, the NLO \order{\alpha} PA works up to fractions of $1\%$ near the resonance.
Motivated and guided by the successful construction of the PA at \order{\alpha}, we describe its concept at \order{\alphas\alpha} and present results on the non-factorizable contributions at this order, which comprise the most delicate contribution to the PA. 
When the invariant mass of the resonance is integrated over, their contribution vanishes~\cite{Fadin:1993dz} at NLO, i.e.\ they only tend to distort the resonance without changing the normalization of the cross section~\cite{Melnikov:1995fx,Beenakker:1997ir,Denner:1997ia}.
Our results will extend this statement, which was formulated for pairs of resonances at NLO, to single-resonance processes at NNLO \order{\alphas\alpha}.

This article is organized as follows: In Sect.~\ref{sec:NLO} we
describe the concept of the pole expansion at NLO, give explicit
results for the electroweak and QCD corrections that will be used as
building blocks in our \order{\alphas\alpha} calculation, and
establish the numerical accuracy of the PA by
comparing it to the full NLO electroweak results. We also briefly review
the effective-field-theory~(EFT) inspired approach to the pole expansion
introduced in Refs.~\cite{Beneke:2003xh,Beneke:2004km}. In
Sect.~\ref{sec:nnlo-nonfact} we develop the pole expansion for the
NNLO \order{\alphas\alpha} corrections and describe the different
contributions to the factorizable and non-factorizable corrections.
We then compute the non-factorizable
\order{\alphas\alpha} corrections explicitly, encountering non-trivial gauge
cancellations that allow to write the corrections in terms of one-loop
EW and QCD results.  This result is obtained using a gauge-invariance
argument, as well as by two independent analytical calculations.  We
describe our treatment of the infrared (IR) singularities in detail and give a
final formula for the non-factorizable corrections that is suitable
for numerical evaluation. In Sect.~\ref{sec:numerics} we present
our numerical results for the non-factorizable corrections. They turn
out to be very small, below the $0.1\%$ level, and therefore
phenomenologically negligible. This shows that the numerically
relevant part of the NNLO \order{\alphas\alpha} corrections in the
resonance region factorizes into contributions from gauge
boson production and decay. Based on our NLO results we expect that
the dominant effect is given by the combination of initial-state QCD
and final-state EW corrections, which we will compute in a future
publication.  Some details in our computation of the NNLO
non-factorizable corrections and explicit expressions for
IR-regularized contributions to the cross sections are relegated
to the appendix.

%%%%%%%%%%%%%%%%%%%%%%%%%%%%%%%%%%%%%%%%%%%%%%%%%%%%%%%%%%%%%%%%%%%%%%
\section{Pole expansion of the NLO \texorpdfstring{\order{\alpha}}{O(a)} corrections to Drell--Yan processes}
\label{sec:NLO}

\begin{figure}[b]
  \centering
  \includegraphics{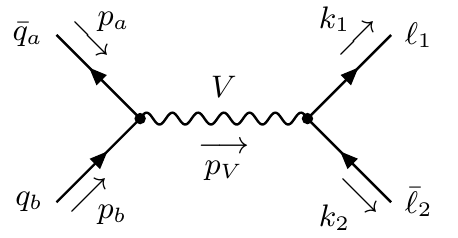}
  \caption{Leading-order diagram to the Drell--Yan process at the LHC.}
 \label{fig:vNLOgraphs:B}
\end{figure}

In this section we discuss the computation of the NLO EW corrections to the processes
\begin{equation}
  \label{eq:dy}
  \Paq_a(p_a) \,+\, \Pq_b(p_b)\to\Pl_1(k_1)\,+\, \Pal_2(k_2)\,+\,X
\end{equation}
in the vicinity of an intermediate vector-boson resonance, $(k_1+k_2)^2\approx \MV^2$ (see Fig.~\ref{fig:vNLOgraphs:B} for the leading-order diagram).
We consider the charged-current process with $\Pl_1\Pal_2=\Pgne\Pep/\Pgngm\Pgmp$ and intermediate vector boson $V=\PWp$, its charge conjugate with $\Pl_1\Pal_2=\Pem\Pagne/\Pgmm\Pagngm$ and $V=\PWm$, as well as the neutral-current process with $\Pl_1\Pal_2=\Pem\Pep/\Pgmm\Pgmp$ and $V=\PZ$.
Quarks are consistently taken as massless ($p_a^2=p_b^2=0$), and small lepton masses are only employed to regularize mass singularities in collinear final-state radiation, otherwise $k_1^2=k_2^2=0$.
We classify and calculate the corrections to the cross section that are enhanced by resonance factors
\begin{equation}
  \label{eq:resonance}
  \Bigl[\,p_V^2-\MV^2+\ri\MV\GV\,\Bigr]^{-1} \sim \order{\frac{1}{\MV\GV}} , 
\end{equation}
where $p_V=k_1+k_2$ and \GV is the decay width of the vector-boson resonance. 
The exact NLO EW corrections to the processes~\eqref{eq:dy} are known~\cite{Baur:1997wa,Zykunov:2001mn,Baur:2001ze,Dittmaier:2001ay,Baur:2004ig,Arbuzov:2005dd,CarloniCalame:2006zq,Zykunov:2005tc,CarloniCalame:2007cd,Arbuzov:2007db,Brensing:2007qm,Dittmaier:2009cr}, but we discuss the pole expansion in some detail in order to introduce the concept employed in the computation of the \order{\alphas\alpha} corrections in Sect.~\ref{sec:nnlo-nonfact} and in order to establish the numerical accuracy of the PA.
This PA is based on the leading term in the expansion of all cross-section contributions about the vector-boson resonance.
In Sect.~\ref{sec:nlo-nonfact} we describe the concept of the pole expansion directly based on a Feynman-diagrammatic approach and, as an alternative for the virtual corrections, also following an EFT approach.
Section~\ref{sec:QCD} briefly reviews the calculation and results for the NLO QCD corrections, which are needed as building block at \order{\alphas\alpha} later.
In Sect.~\ref{sec:nlo-num} we compare the numerical results of the PA to the full NLO result, revealing agreement at the level of fractions of $1\%$ in the resonance regions.
We, therefore, can expect an approximation of the full \order{\alphas\alpha} corrections by the PA at the level of $\sim0.1\%$, which is more than sufficient for phenomenology.

%%%%%%%%%%%%%%%%%%%%%%%%%%%%%%%%%%%%%%%%%%%%%%%%%%
\subsection{Definition of factorizable and non-factorizable corrections}
\label{sec:nlo-nonfact}

%%%%%%%%%%%%%%%%%%%%%%%%%%%%%%
\subsubsection{Concept of the pole expansion}
\label{sec:pole-expansion}

The general idea~\cite{Stuart:1991xk,Aeppli:1993rs} of a PA for any Feynman diagram with a single resonance is the systematic isolation of all parts that are enhanced by a resonance factor~\eqref{eq:resonance}. 
For \PW~production different variants of PAs have been suggested and discussed at NLO already in Refs.~\cite{Wackeroth:1996hz,Dittmaier:2001ay,Baur:1998kt}.  
For the virtual corrections we follow the PA approach of Ref.~\cite{Dittmaier:2001ay}.  
In that reference the real-emission corrections were evaluated on the basis of the full amplitudes without further approximations.
After a proper cancellation of IR singularities between the real and virtual corrections, the PA was then applied only to the finite remainder of the virtual corrections. 
In order to be able to calculate and discuss the different contributions of factorizable (production and decay) and non-factorizable corrections separately, a consistent application of the PA also in the real corrections is required.
Here, we will follow Ref.~\cite{Denner:1997ia} for the pole expansion for the real corrections in order to prepare for the extension to the mixed QCD--EW corrections.
We do not apply the PA to the LO cross section, which is kept without approximation.

Schematically, each transition amplitude for the processes~\eqref{eq:dy} has the form
\begin{equation}
  \M=\frac{W(p_V^2)}{p_V^2-\MV^2+\Sigma(p_V^2)}+N(p_V^2),
\end{equation}
with functions $W$ and $N$ describing resonant and non-resonant parts, respectively, and $\Sigma$ denoting the self-energy of $V$.
The resonant contributions of $\M$ are isolated in a gauge-invariant way as follows,
\begin{equation}
  \M =
  \frac{W(\mu_V^2)}{p_V^2-\mu_V^2}\,\frac{1}{1+\Sigma'(\mu_V^2)}
  +\left[\frac{W(p_V^2)}{p_V^2-\MV^2+\Sigma(p_V^2)}
  - \frac{W(\mu_V^2)}{p_V^2-\mu_V^2}\,\frac{1}{1+\Sigma'(\mu_V^2)} \right] + N(p_V^2) ,
  \label{eq:PA}
\end{equation}
where 
\begin{equation}
  \mu_V^2=\MV^2-\ri\MV\GV
\end{equation}
is the gauge-invariant~\cite{Sirlin:1991fd,Gambino:1999ai,Grassi:2001bz} location of the propagator pole in the complex $p_V^2$ plane.
Equation~\eqref{eq:PA} can serve as a basis for the gauge-invariant introduction of the finite decay width in the resonance propagator, thereby defining the so-called \emph{pole scheme}. 
In this scheme the term in square brackets is perturbatively expanded in the coupling \alpha including terms up to \order{\alpha}, while the full $p_V^2$ dependence is kept.
An application of this scheme to Z-boson production is, e.g., described in Ref.~\cite{Dittmaier:2009cr} in detail.

\begin{figure}
  \centering
  \begin{subfigure}[c]{0.3\linewidth}
    \centering
    \includegraphics[scale=.75]{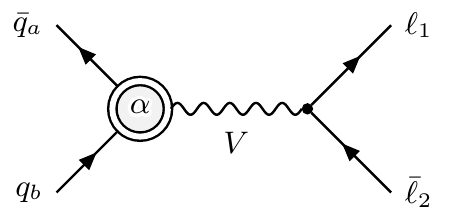}
    \subcaption{a}{}
    %\label{fig:vNLOgraphs:pro}
  \end{subfigure}
  \begin{subfigure}[c]{0.3\linewidth}
    \centering
    \includegraphics[scale=.75]{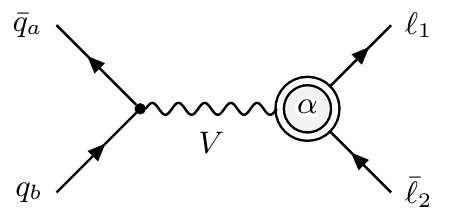}
    \subcaption{b}{}
    %\label{fig:vNLOgraphs:dec}
  \end{subfigure}
  \begin{subfigure}[c]{0.3\linewidth}
    \centering
    \includegraphics[scale=.75]{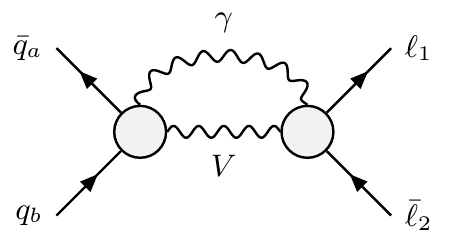}
    \subcaption{c}{}
    %\label{fig:vNLOgraphs:nf}
  \end{subfigure}
  \caption{Generic diagrams 
  for the EW virtual NLO factorizable corrections to production~(a) and decay~(b), as well as for virtual non-factorizable corrections~(c), where the empty blobs stand for all relevant tree structures and the ones with ``\alpha'' inside for one-loop corrections of \order{\alpha}.}
  \label{fig:vNLOgraphs}
\end{figure}

\begin{figure}
  \centering
  \begin{subfigure}[c]{0.28\linewidth}
    \centering
    $\left\lvert\vphantom{\rule{1cm}{1cm}}
    \Vcenter{\includegraphics[scale=0.75]{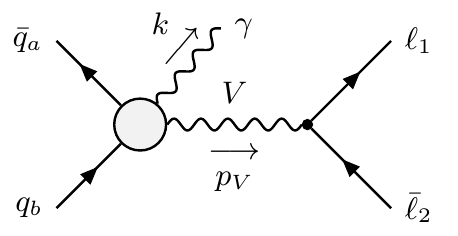}}
    \right\rvert^2$
    \subcaption{a}{}
    %\label{fig:rNLOgraphs:pro}
  \end{subfigure}
  \begin{subfigure}[c]{0.28\linewidth}
    \centering
    $\left\lvert\vphantom{\rule{1cm}{1cm}}
    \Vcenter{\includegraphics[scale=0.75]{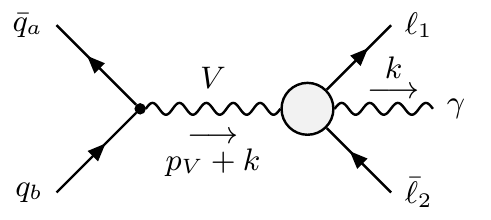}}
    \right\rvert^2$
    \subcaption{b}{}
    %\label{fig:rNLOgraphs:dec}
  \end{subfigure}
  \begin{subfigure}[c]{0.4\linewidth}
    \centering
    \includegraphics[scale=0.75]{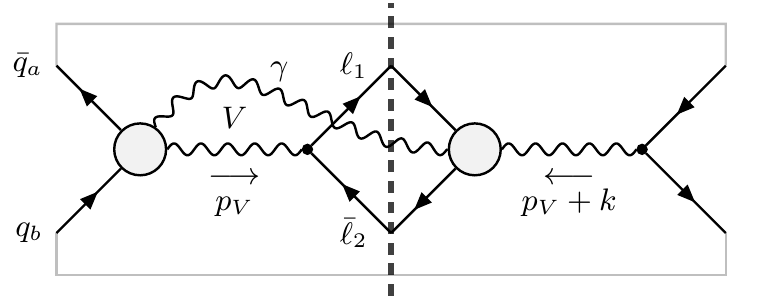}
    \subcaption{c}{}
    %\label{fig:rNLOgraphs:nf}
  \end{subfigure}
  \caption{Generic diagrams for the real photonic NLO factorizable corrections to production~(a) and decay~(b), as well as for real non-factorizable corrections~(c), where the blobs stand for all relevant tree structures.}
  \label{fig:rNLOgraphs}
\end{figure}

The PA for the amplitude results from the r.h.s.\ of Eq.~\eqref{eq:PA} upon neglecting the last, non-resonant term and asymptotically expanding the term in square brackets in $p_V^2$ about the point $p_V^2=\mu_V^2$, where only the leading, resonant term of the expansion is kept.
The first term on the r.h.s.\ of~\eqref{eq:PA} defines the so-called \emph{factorizable} corrections in which on-shell production and decay amplitudes for $V$ are linked by the off-shell propagator; these contributions are illustrated by diagrams~(a) and (b) of Fig.~\ref{fig:vNLOgraphs}.
The term on the r.h.s.\ of Eq.~\eqref{eq:PA} in square brackets contains the so-called \emph{non-factorizable} corrections which receive resonant contributions from all diagrams where the limit $p_V^2\to\mu_V^2$ in $W(p_V^2)$ or $\Sigma'(p_V^2)$ would lead to (infrared) singularities. 
At NLO, this happens if a soft photon of energy $E_\Pgg\lesssim\GV$ is exchanged between the production process, the decay part, and the intermediate $V$ bosons; a generic loop diagram is shown in Fig.~\ref{fig:vNLOgraphs}(c).  
Figure~\ref{fig:rNLOgraphs} shows the real-photon emission counterparts of the
 one-loop corrections displayed in Fig~\ref{fig:vNLOgraphs}. 

%%%%%%%%%%%%%%%%%%%%%%%%%%%%%%
\subsubsection{Virtual corrections}
\label{sec:virt-nonfact}

We now consider the PA for the virtual EW NLO corrections in more detail.
The \emph{factorizable} virtual corrections are defined as the product of the on-shell matrix elements of $V$-boson production and decay times the off-shell $V$-boson propagator, as illustrated in Figs.~\ref{fig:vNLOgraphs}(a),(b),
\begin{equation}
  \label{eq:nlo-fact-virt}
  \delta\M_{\Vew,\fact}^{\Paq_a\Pq_b\to\Pl_1\Pal_2}=
  \sum_\lambda 
  \frac{
    \delta\M_\Vew^{\Paq_a\Pq_b\to V}(\lambda)\;
    \M_0^{V\to\Pl_1\Pal_2}(\lambda)
    +\M_0^{\Paq_a\Pq_b\to V}(\lambda)\;
    \delta\M_\Vew^{V\to\Pl_1\Pal_2}(\lambda)
  }{p_V^2-\mu_V^2} ,
\end{equation}
where the sum over the physical polarization states $\lambda$ of the
vector boson $V$ encodes the proper spin correlation.  Here $\M_0$ and
$\delta\M_\Vew$ denote tree-level and EW one-loop amplitudes,
respectively.  The expressions for the virtual corrections to
vector-boson production and decay can be found, e.g., in
Refs.~\cite{Dittmaier:2001ay,Dittmaier:2009cr}.  Care has to be taken
that the subamplitudes appearing on either side of the $V$ resonance are
evaluated for on-shell $V$ bosons, otherwise gauge invariance cannot
be guaranteed.  
For the \order{\alpha} approximation, the use of the
problematic complex value $p_V^2=\mu_V^2$ in the on-shell condition of the
subamplitudes is not
necessary, so that we have set $p_V^2=\MV^2$ in the numerator of the
first term of Eq.~\eqref{eq:PA} in order to obtain
Eq.~\eqref{eq:nlo-fact-virt}.  Similarly, the conventional on-shell
renormalization scheme (see, e.g., Ref.~\cite{Denner:1991kt}) can be
used where the real part of the residue of the propagator is
normalized to one at $p_V^2=\MV^2$.  Furthermore, the gauge-boson
width \GV can be set to zero in the self-energy in the
\order{\alpha} approximation, so that the residue correction
$(1+\Sigma')^{-1}$ in Eq.~\eqref{eq:PA} reduces to one in the 
on-shell renormalization scheme and does not
appear in Eq.~\eqref{eq:nlo-fact-virt}.  The \emph{on-shell
  projection} $p_V^2\to \MV^2$ of the subamplitudes $\M_0$ and
$\delta\M_{\Vew}$ in Eq.~\eqref{eq:nlo-fact-virt} has to be defined
carefully and is not unique, because the phase space is parametrized
by more than one variable.  Different variants may lead to results
that differ within the intrinsic uncertainty of the PA, which is of
\order{\alpha/\pi\times\GV/\MV} in the resonance region when
applied to \order{\alpha} corrections.  Despite this freedom in the
choice of the specific on-shell projection, care has to be taken that
virtual and real corrections still match properly in the (soft and
collinear) infrared limits in order to guarantee the cancellation of
the corresponding singularities.  Note that in case of the
neutral-current process the exchange of an off-shell photon is
disregarded, as it does not lead to a resonant contribution.

\begin{figure}
  \centering
  \begin{subfigure}[m]{.49\linewidth}
    \centering
    \includegraphics{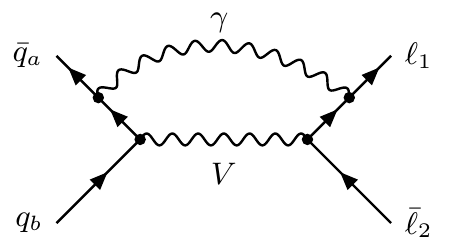}
    \subcaption{a}{Manifestly non-factorizable corrections}
    %\label{fig:nlo-nf-virt:mani}
  \end{subfigure}
  \begin{subfigure}[m]{.49\linewidth}
    \centering
    \includegraphics{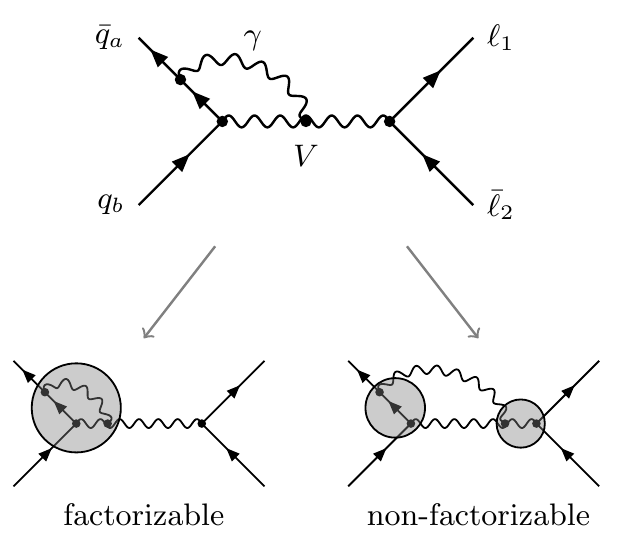}
    \subcaption{b}{Not manifestly non-factorizable corrections}
    %\label{fig:nlo-nf-virt:notmani}
  \end{subfigure}
  \caption{Example diagrams illustrating the manifestly non-factorizable corrections~(a), and contributions that contain both, factorizable and non-factorizable corrections~(b).}
  \label{fig:nlo-nf-virt}
\end{figure}

The \emph{non-factorizable} corrections involve contributions for which the production and decay do not proceed independently.  
This is illustrated in Fig.~\ref{fig:vNLOgraphs}(c) for the case of virtual corrections.  
The manifestly non-factorizable corrections, for instance the diagram in Fig.~\ref{fig:nlo-nf-virt}(a), do not contain an explicit propagator factor $(p_V^2-\mu_V^2)^{-1}$ before the loop integration.
As shown in Ref.~\cite{Denner:1997ia} by power-counting arguments, a
resonant contribution of such diagrams is connected to the soft-photon
exchange between the production and decay subprocesses and involves IR
singularities, while the exchange of massive particles (such as \PZ
bosons) or highly-energetic photons between production and decay leads
to non-resonant corrections only.  Diagrams with photons coupled to
the intermediate vector boson $V$ contribute to both the factorizable
and non-factorizable corrections.  The factorizable part of those
diagrams is obtained upon setting the vector-boson momentum $p_V$ on
its mass shell in the loop containing the photon.  This attributes
this contribution either to on-shell production or on-shell decay of
the $V$ boson, and these initial- or final-state factorizable
corrections become gauge invariant.  
The difference between the full
amplitude of diagrams like Fig.~\ref{fig:nlo-nf-virt}(b) and its
factorizable part~\eqref{eq:nlo-fact-virt} defines their
non-factorizable contributions, which result
from the fact that setting $p_V$ on shell in the loop creates
artificial soft singularities,
\begin{equation}
  \label{eq:def-nonfact}
  \delta\M_{\Vew,\nf}^{\Paq_a\Pq_b\to\Pl_1\Pal_2}=
  \left\{ \delta\M_\Vew^{\Paq_a\Pq_b\to\Pl_1\Pal_2}
  - \delta\M_{\Vew,\fact}^{\Paq_a\Pq_b\to\Pl_1\Pal_2}
  \right\}_{p_V^2\to \MV^2} .
\end{equation}
In fact the whole non-factorizable part receives only resonant contributions from soft-photon exchange and entirely results from the non-commutativity of the on-shell and soft-photon limits.
The explicit evaluation of the non-factorizable corrections proceeds by applying the so-called ``extended soft-photon approximation'' (ESPA).
It differs from the usual soft-photon approximation by keeping the exact dependence on the photon momentum $q$ in propagators intact where $q\to0$ would lead to further singularities, but setting $q=0$ in all other regular factors.
The complex mass $\mu_V$ is used in the gauge-boson propagators throughout, but the limits $p_V^2,\mu_V^2\to\MV^2$ are taken whenever they do not lead to divergences. In particular, the soft photon momentum cannot be neglected in the resonant gauge-boson propagators since the scalar product $(p_V\cdot q)$ is of the same order as the virtuality $(p_V^2-\mu_V^2)\sim\MV\GV$.
In the numerators, all photon momenta are negligible, 
reducing the occurrence of loop integrals to scalar integrals only.
The non-factorizable corrections factorize from the lower-order amplitude, as a direct consequence of the property of the corrections in the soft limit,
\begin{equation}
  \delta\M_{\Vew,\nf}^{\Paq_a\Pq_b\to\Pl_1\Pal_2}
  \,=\,\delta_{\Vew,\nf}^{\Paq_a\Pq_b\to\Pl_1\Pal_2}\, \M_{0,\PA}^{\Paq_a\Pq_b\to\Pl_1\Pal_2} .
\end{equation} 
Here the subscript $\PA$ on the lower-order matrix element indicates
that the non-resonant photon-exchange diagrams are not
included in case of the neutral-current Drell-Yan process.
Employing regularization in $D=4-2\epsilon$ dimensions, the relative correction factor is given by
\begin{equation}
  \label{eq:nf-virt}
  \begin{aligned}
    \delta_{\Vew,\nf}^{\Paq_a\Pq_b\to\Pl_1\Pal_2} 
    =& - \frac{\alpha}{2\pi}
    \sum_{\substack{i=a,b\\f=1,2}}\eta_i Q_i \eta_f Q_f \Biggl\lbrace
    2 +\Li_{2}\left(1+\frac{\MV^2}{t_{if}}\right) \\
    &\quad+\left[ \frac{c_\epsilon}{\epsilon} - 2\ln\left(\frac{\mu_V^2-s_{12}}{\mu \MV}\right) \right]
    \left[ 1-\ln\left(\frac{\MV^2}{-t_{if}}\right) \right]
    \Biggr\rbrace ,
  \end{aligned}
\end{equation}
where $c_\epsilon=(4\pi)^\epsilon\Gamma(1+\epsilon)$, $s_{12}=(k_1+k_2)^2$, $t_{if}=(p_i-k_f)^2$, and $\eta_i=1$ for incoming particles and outgoing antiparticles and $\eta_i=-1$ for incoming antiparticles and outgoing particles.  
For \PW production this result was already given in Ref.~\cite{Dittmaier:2001ay}.
Note that all collinear singularities cancel in the non-factorizable corrections and only a soft singularity remains.  
In case of neutral-current processes, all terms in Eq.~\eqref{eq:nf-virt} that do not depend on the indices $i$ or $f$ vanish due to charge conservation,
\begin{equation}
  Q_{V=\PZ} = \sum_{i=a,b}\eta_iQ_i = -\sum_{f=1,2}\eta_fQ_f=0.
\end{equation}

%%%%%%%%%%%%%%%%%%%%%%%%%%%%%%
\subsubsection{Real corrections}
\label{sec:real-nonfact}

In the previous section we have already encountered the subtlety in the separation of the factorizable from the non-factorizable virtual corrections in the case where the virtual photon is attached to the internal $V$ propagator.
The real corrections involve the additional complication that photon emission off a $V$ propagator leads to two $V$-boson resonances in phase space which overlap if the photon is not very hard, i.e.\ for $E_\Pgg=k^0\lesssim\GV$.
To disentangle the two adjacent resonance factors, the following identity can be used,
\begin{alignat}{3}
  \label{eq:prop-id}
  & \frac{1}{(p_V+k)^2-\mu_V^2}\cdot\frac{1}{p_V^2-\mu_V^2} 
  &\quad=\quad& \frac{1}{2 p_V\cdot k}\Biggl[ \frac{1}{p_V^2-\mu_V^2} 
  &\quad-\quad& \frac{1}{(p_V+k)^2-\mu_V^2} \Biggr], \\
  & \raisebox{-22pt}{\includegraphics{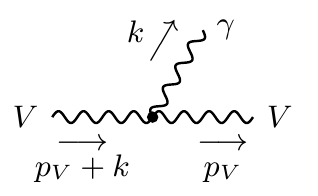}}
  &\quad=\quad& \raisebox{-5pt}{\includegraphics{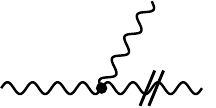}}
  &\quad+\quad& \raisebox{-5pt}{\includegraphics{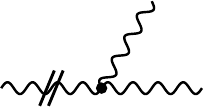}} . \nonumber
\end{alignat}
Note that the factor $2p_V\cdot k$ resembles a propagator denominator
for an on-shell $V$ boson, e.g.\ $2p_V\cdot k=(p_V+k)^2-\mu_V^2$ for
$p_V^2=\mu_V^2$ and $k^2=0$.
The double slash on a propagator line indicates which momentum is set on its mass shell in the rest of the diagram (but not on the slashed line itself).
Similar to the virtual corrections, the 
use of the real on-shell conditions  $p_V^2= M_V^2$ and $(p_V+k)^2= M_V^2$
in the \order{\alpha} corrections is possible in \order{\alpha} precision.
We will employ this in the following to avoid the unnecessary
complication by complex momenta.

\begin{figure}
  \centering
  $\left. \Vcenter{\includegraphics{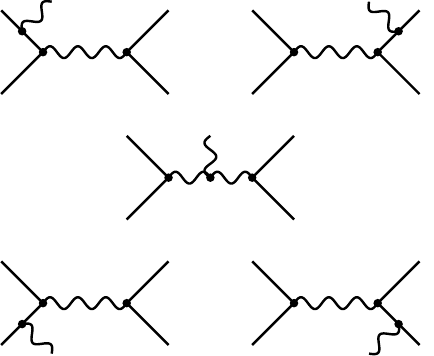}} \right\} 
  \longrightarrow
  \raisebox{-30pt}{\includegraphics{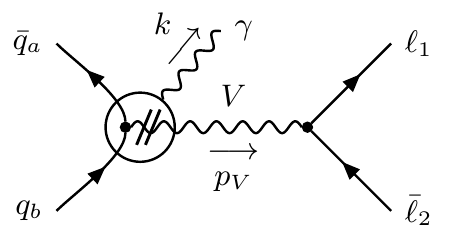}} + 
  \raisebox{-30pt}{\includegraphics{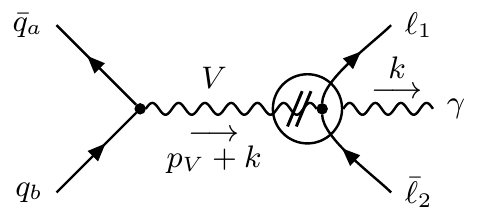}}$
  \caption{Real-emission diagrams divided into initial-state and final-state contributions, where the compact diagrammatic notation defined in Eq.~\eqref{eq:on-shell-blob} is used. Double lines on the $V$ propagator indicate on-shellness.}
  \label{fig:nlo-real}
\end{figure}

Figure~\ref{fig:nlo-real} illustrates the decomposition that is obtained by applying identity~\eqref{eq:prop-id} to the real-emission diagrams. 
Here we have introduced a compact diagrammatic notation, where a particle attached to a circle represents all diagrams where the particle is attached in all possible ways to the encircled subdiagram.
Furthermore, the double-line at the $V$ propagator inside the circle indicates that the $V$ boson should be considered on shell inside the blob, 
\begin{equation}
  \label{eq:on-shell-blob}
  \raisebox{-20pt}{\includegraphics{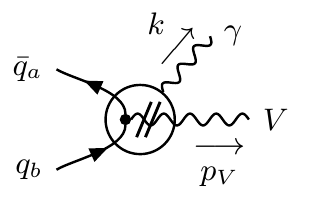}} \quad = \quad
  \raisebox{-12pt}{\includegraphics{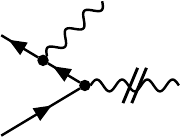}} \quad + \quad
  \raisebox{-22pt}{\includegraphics{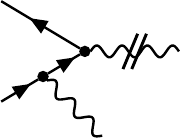}} \quad + \quad
  \raisebox{-12pt}{\includegraphics{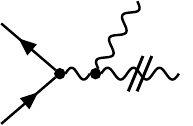}} , 
\end{equation}
where the last diagram corresponds to the contribution coming from the first term on the r.h.s.\ of Eq.~\eqref{eq:prop-id}.

\begin{figure}
  \centering
  \begin{subfigure}[m]{.48\linewidth}
    \centering
    $\left\vert\Vcenter{\includegraphics{images/diags/DY_NLO_Rfac_p_OS}}\right\rvert^2$
    \subcaption{a}{Factorizable initial-state corrections}
    %\label{fig:nlo-pa-real:pro}
  \end{subfigure}
  \begin{subfigure}[m]{.48\linewidth}
    \centering
    $\left\vert\Vcenter{\includegraphics{images/diags/DY_NLO_Rfac_d_OS}}\right\rvert^2$
    \subcaption{b}{Factorizable final-state corrections}
    %\label{fig:nlo-pa-real:dec}
  \end{subfigure}
  \\[1.5em]
  \begin{subfigure}[m]{.98\linewidth}
    \centering
    \includegraphics{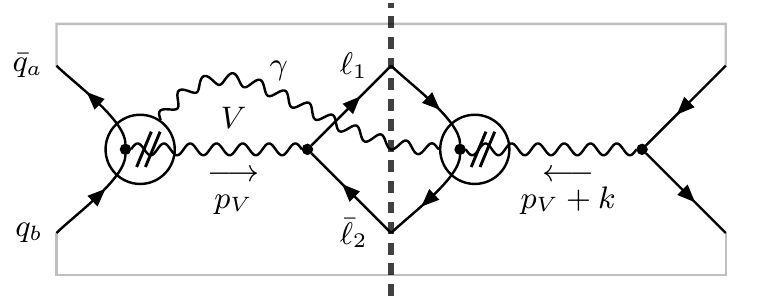}
    \subcaption{c}{Non-factorizable corrections}
    %\label{fig:nlo-pa-real:nf}
  \end{subfigure}
  \caption{Factorizable real photonic corrections to production~(a) and decay~(b), as well as non-factorizable real photonic corrections~(c).}
  \label{fig:nlo-pa-real}
\end{figure}

The \emph{factorizable} real corrections consist of the two separately gauge-invariant parts induced by photon emission during the $V$ production and $V$ decay. 
The respective squared amplitudes $\lvert\M_{\Rew,\fini}^{\Paq_a\Pq_b\to\Pl_1\Pal_2\Pgg}\rvert^2$ and $\lvert\M_{\Rew,\ffin}^{\Paq_a\Pq_b\to\Pl_1\Pal_2\Pgg}\rvert^2$ are derived from the on-shell matrix elements for vector-boson production with an additional photon emitted from the initial or final state,
\begin{align}
  \label{eq:nlo-fact-real:pro}
  \M_{\Rew,\fini}^{\Paq_a\Pq_b\to\Pl_1\Pal_2\Pgg}
  &=\sum_\lambda \frac{\M_{\Rew}^{\Paq_a\Pq_b\to V\Pgg}(\lambda)\;
  \M_0^{V\to \Pl_1\Pal_2}(\lambda)}{p_V^2-\mu_V^2} , \\
  \label{eq:nlo-fact-real:dec}
  \M_{\Rew,\ffin}^{\Paq_a\Pq_b\to\Pl_1\Pal_2\Pgg}
  &=\sum_\lambda \frac{\M_0^{\Paq_a\Pq_b\to V}(\lambda)\;
  \M_{\Rew}^{V\to\Pl_1\Pal_2\Pgg}(\lambda)}{(p_V+k)^2-\mu_V^2} ,
\end{align}
as illustrated in Figs.~\ref{fig:nlo-pa-real}(a),(b).
Similarly to the virtual case, a projection of the kinematics onto the on-shell phase space is necessary to render the subamplitudes gauge invariant.
Note that for the real corrections, however, two different projections of the $V$ momentum $p_V=k_1+k_2$ are required, with $p_V^2\to \MV^2$ and $(p_V^2+k)^2\to \MV^2$, for the corrections to the decay and the production subprocesses, respectively.
In the actual numerical evaluation of the initial- and final-state radiation effects we employ the dipole subtraction approach for photon radiation~\cite{Dittmaier:1999mb,Dittmaier:2008md} to separate soft and collinear singularities.
To this end, we had to adapt this formalism to cover decay kinematics, a subject that will be discussed in another publication.

The \emph{non-factorizable} corrections emerge from the interference terms illustrated in Fig.~\ref{fig:nlo-pa-real}(c).
This definition is analogous to the virtual case, where the non-factorizable corrections are defined through the difference of the full squared matrix element and the factorizable corrections.
For hard-photon emission the two $V$ propagators of the interference term shown in Fig.~\ref{fig:nlo-pa-real}(c) are widely separated and do not overlap, so that no resonant contribution results.
Thus, only the soft-photon region $k^0\lesssim\GV$ is relevant for resonant corrections, where the ESPA can be applied.
The real non-factorizable corrections then can be written as a correction to the Born cross section,
\begin{align}
  \label{eq:nlo-nf-real}
  \left\lvert\M_{\Rew}^{\Paq_a\Pq_b\to\Pl_1\Pal_2\Pgg}\right\rvert^2_\nf
  \;=&\; \delta_{\Rew,\nf}^{\Paq_a\Pq_b\to\Pl_1\Pal_2\Pgg} 
  \left\lvert\M_{0,\PA}^{\Paq_a\Pq_b\to\Pl_1\Pal_2}\right\rvert^2  \\
\intertext{with}
  \label{eq:nlo-nf-real-delta}
  \delta_{\Rew,\nf}^{\Paq_a\Pq_b\to\Pl_1\Pal_2\Pgg}
  \;=&\; -2 \Re\Bigl\lbrace \mathcal{J}_{\pro,\mu}(p_a,p_b,p_a+p_b)
  \bigl( \mathcal{J}_{\dec}^{\mu}(k_1,k_2) \bigr)^* \Bigr\rbrace .
\end{align}
The emission of the photon is described by the following modified eikonal currents~\cite{Denner:1997ia},
\begin{subequations}
\label{eq:eik}
\begin{align}
  \label{eq:eik-pro}
  \mathcal{J}_{\pro}^{\mu}(p_a,p_b,p_V)
  &= e \left[ Q_a\frac{p_a^\mu}{p_a\cdot k} -Q_b\frac{p_b^\mu}{p_b\cdot k} 
  -(Q_a-Q_b)\frac{p_V^\mu}{p_V\cdot k} \right], \\
  \label{eq:eik-dec}
  \mathcal{J}_{\dec}^{\mu}(k_1,k_2)
  &= e \left[ Q_1\frac{k_1^\mu}{k_1\cdot k} -Q_2\frac{k_2^\mu}{k_2\cdot k} 
    -(Q_1-Q_2)\frac{k_1^\mu+k_2^\mu}{(k_1+k_2)\cdot k} \right]
  \frac{(k_1+k_2)^2-\mu_V^2}{(k_1+k_2+k)^2 -\mu_V^2},
\end{align}
\end{subequations}
where the ratio of propagators in Eq.~\eqref{eq:eik-dec} is required to restore the correct momenta in the $V$ propagators of the interference term shown in Fig.~\ref{fig:nlo-pa-real}(c).

The actual phase-space integration of $\lvert\M_{\Rew}^{\Paq_a\Pq_b\to\Pl_1\Pal_2\Pgg}\rvert^2_\nf$ in Eq.~\eqref{eq:nlo-nf-real} can be performed in two different ways, which are equally good in PA accuracy.
One possibility, of course, is to evaluate the phase-space integral 
\begin{equation}
  \label{eq:lips-photon-exact}
  \int\rd\Phi_{\Pl_1\Pl_2\Pgg} = \int\frac{\rd^3\mathbf{k}_1}{2k_1^0(2\pi)^3}
  \int\frac{\rd^3\mathbf{k}_2}{2k_2^0(2\pi)^3}\int\frac{\rd^3\mathbf{k}}{2k^0(2\pi)^3}
  \;(2\pi)^4 \delta(p_a+p_b-k_1-k_2-k)
  \biggr\rvert_{\substack{k^0=\lvert\mathbf{k}\rvert, \\ k^0_i=\lvert\mathbf{k}_i\rvert}}
\end{equation}
without any approximation.
Within the PA, however, the photon momentum $k$ can be neglected in the $\delta$ function, i.e.\ the phase space can be factorized according to

\begin{align}
  \label{eq:lips-photon-fact}
  &\int\rd\Phi_{\Pl_1\Pl_2\Pgg} \to \int\rd\hat\Phi_{\Pl_1\Pl_2\Pgg}
  = \int\rd\Phi_{\Pl_1\Pl_2}\int\frac{\rd^3\mathbf{k}}{2k^0(2\pi)^3} \biggr\rvert_{k^0=\lvert\mathbf{k}\rvert} , \\
  &\int\rd\Phi_{\Pl_1\Pl_2} = 
  \int\frac{\rd^3\mathbf{k}_1}{2k_1^0(2\pi)^3}\int\frac{\rd^3\mathbf{k}_2}{2k_2^0(2\pi)^3}
  \;(2\pi)^4 \delta(p_a+p_b-k_1-k_2) \biggr\rvert_{k^0_i=\lvert\mathbf{k}_i\rvert} ,
\end{align}
leading to some technical simplifications.
Note that the upper limit of $k^0_\max$ of the photon energy $k^0$ is not fixed by the process kinematics anymore.
Reasonable values for $k^0_\max$ cover the whole range of photons with $k^0\lesssim\GV$, without introducing artificially large scales.
Although the introduction of $k^0_\max$ even cuts off an artificially created UV singularity at $k^0\to\infty$, it only enters suppressed non-resonant terms that are beyond PA accuracy.
In practice, we set $k^0_\max$ to some value within $(10-20)\times\GV$.
We have checked that both variants of integrating the non-factorizable corrections yield identical results with an accuracy below the $0.1\%$ level.

Similarly to the virtual non-factorizable correction, collinear singularities cancel when the non-factorizable corrections are integrated over the photon phase space, while a soft singularity remains. 
Here we regularize this singularity using soft slicing.
The integral over the photon momentum is split into two contributions according to $E_\Pgg<\Delta E$ and $E_\Pgg>\Delta E$, where $E_\Pgg=k^0$ denotes the energy of the photon in the partonic centre-of-mass frame. 
The cutoff has to be chosen much smaller than any relevant scale of the process, i.e.\ $\Delta E\ll\GV$.
The cross section with NLO non-factorizable corrections can then be written in the form
\begin{equation}
  \sighat_\nf^{\NLO_\rew}
  = \iint\limits_{2+\Pgg} \rd\sigma_\nf^{\Rew}
  + \int_2 \rd\sigma_\nf^{\Vew}
  = \iint\limits_{\substack{2+\Pgg\\E_\Pgg>\Delta E}} \rd\sigma_\nf^{\Rew}
  + \iint\limits_{\substack{2+\Pgg\\E_\Pgg<\Delta E}} \rd\sigma_\nf^{\Rew}
  + \int_2 \rd\sigma_\nf^{\Vew} .
\end{equation}
Here the symbol $\int_m$ denotes the integration over the $m$-particle phase space.
Below the cutoff ($E_\Pgg<\Delta E$), the photon momentum can be also neglected in the resonant $V$ propagators, so that the ratio of propagators in the decay current~\eqref{eq:eik-dec} cancels and the ESPA reduces to the usual eikonal approximation. 
The resulting integrals can be performed analytically using results from Refs.~\cite{Beenakker:1988bq,Denner:1991kt,Harris:2001sx},
\begin{align}
  \label{eq:nf-slice}
  \delta_{\soft}^{\Paq_a\Pq_b\to\Pl_1\Pal_2} (\Delta E)
  =& - \mu^{2\epsilon} \int\limits_{\mathclap{\lvert\mathbf{k}\rvert<\Delta E}}
  \frac{\rd^{D-1}\mathbf{k}}{(2\pi)^{D-1}2k^0}  \biggr.
  \;2 \Re\left\lbrace \mathcal{J}_{\pro,\mu}
  \left( \mathcal{J}_{\dec}^{\mu} \right)^* \right\rbrace 
  \biggr\rvert_{k^0=\lvert\mathbf{k}\rvert} \nonumber\\
  =&\;\frac{\alpha}{\pi}
  \sum\limits_{\substack{i=a,b\\f=1,2}}\eta_i Q_i \eta_f Q_f
  \Biggl\lbrace 2 +\Li_{2}\left(1+\frac{\MV^2}{t_{if}}\right) \nonumber\\
  &\quad+\left[ \frac{c_\epsilon}{\epsilon} -2\ln\left(\frac{2\Delta E}{\mu}\right) \right]
  \left[ 1-\ln\left(\frac{\MV^2}{-t_{if}}\right) \right] 
  \Biggr\rbrace .
\end{align}

We then obtain our final form for the non-factorizable corrections to the cross section
\begin{equation}
  \sighat_\nf^{\NLO_\rew}
  = \iint\limits_{\substack{2+\Pgg\\E_\Pgg>\Delta E}} \rd\sigma^0_\PA
  \; \delta_{\Rew,\nf}^{\Paq_a\Pq_b\to\Pl_1\Pal_2\Pgg}
  + \int_2 \rd\sigma^0_\PA \left[ 
  2\Re\left\{\delta_{\Vew,\nf}^{\Paq_a\Pq_b\to\Pl_1\Pal_2}\right\}
  + \delta_{\soft}^{\Paq_a\Pq_b\to\Pl_1\Pal_2}(\Delta E) \right] ,
\end{equation}
where the subscript PA in the leading-order cross section indicates the omission of non-resonant diagrams.
The poles in $\epsilon$ cancel between the virtual non-factorizable corrections and the integrated soft slicing terms.
The remaining integral over the real-photon phase space with the cutoff $E_\Pgg>\Delta E$ is performed numerically. 

The photon-induced processes $\Pgg\Pq_b\to\Pl_1\Pal_2\Pq_a$,
$\Pgg\Paq_a\to\Pl_1\Pal_2\Paq_b$, which appear in
\order{\alpha} as well, and the process
$\Pgg\Pgg\to\Pl_1\Pal_2$, which appears for the neutral-current
process already at LO, were not considered here.  As discussed in
Refs.~\cite{CarloniCalame:2007cd,Brensing:2007qm,Dittmaier:2009cr}
they deliver only small corrections at \order{\alpha}. Therefore their
contributions in \order{\alphas\alpha} can be safely neglected.
However, applying a PA would not lead to any simplification
here.  Note that the photon-induced processes do not contribute to the
non-factorizable corrections.

%%%%%%%%%%%%%%%%%%%%%%%%%%%%%%
\subsubsection{Effective-field-theory inspired approach}
\label{sec:nlo-eft}

Alternatively to the method discussed in
Sect.~\ref{sec:virt-nonfact} we  also employ an approach that is
inspired by the construction of an EFT for
unstable particle production~\cite{Beneke:2003xh,Beneke:2004km}, which
can be viewed as an application of the method of
regions~\cite{Beneke:1997zp}. 
We discuss this method here for the computation of the virtual
corrections, whereas the factorizable and
non-factorizable real corrections will be defined in the same way
as in Sect.~\ref{sec:real-nonfact}.%
\footnote{See also Ref.~\cite{Falgari:2013gwa} for a similar application of the EFT approach to the non-factorizable corrections to top-quark production at hadron colliders.} 
In the EFT approach, loop integrals contributing to the process~\eqref{eq:dy} near resonance are split into contributions from different momentum regions. The regions contributing in the case at hand are characterized by \emph{hard}, \emph{soft}, or \emph{collinear} photon momenta.  The corresponding momentum scalings
in the centre-of-mass frame are:
\begin{equation}
  \label{eq:momentum-scalings}
  \begin{aligned}
  \text{hard } (h) &: \quad q^0\sim |\mathbf{q}| \sim \MV\,, \\
  \text{soft } (s)&:  \quad q^0\sim |\mathbf{q}|\sim  \GV\,, \\
  \text{$n$-collinear } (c) &:   \quad q_- \sim \MV,\, q_+\sim \GV,\,
  q_\perp \sim \sqrt{\GV \MV}\,, \\
  \text{$\bar n$-collinear } (\bar c) &: \quad q_+ \sim \MV,
  \, q_-\sim \GV,\, q_\perp \sim \sqrt{\GV \MV} \,.
  \end{aligned}
\end{equation}
For the collinear momenta the light-cone decomposition
\begin{equation}
\label{eq:light-cone}
  q^\mu=\frac{q_-}{2}\,n^\mu+\frac{q_+}{2}\,\bar n^\mu+q_\perp^\mu
\end{equation}
with two light-like vectors $n$ and $\bar n$ satisfying $n\cdot \bar
n=2$ and $q_\perp\cdot n=q_\perp\cdot\bar n=0$ have been introduced. 
We align the light-like vectors in the
decomposition~\eqref{eq:light-cone} with
the momenta of the incoming partons, $p^\mu_a=\sqrt{\hat s}\, n^\mu/2$
and $p^\mu_b=\sqrt{\hat s}\, \bar n^\mu/2$.  For the decay products of
the $V$ boson two analogous collinear regions have to be introduced
which we have not explicitly written down.

\begin{figure}
  \centering
  \includegraphics{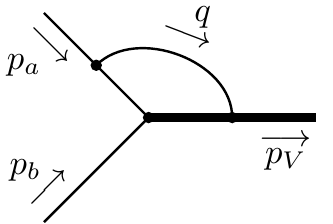}
  \caption{Sample one-loop scalar integral leading to hard and
    soft contributions in the EFT approach. The momentum $p_V$ is
    slightly off shell with virtuality $|p_V^2-\mu_V^2|\sim
    \MV\GV$.}
  \label{fig:nlo-eft}
\end{figure}

In the full-fledged EFT treatment~\cite{Beneke:2003xh,Beneke:2004km},
an effective Lagrangian is constructed for the resonant $V$ particle,
the collinear modes, and the soft photons. Particles with hard momenta
(highly-energetic photons, non-resonant $V$ bosons, other massive
particles) are integrated out and do not represent dynamical degrees of 
freedom in the EFT.
Their effect is included through the Wilson coefficients of the operators describing
$V$ production and decay. Here we will not make use of an effective
Lagrangian, but rather compute the contributions of the different
momentum regions in Eq.~\eqref{eq:momentum-scalings} to loop integrals
directly using the expansion by regions. In this method, the
integrands of the loop integrals are expanded up to the desired order
in the ratio $\GV/\MV$ assuming the loop momentum is in the hard,
soft, or collinear region. The expansion of the full loop integral is
obtained by adding the contributions from the different regions,
integrated over the whole range of the loop momentum in dimensional
regularization.  This procedure creates artificial IR and UV
divergences in the integrals over the different regions, which cancel
in the sum.  The EFT method is closely related to the PA and, in fact,
the leading contributions from the hard (soft) region reproduce
the factorizable (non-factorizable) corrections defined in
Sect.~\ref{sec:nlo-nonfact}, while the collinear regions do not
contribute at NLO.  As an example consider the following scalar vertex
integral (see Fig.~\ref{fig:nlo-eft}),
\begin{equation}
  I= \frac{(2\pi\mu)^{4-D}}{\ri\pi^2}
  \int \rd^D q\, \frac{1}{(q^2+\ri0)
  \left[(p_V-q)^2-\mu_V^2\right]
  \left[(p_a-q)^2+\ri0\right]}.
\end{equation}

%%%%%%%%%%
\paragraph{Hard region} In the hard region, the virtuality of the
vector boson is much smaller than the loop momentum,
$|p_V^2-\mu_V^2|\ll |q^2|\, , |p_V\cdot q|$. The expansion in the hard
region therefore induces a Taylor expansion of the vector-boson
propagator,
\begin{equation}
  \label{eq:hard-exp}
  \frac{1}{(p_V-q)^2-\mu_V^2}
  = \frac{1}{\left(q^2-2p_V\cdot q\right)}- \frac{(p_V^2-\mu_V^2)}{
  \left(q^2-2p_V\cdot q\right)^2} +\dots,
\end{equation}
while no simplification can be made in the other propagators.  The leading term in the hard region is therefore given by the integral evaluated for an  on-shell  vector boson: 
\begin{equation}
  I_{\text{hard}}= \frac{(2\pi\mu)^{4-D}}{\ri\pi^2} \int \rd^D q \left.\frac{1}{(q^2+\ri0)
  \left(q^2-2p_V\cdot q \right)
  \left[(p_a-q)^2+\ri0\right]}\right|_{p_V^2=\MV^2}.
\end{equation}
As for the factorizable corrections in the pole
decomposition~\eqref{eq:PA}, gauge invariance demands that
``on-shell'' is defined with respect to the complex pole
$p_V^2=\mu_V^2$. However, in the hard region the decay width can be
treated perturbatively so that at NLO accuracy the complex pole
location has been replaced by its real part, $\mu_V^2\to
\MV^2$. Therefore the result for the leading term of the expansion
of the integral in the hard region precisely coincides with the
factorizable contribution in the terminology of
Sect.~\ref{sec:virt-nonfact}.  The subleading terms in the
expansion in Eq.~\eqref{eq:hard-exp} contribute to the non-resonant
terms $N(p_V^2)$ in the pole decomposition~\eqref{eq:PA}, which we do
not consider in this work.

%%%%%%%%%%
\paragraph{Soft region} 
In the soft region, the components of the loop momentum are of the same order as the virtuality of the vector boson, $q^0\sim |\mathbf{q}|\sim (p_V^2-\MV^2)/\MV$. Expanding the integral to leading order in the soft region yields the result
\begin{equation}
  \label{eq:soft-int}
  I_{\text{soft}}= \frac{(2\pi\mu)^{4-D}}{\ri\pi^2}
  \int \rd^D q\, \frac{1}{(q^2+\ri0)
  \left[- 2 (p_V\cdot q)+p_V^2-\mu_V^2 \right]
  \left[-2(p_a\cdot q)+\ri0\right]}.
\end{equation}
In the soft region it is mandatory to resum the decay width in the propagator, since it is of the same order as the soft  loop momentum. 
Applying the soft expansion to the full diagrams including the numerator structure, the usual soft-photon manipulations show that the soft contributions factorize from the lower-order diagrams  and are given by the same expressions as in the ESPA described in Sect.~\ref{sec:virt-nonfact}.

In the computation of the resulting loop integrals it is useful to combine the eikonal propagators $\ri/(p_i\cdot q)$ with the remaining propagators by successively employing a modified Feynman para\-met\-ri\-zation, 
\begin{equation}
  \frac{1}{a_1^{m_1}a_2^{m_2}}=
  \frac{\Gamma(m_1+m_2)}{\Gamma(m_1)\Gamma(m_2)}\int_0^\infty \rd x\, 
\frac{x^{m_2-1}}{(a_1+a_2x)^{m_1+m_2}}.
\end{equation}
The resulting integrals can then be performed using standard methods.
In the above example one obtains
\begin{align}
  I_{\text{soft}}&=
  \frac{(2\pi\mu)^{4-D}}{\ri\pi^2}
  \int_0^\infty \rd x\, \int_0^\infty \rd y \int \rd^D q 
  \frac{y}{\left[q^2 -2y q\cdot (p_a+x p_V)+yx (p_V^2-\mu_V^2)\right]^3}
  \nonumber\\
  &=c_\epsilon  \frac{\Gamma(-\epsilon)\,\Gamma(2\epsilon) }{\Gamma(1+\epsilon)}
  \frac{1}{ -2(p_a\cdot p_V )}\left(\frac{\mu_V^2-p_V^2}{p_V^2}\right)^{-2\epsilon}
  \left(\frac{p_V^2}{\mu^2}\right)^{-\epsilon}.
\end{align}
All the integrals appearing in the calculation of the soft corrections
to the Drell--Yan type processes can be found, e.g., in Appendix~D of
Ref.~\cite{Falgari:2009zz}. The sum of all virtual soft-photon
corrections then leads to the same result as for the non-factorizable
corrections given in Eq.~\eqref{eq:nf-virt}.  The split of the diagram in
Fig.~\ref{fig:nlo-eft} into hard and soft contributions therefore
corresponds to the decomposition of not manifestly non-factorizable
diagrams into factorizable and non-factorizable contributions,
as illustrated in Fig.~\ref{fig:nlo-nf-virt}(b).

%%%%%%%%%%
\paragraph{Collinear regions}
Expanding the integrand to leading order in a $n$-collinear loop momentum results in the integral
\begin{equation}
  I_{c}= \frac{(2\pi\mu)^{4-D}}{2\,\ri\pi^2}
  \int
  \frac{\rd q_+\,\rd q_-\, \rd^{D-2} q_\perp}{(q_-q_++q_\perp^2+\ri0)
  \left(-p_{V+}q_-+\ri0\right)
  \left[(q_--p_{a-})q_++q_\perp^2+\ri0\right]}.
\end{equation}
This integral vanishes, as can be seen as follows. 
Both poles in $q_+$
lie in the same half-plane unless $0<q_-<p_{a-}$. In this case, closing the integral contour in the lower half of the complex $q_+$ plane picks up the pole at  $q_+=-\frac{q_\perp^2+\ri\epsilon}{q_-}$. In the resulting expression the $q_\perp^2$ integral is scaleless so that the contribution from the $n$-collinear region  vanishes. 

Expanding the integral in the $\bar n$-collinear region leads to the expression
\begin{equation}
  \label{eq:nbar-coll}
  I_{\bar c}= \frac{(2\pi\mu)^{4-D}}{2\,\ri\pi^2}
  \int  \frac{\rd q_+\,\rd q_-\,\rd^{D-2} q_\perp}{(q^2+\ri0)
  \left(-p_{V-}q_++\ri0\right)\left(p_{a-}q_++\ri0\right)}.
\end{equation}
Again the integral is scaleless and vanishes in dimensional
regularization.  The same is true for all other collinear diagrams for
external on-shell massless particles.\footnote{Note that the
introduction of light-fermion masses would introduce further
momentum regions instead of modifying the collinear integrals, see
e.g.\ Ref.~\cite{Beneke:2007zg}.}  Since only the soft and hard
regions give non-vanishing contributions, the soft corrections agree
with the non-factorizable corrections in the definition of
Sect.~\ref{sec:virt-nonfact} as difference of the full integral and
the factorizable (hard) corrections. In the computation of the
two-loop \order{\alphas\alpha} corrections and the one-loop
QCD corrections to the process with real-photon emission we will
obtain non-vanishing collinear contributions from individual diagrams
which, however, cancel after summing all diagrams (see
Paragraph~\ref{sec:vv-eft} in Sect.~\ref{sec:VV}).

%%%%%%%%%%%%%%%%%%%%%%%%%%%%%%%%%%%%%%%%%%%%%%%%%%
\subsection{QCD corrections}
\label{sec:QCD}

Since the NLO QCD corrections to process~\eqref{eq:dy} will also appear as building blocks in the calculation of non-factorizable \order{\alphas\alpha} corrections below, we here review their calculation and recite the well-known results.
The virtual QCD corrections to vector-boson production cross sections are given by 
\begin{equation}
  \label{eq:sig-qcd}
  \rd\sigma_{\Paq_a\Pq_b}^{\Vs}
  =2\Re\left\{\delta_{\Vs}^{\Paq_a\Pq_b\to\Pl_1\Pal_2} (s_{ab}) \right\}\,\rd\sigma_{\Paq_a\Pq_b}^0 ,
\end{equation}
with the correction factor
\begin{equation}
  \label{eq:nlo-qcd}
  \begin{aligned}
  \delta_{\Vs}^{\Paq_a\Pq_b\to\Pl_1\Pal_2}(s_{ab})
  =&-\frac{\alphas}{2\pi}C_\rF
  \biggl\lbrace\,\frac{c_\epsilon}{\epsilon^2} + \frac{c_\epsilon}{\epsilon}
  \left[\ln\left(\frac{\mu^2}{-s_{ab}-\ri0}\right) + \frac{3}{2}\right] \\
  &\quad + \frac{1}{2}\ln^2\left(\frac{\mu^2}{-s_{ab}-\ri0}\right) 
  + \frac{3}{2}\ln\left(\frac{\mu^2}{-s_{ab}-\ri0}\right) 
  -\frac{\pi^2}{6} + 4\biggr\rbrace ,
  \end{aligned}
\end{equation}
with $s_{ab}=(p_a+p_b)^2$.
In addition to the quark--antiquark induced processes, new gluon-induced channels become available in the real-correction contributions:
\begin{subequations}
\label{eq:rQCD}
\begin{align}
  \Paq_a(p_a) \,+\, \Pq_b(p_b) &\,\to\, \Pl_1(k_1) \,+\, \Pal_2(k_2) \,+\, \Pg(k_\Pg) ,\\
  \Pg(p_\Pg) \,+\, \Pq_b(p_b) &\,\to\, \Pl_1(k_1) \,+\, \Pal_2(k_2) \,+\, \Pq_a(k_a) ,\\
  \Pg(p_\Pg) \,+\, \Paq_a(p_a) &\,\to\, \Pl_1(k_1) \,+\, \Pal_2(k_2) \,+\, \Paq_b(k_b) .
\end{align}
\end{subequations}

To regularize the phase-space integral of the real corrections we employ the dipole subtraction formalism~\cite{Catani:1996vz}.
In this framework, the hard partonic cross section can be schematically written as
\begin{equation}
\label{eq:nlo-dipole}
  \begin{aligned}
    \sighat^{\NLO_\rs}
    =&\int_3\rd\sigma^{\Rs} + \int_2\rd\sigma^{\Vs} + \int_2\rd\sigma^{\Cs} \\
    =&\int_3\;\biggl[ \left(\rd\sigma^{\Rs}\right)_{\epsilon=0}  
    - \left(\rd\sigma^{\As}\right)_{\epsilon=0} \biggr] 
    + \int_2\;\biggl[ \rd\sigma^{\Vs} + \rd\sigma^{\Cs}
    + \int_1\rd\sigma^{\As} \biggr]_{\epsilon=0} .
  \end{aligned}
\end{equation}
Here, $\rd\sigma^{\Rs}$ comprises the real corrections to the cross section and the quantity $\rd\sigma^{\Cs}$ is the collinear counterterm that is subtracted from the NLO cross section to absorb collinear divergences into the NLO PDFs.
The subtraction term $\rd\sigma^{\As}$ is constructed from the LO amplitudes and the dipole operators $\CSV$, which induce further colour and helicity correlations. 
The dipole operators can be integrated analytically over the singular one-particle subspace, which in turn cancels all the IR divergences in the virtual corrections $\rd\sigma^{\Vs}$ and $\rd\sigma^{\Cs}$.
As a result, the cross section can be written in the form
\begin{align}
 \label{eq:dipole-nlo}
  \sighat^{\NLO_\rs}
  =& \int_3\;\biggl[ \biggl(\rd\sigma^{\Rs}\biggr)_{\epsilon=0}  
  - \biggl(\sum_\text{dipoles}\;\rd\sigma^0\!\otimes\!\CSV\biggr)_{\epsilon=0} \biggr] \nonumber\\
  &+\int_2\;\biggl[ \rd\sigma^{\Vs}
  + \rd\sigma^0\!\otimes\!\CSI \biggr]_{\epsilon=0}  
  + \int_0^1\rd x\int_2\;\biggl[\rd\sigma^0\!\otimes\!\left(\CSK+\CSP\right)\biggr]_{\epsilon=0} ,
\end{align}
where $\otimes$ encodes the possible colour and helicity correlations
between $\rd\sigma^{0}$ and the dipole operators, and the integration
over $x$ corresponds to a convolution over the momentum fraction
carried away by collinear parton emission of the initial state.  The
insertion operators $\CSI$, $\CSK$, and $\CSP$ emerge from the
collinear counterterm and the integrated dipoles.  More details and
the explicit expressions for these insertion operators can be found in
Ref.~\cite{Catani:1996vz}. Using this formalism, the
contributions of the virtual corrections, Eq.~\eqref{eq:sig-qcd}, and
the real-correction subprocesses in Eq.~\eqref{eq:rQCD} to the
partonic NLO QCD cross section read
\begin{subequations}
\label{eq:dipole-nlo-ch}
\begin{align}
  %%%%%
  \label{eq:dipole-nlo-qq}
  \sighat_{\Paq_a\Pq_b}^{\NLO_\rs}(p_a,p_b)
  =& \int_3\;\biggl[ \rd\sigma_{\Paq_a\Pq_b}^{\Rs}(p_a,p_b)  
  - \rd\sigma_{\Paq_a\Pq_b}^0\bigl(\widetilde\Phi_{2,(\Paq_a\Pg)\Pq_b}\bigr)
  \!\otimes\!\CSV^{\Paq_a, \Paq_a} 
  - \rd\sigma_{\Paq_a\Pq_b}^0\bigl(\widetilde\Phi_{2,(\Pq_b\Pg)\Paq_a}\bigr)
  \!\otimes\!\CSV^{\Pq_b, \Pq_b} \biggr] \nonumber\\
  &+ \int_2\;\biggl[ \rd\sigma_{\Paq_a\Pq_b}^{\Vs}(p_a,p_b)
  + \rd\sigma_{\Paq_a\Pq_b}^0(p_a,p_b)\!\otimes\!\CSI \biggr] \nonumber\\
  &+\int_0^1\rd x\int_2\rd\sigma_{\Paq_a\Pq_b}^0(xp_a,p_b)\!\otimes\! 
  \left(\CSK+\CSP\right)^{\Paq_a,\Paq_a}\nonumber\\
  &+\int_0^1\rd x\int_2\rd\sigma_{\Paq_a\Pq_b}^0(p_a,xp_b)\!\otimes\!
  \left(\CSK+\CSP\right)^{\Pq_b,\Pq_b}  , \\
  %%%%%
  \label{eq:dipole-nlo-gq}
  \sighat_{\Pg\Pq_b}^{\NLO_\rs}(p_\Pg,p_b)
  =& \int_3\;\biggl[ \rd\sigma_{\Pg\Pq_b}^{\Rs}(p_\Pg,p_b)
  - \rd\sigma_{\Paq_a\Pq_b}^0\bigl(\widetilde\Phi_{2,(\Pg\Paq_a)\Pq_b}\bigr)
  \!\otimes\!\CSV^{\Pg,\Paq_a}\biggr] \nonumber\\
  &+\int_0^1\rd x\int_2\rd\sigma_{\Paq_a\Pq_b}^0(xp_\Pg,p_b)\!\otimes\!
  \left(\CSK+\CSP\right)^{\Pg,\Paq_a} , 
\end{align}
\end{subequations}
and an analogous expression for the gluon--antiquark induced channel.
The dipole phase space $\widetilde\Phi_{n,(ai)b}(\widetilde p_{ai}^\mu,\widetilde p_b;\widetilde k_1,\widetilde k_2)$ is constructed from the real-emission phase space $\Phi_{n+1}(p_a,p_b;k_1,k_2,k_i)$ as follows,
\begin{subequations}
\label{eq:dipole_ii}
  \begin{align}
    \label{eq:dipole_ii_a}
    \widetilde p_{ai}^\mu &= x_{i,ab} \; p_a^\mu ,
    & x_{i,ab} &= \frac{p_a\cdot p_b-k_i\cdot p_a-k_i\cdot p_b}{p_a\cdot p_b} ,\\
    \label{eq:dipole_ii_b}
    \widetilde p_b^\mu &= p_b^\mu ,
    & \widetilde k_j^\mu &= \Lambda^\mu{}_\nu \; k_j^\nu \quad (j=1,2) ,
  \end{align}
\end{subequations}
where $k_i$ is the momentum of the additional gluon or (anti)quark in the final state.
The explicit Lorentz transformation matrix $\Lambda^\mu{}_\nu$ for the final-state momenta can be found in Ref.~\cite{Catani:1996vz}.
Note that the dipole kinematics corresponds to a collinear splitting of an initial-state parton $ai(p_a)\rightarrow a(x_{i,ab}p_a) + i((1-x_{i,ab})p_a)$ with $i=\Pg,\Pq,\Paq$ and that its partonic rest frame for the momentum $\widetilde p_{ai}+\widetilde p_b$ is Lorentz boosted along the beam axis with respect to the rest frame of the real-emission kinematics with the momentum $p_a+p_b$.

%%%%%%%%%%%%%%%%%%%%%%%%%%%%%%%%%%%%%%%%%%%%%%%%%%
\subsection{Numerical results at NLO}
\label{sec:nlo-num}

In this section we present the numerical results for the NLO corrections to the Drell--Yan process at the LHC for a centre-of-mass energy of $\sqrt{s}=14~\TeV$, where we restrict ourselves to the discussion of the two specific processes 
\begin{align}
  \label{eq:nlo-num-procs}
  \Pp+\Pp \;\to&\; \PWp \;\to\; \Pgngm+\Pgmp , \\
  \Pp+\Pp \;\to&\; \PZ \;(\Pgg^*) \;\to\; \Pgmm+\Pgmp ,
\end{align}
with muons in the final state.
We compare the results obtained by the full NLO EW calculation with those obtained by applying the PA as described in the previous section.

%%%%%%%%%%%%%%%%%%%%%%%%%%%%%%
\subsubsection{Input parameters and setup}
\label{sec:input}

For the numerical evaluation we use the following set of input parameters%
\footnote{Note that we take the experimental values of the \PW- and \PZ-boson widths as input parameters instead of calculating the decay widths in the respective order.}
\begin{equation}
\label{eq:params}
\begin{aligned}
  G_\mu \;=&\; 1.1663787\times 10^{-5} ~\GeV^{-2} ,
  &\alphas(\MZ) \;=&\; 0.119 , \\
  \MW^\OS \;=&\; 80.385~\GeV ,
  &\GW^\OS \;=&\; 2.085~\GeV , \\
  \MZ^\OS \;=&\; 91.1876~\GeV ,
  &\Gamma_\PZ^\OS \;=&\; 2.4952~\GeV , \\
  M_\PH \;=&\; 125.9~\GeV , 
  &m_\Pqt \;=&\; 173.07~\GeV ,
\end{aligned}
\end{equation}
which essentially follow Ref.~\cite{Beringer:1900zz}.
We employ the $G_\mu$ scheme, where the electroweak coupling constant is derived from the Fermi constant via the relation
\begin{equation}
  \label{eq:G_mu-scheme}
  \alpha_{G_\mu} \;=\; \frac{\sqrt{2}}{\pi} G_\mu \MW^2 
  \left( 1-\frac{\MW^2}{\MZ^2} \right) ,
\end{equation}
which avoids large logarithms of the light fermion masses induced by the running of the coupling constant $\alpha(Q^2)$ from the Thomson limit ($Q^2=0$) to the electroweak scale ($Q^2\sim\MZ^2$).
The masses of the light quark flavours (\Pqu, \Pqd, \Pqc, \Pqs, \Pqb) and of the leptons are neglected throughout, with the only exception in case of non-collinear-safe observables, where the final-state collinear singularity is regularized by the finite physical mass of the muon
\begin{equation}
  \label{eq:muon-mass}
  m_\Pgm \;=\; 105.658369~\MeV .
\end{equation}
The on-shell masses and widths of the gauge bosons in
Eq.~\eqref{eq:params} are converted to the corresponding pole masses
and widths via the following
relation~\cite{Bardin:1988xt,Sirlin:1991fd,Beenakker:1996kn},
\begin{equation}
  \label{eq:os-to-pole}
  \MV\;=\;\frac{\MV^\OS}{c_V} , \qquad 
  \GV\;=\;\frac{\GV^\OS}{c_V} , \qquad
  c_V\;=\;\sqrt{ 1+\left(\frac{\GV^\OS}{\MV^\OS}\right)^2 } .
\end{equation}
The CKM matrix is chosen diagonal in the third generation and the mixing between the first two generations is parametrized by the following values for the entries of the quark-mixing matrix,
\begin{equation}
  \label{eq:ckm}
  \lvert V_{\Pqu\Pqd} \rvert \,=\, 
  \lvert V_{\Pqc\Pqs} \rvert \,=\, 0.974, \qquad
  \lvert V_{\Pqc\Pqd} \rvert \,=\, 
  \lvert V_{\Pqu\Pqs} \rvert \,=\, 0.227. 
\end{equation}
For the PDFs we consistently use the NNPDF2.3 sets~\cite{Ball:2012cx}, where the NLO corrections are evaluated using the NNPDF2.3QED NLO set~\cite{Ball:2013hta}, which also includes \order{\alpha} corrections.
The value of the strong coupling $\alphas(\MZ)$ quoted in Eq.~\eqref{eq:params} is dictated by the choice of these PDF sets.
The renormalization and factorization scales are set equal, with a fixed value given by the respective gauge-boson mass, 
\begin{equation}
  \label{eq:scale}
  \mur \;=\; \muf \;\equiv\; \mu \;=\; \MV ,
\end{equation}
of the process under consideration.

%%%%%%%%%%%%%%%%%%%%%%%%%%%%%%
\subsubsection{Phase-space cuts and event selection}
\label{sec:cuts}

For the experimental identification of the Drell--Yan process we impose the following cuts on the charged leptons
\begin{equation}
  \label{eq:cut-lep}
  p_{\rT,\Plpm} > 25~\GeV , \quad
  \lvert \eta_{\Plpm} \rvert < 2.5 , 
\end{equation}
and additionally
\begin{equation}
  \label{eq:cut-miss}
  E_\rT^\miss > 25~\GeV ,
\end{equation}
which is only relevant in case of the charged-current process.
For the neutral-current process we further require the cut on the invariant mass of the lepton pair
\begin{equation}
  \label{eq:cut-mll}
  M_{\Pl\Pl} > 50~\GeV ,
\end{equation}
in order to avoid the photon pole at $M_{\Pl\Pl}\to0$.

The event selection described so far is not collinear safe with respect to the emission of photons from the charged leptons and will in general lead to corrections that involve large logarithms of the small lepton mass.
While the experimental isolation of a collinear lepton--photon configuration is feasible in case of muons (``bare muons''), for electrons such collinear configurations need to be treated inclusively, as can be achieved by so-called photon recombination.
In this paper we show only results on bare muons, but our results hold for the case of photon recombination in an analogous way.

%%%%%%%%%%%%%%%%%%%%%%%%%%%%%%
\subsubsection{Results}
\label{sec:nlo-result}

\begin{figure}
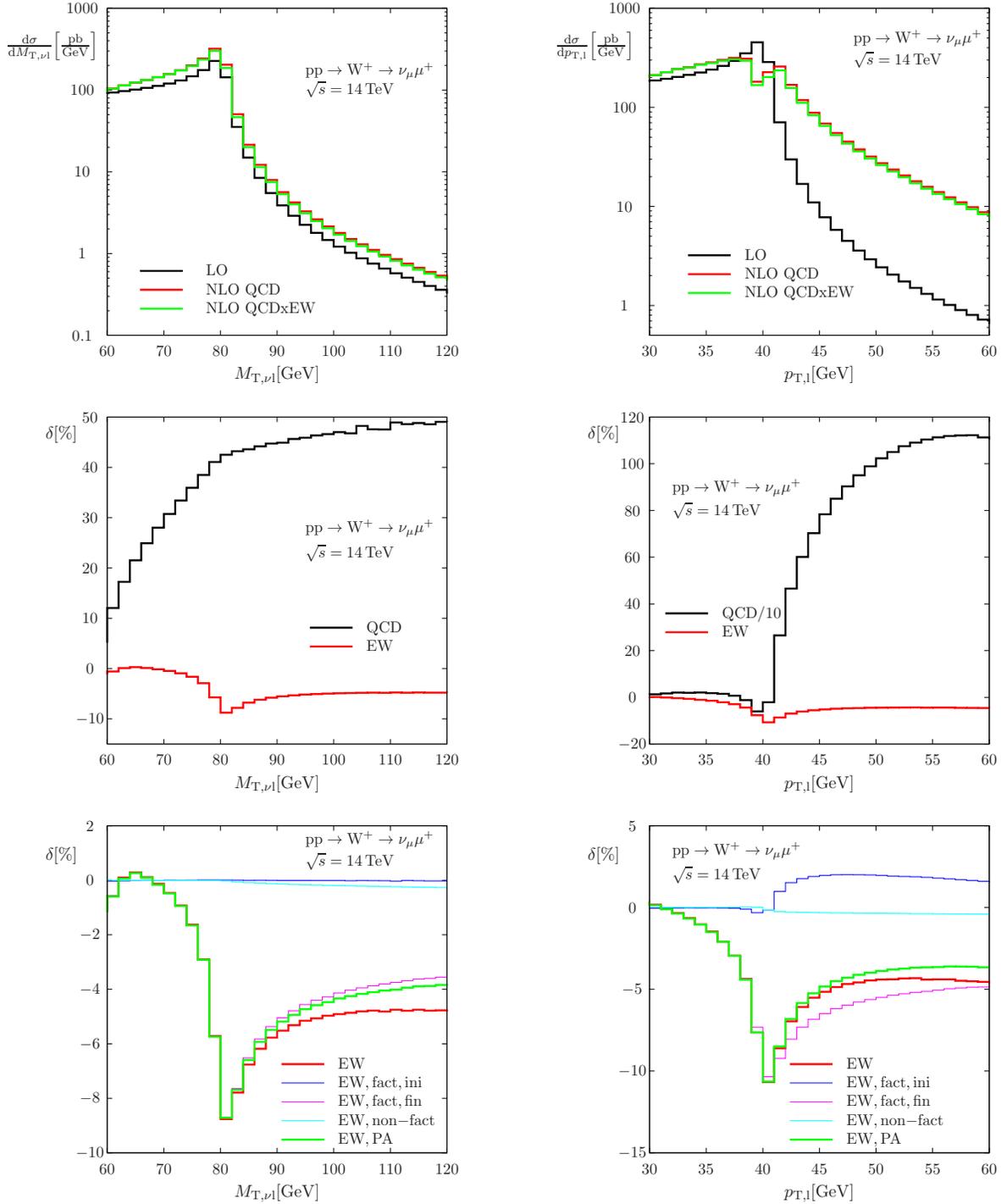

  \centering
  \begin{tabular}{r@{\hspace{1.5cm}}r}
    \includegraphics[scale=0.7]{{{images/plots/Wp.mtll.abs}}}
    &\includegraphics[scale=0.7]{{{images/plots/Wp.ptl.abs}}} \\[.5em]
    \includegraphics[scale=0.7]{{{images/plots/Wp.mtll.rel}}}
    &\includegraphics[scale=0.7]{{{images/plots/Wp.ptl.rel}}} \\[.5em]
    \includegraphics[scale=0.7]{{{images/plots/Wp.mtll.rel.PA}}} 
    &\includegraphics[scale=0.7]{{{images/plots/Wp.ptl.rel.PA}}} 
  \end{tabular}
  \caption{Distributions in the transverse-mass~(left) and transverse-lepton-momentum~(right)
  for \PWp production at the LHC, with the upper plot showing the absolute distributions,
  the middle plots the full relative NLO QCD and EW corrections, and the lower plots the relative NLO 
  EW corrections in PA broken up into their factorizable and non-factorizable parts.}
  \label{fig:distNLO-W}
\end{figure}
\begin{figure}
  \centering
  \begin{tabular}{r@{\hspace{1.5cm}}r}
    \includegraphics[scale=0.7]{{{images/plots/Z.mll.abs}}}
    &\includegraphics[scale=0.7]{{{images/plots/Z.ptl.abs}}} \\[.5em]
    \includegraphics[scale=0.7]{{{images/plots/Z.mll.rel}}}
    &\includegraphics[scale=0.7]{{{images/plots/Z.ptl.rel}}} \\[.5em]
    \includegraphics[scale=0.7]{{{images/plots/Z.mll.rel.PA}}} 
    &\includegraphics[scale=0.7]{{{images/plots/Z.ptl.rel.PA}}} 
  \end{tabular}
  \caption{Distributions in the invariant-mass~(left) and transverse-lepton-momentum~(right)
  for \PZ production at the LHC, with the upper plot showing the absolute distributions,
  the middle plots the full relative NLO QCD and EW corrections, and the lower plots the relative NLO 
  EW corrections in PA broken up into their factorizable and non-factorizable parts.}
  \label{fig:distNLO-Z}
\end{figure}

To illustrate the structure and quality of the PA applied to the EW corrections, we study two crucial observables for \PWp production at the LHC: the transverse-mass $M_{\rT,\Pgn\Pl}$ and the transverse-lepton-momentum $p_{\rT,\Pl}$ distribution, which are shown in Fig.~\ref{fig:distNLO-W}.
The transverse mass is defined as $M_{\rT,\Pgn\Pl} = \sqrt{ 2\; p_{\rT,\Pl}\;E_\rT^\miss (1-\cos\phi_{\Pgn\Pl}) }$, where $\phi_{\Pgn\Pl}$ denotes the  angle between the lepton and the missing momentum in the transverse plane.
The distributions in the upper panels of Fig.~\ref{fig:distNLO-W} exhibit the well-known Jacobian peaks at $M_{\rT,\Pgn\Pl}\approx\MW$ and $p_{\rT,\Pl}\approx\MW/2$, respectively, which play a central role in the measurement of the \PW-boson mass \MW at hadron colliders. In addition to the LO result, these plots include the NLO QCD and the full NLO EW predictions (without photon-induced processes) based on results of Refs.~\cite{Dittmaier:2001ay,Dittmaier:2009cr,Brensing:2007qm}, where we have naively multiplied the two latter corrections in the curve labelled as $\QCD\times\EW$. 
The middle plots of Fig.~\ref{fig:distNLO-W} show the relative corrections with respect to the LO prediction for the NLO QCD and the full EW corrections. 
It can be seen that the
EW corrections significantly distort the distributions and shift the
peak position.  While the QCD corrections are moderate for the
$M_{\rT,\Pgn\Pl}$ distribution, they become extremely large above the threshold
in the $p_{\rT,\Pl}$ distribution. This effect is induced by the
recoil of the \PW~boson against the hard jet in the QCD real-emission corrections.

In order to assess the quality of the PA, in the lower panels of
Fig.~\ref{fig:distNLO-W} we perform a comparison of the results obtained using the PA
against the full NLO EW
corrections. We also break down the result of the PA further into
factorizable corrections to the initial/final state and into the
non-factorizable contributions.  The accuracy of the PA near the
resonance is excellent, of the order of some $0.1\%$.  Above the
Jacobian peaks, the difference to the full EW correction grows to the
percent level. The PA remains accurate below the peaks where the
distributions are still dominated by resonant \PW-production.  Turning
to the relative importance of the different contributions in the PA,
it is seen that the impact of the non-factorizable corrections is
suppressed to the $0.1\%$ level and, thus, phenomenologically
negligible.  The factorizable initial-state corrections are also very
small for the case of the $M_{\rT,\Pgn\Pl}$ spectrum and for the
$p_{\rT,\Pl}$ distribution below the peaks. In the
latter case, the relative correction with respect to the LO prediction
becomes more sizeable above threshold, however, it should be
taken into account that the EW corrections to the $p_{\rT,\Pl}$
distribution are overwhelmed by the QCD corrections in this region.
Thus, the relevant part of the NLO EW corrections near the
threshold almost entirely results from the factorizable final-state corrections,
where the bulk originates from collinear final-state radiation from
the charged leptons.

Figure~\ref{fig:distNLO-Z} shows the respective results for the 
lepton-invariant-mass ($M_{\Pl\Pl}$) and transverse-lepton-momentum ($p_{\rT,\Pl}$) 
distributions in case of the neutral-current Drell--Yan process.
The invariant-mass spectrum shows much larger EW corrections than the other distributions, which is a well known effect of photonic final-state corrections that shift the peak location to lower values. The PA works well for the invariant-mass range considered in the plot and is again completely dominated by the final-state factorizable corrections. 
Due to the additional charged lepton in the final-state, the EW corrections to the neutral-current Drell--Yan process are larger compared to the corrections for the charged-current case.

%%%%%%%%%%%%%%%%%%%%%%%%%%%%%%%%%%%%%%%%%%%%%%%%%%%%%%%%%%%%%%%%%%%%%%
\section{Pole expansion for mixed QCD--electroweak corrections}
\label{sec:nnlo-nonfact}

In this section we set up the pole expansion for the NNLO
\order{\alphas\alpha} corrections to the Drell--Yan
process~\eqref{eq:dy}.  In this work we focus on the
\emph{non-factorizable} corrections which we calculate explicitly,
leaving the calculation of the factorizable corrections to future
work.  In Sect.~\ref{sec:nnlo-classification} we outline the structure
of the pole expansion and describe the various ingredients of the
non-factorizable corrections. The computation of the different
contributions to the non-factorizable corrections is performed in
Sect.~\ref{sec:calc-nf-nnlo}. In Sect.~\ref{sec:master} we combine these
building blocks into a master formula suitable for numerical
evaluation and discuss our treatment of IR singularities.

%%%%%%%%%%%%%%%%%%%%%%%%%%%%%%%%%%%%%%%%%%%%%%%%%%
\subsection{Concept and classification of contributions}
\label{sec:nnlo-classification}

As in the NLO EW corrections discussed in Sect.~\ref{sec:NLO} our aim is to calculate the leading corrections in the PA, i.e.\ the expansion around the resonance pole $p_V^2\approx\MV^2$.
The full NNLO  \order{\alphas\alpha} corrections to the processes~\eqref{eq:dy} consist of four types of contributions:
\begin{enumerate}[{\bf (a)}]
%%%%%%%%%%
\item {\bf Double-virtual corrections:} 
  The virtual \order{\alphas\alpha} corrections are given by two-loop corrections to the amplitude of the hard process~\eqref{eq:dy}
  and interferences of one-loop amplitudes of \order{\alphas} and \order{\alpha}.
  In the PA, these corrections consist of factorizable and non-factorizable parts.  
  The factorizable part is obtained from the factorizable virtual \order{\alpha} corrections in Eq.~\eqref{eq:nlo-fact-virt} by adding a QCD loop to the production and decay matrix elements.  
  The result consists of the two-loop \order{\alphas\alpha} corrections to on-shell $V$-boson production shown in Fig.~\ref{fig:NNLOcontrib}(a), the \order{\alphas} corrections to production multiplied by the \order{\alpha} corrections to the decay, as shown in Fig.~\ref{fig:NNLOcontrib}(b), and the \order{\alphas\alpha} corrections to the decay shown in Fig~\ref{fig:NNLOcontrib}(c).
\begin{figure}
  \centering
  \begin{subfigure}[m]{.48\linewidth}
    \centering
    \includegraphics{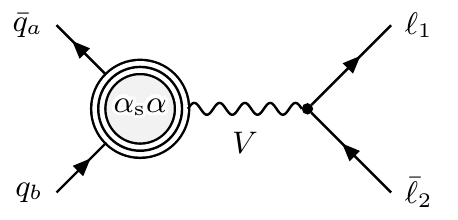}
    \subcaption{a}{Factorizable initial$\times$initial corrections}
    %\label{fig:NNLOcontrib:ii}
  \end{subfigure}
  \begin{subfigure}[m]{.48\linewidth}
    \centering
    \includegraphics{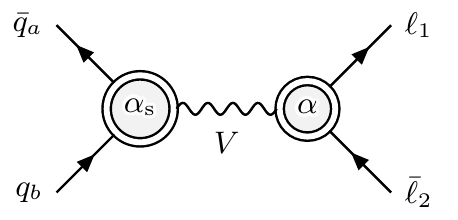}
    \subcaption{b}{Factorizable initial$\times$final corrections}
    %\label{fig:NNLOcontrib:if}
  \end{subfigure}
  \\[1.5em]
  \begin{subfigure}[m]{.48\linewidth}
    \centering
    \includegraphics{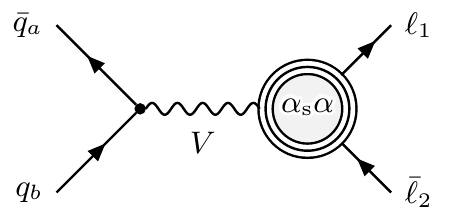}
    \subcaption{c}{Factorizable final$\times$final corrections}
    %\label{fig:NNLOcontrib:ff}
  \end{subfigure}
  \begin{subfigure}[m]{.48\linewidth}
    \centering
    \includegraphics{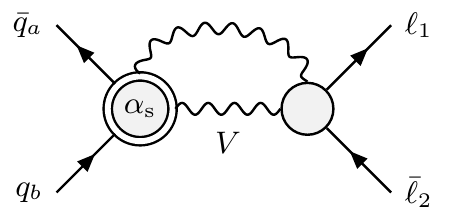}
    \subcaption{d}{Non-factorizable corrections}
    %\label{fig:NNLOcontrib:nf}
  \end{subfigure}
  \caption{The four types of corrections that contribute to the mixed QCD--EW corrections in the PA illustrated in terms of generic two-loop amplitudes.}
  \label{fig:NNLOcontrib}
\end{figure}
  The latter are given only by a pure counterterm contribution involving the QCD corrections to the $V$-boson self-energies in the renormalization constants of the electroweak couplings.  
  As in the NLO corrections, the residue of the $V$-boson propagator is normalized to one so that no propagator corrections arise.
  A generic diagram for the non-factorizable two-loop corrections is shown in Fig.~\ref{fig:NNLOcontrib}(d). 
  In analogy to Eq.~\eqref{eq:def-nonfact}, they are defined by the limit $p_V^2\to \MV^2$ of the difference of the full two-loop matrix element to the process~\eqref{eq:dy} and its factorizable part. 
 The result involves a gluon loop correction to the initial quark--antiquark pair and soft photons connecting the initial state, the intermediate vector boson, and the final-state leptons. 
 Explicit interference diagrams for double-virtual non-factorizable corrections to the cross section that also include the interference of one-loop \order{\alphas} and non-factorizable \order{\alpha} corrections are shown in Fig.~\ref{fig:NNLO-nf-graphs}(a). 
 These corrections are calculated in Sect.~\ref{sec:VV}.
\begin{figure}
  \centering
  \begin{subfigure}[m]{\linewidth}
    \centering
    \includegraphics[scale=0.95]{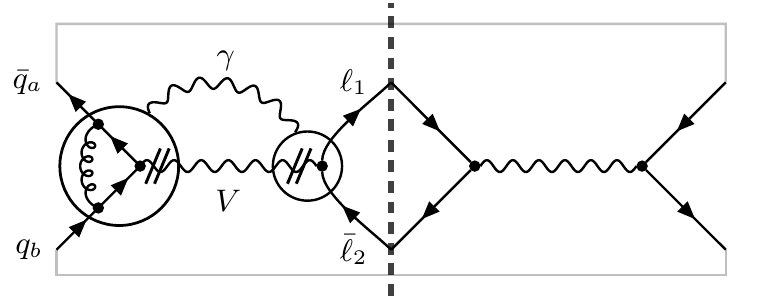}
    \qquad
    \includegraphics[scale=0.95]{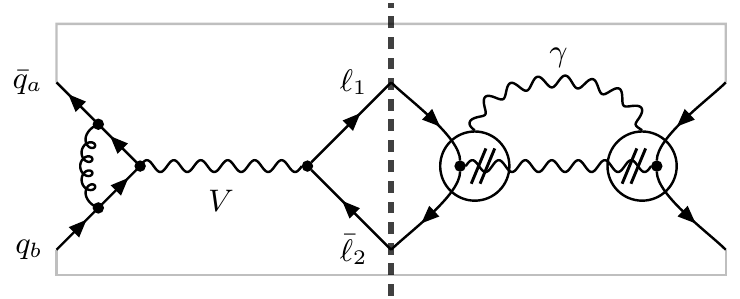}
    \subcaption{a}{Non-factorizable double-virtual corrections}
    %\label{fig:NNLO-nf-graphs-vv}
  \end{subfigure}
  \\[1.5em]
  \begin{subfigure}[m]{\linewidth}
    \centering
    \includegraphics[scale=0.95]{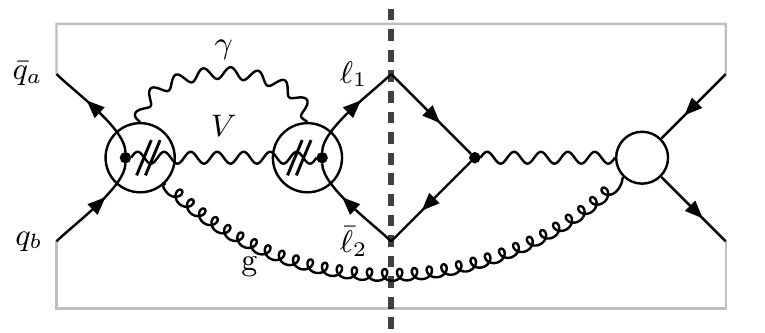}
    \qquad
    \includegraphics[scale=0.95]{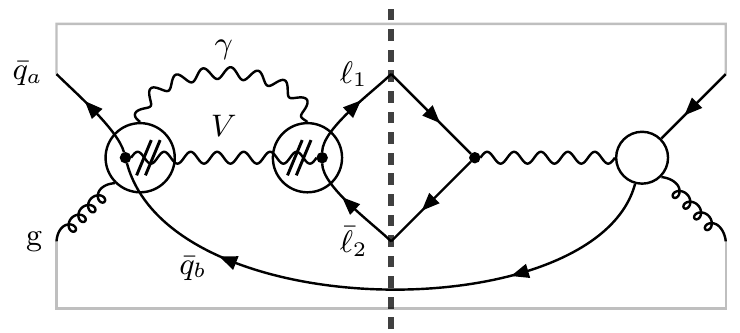}
    \subcaption{b}{Non-factorizable real QCD $\times$ virtual EW corrections}
    %\label{fig:NNLO-nf-graphs-rv}
  \end{subfigure}
  \\[1.5em]
  \begin{subfigure}[m]{\linewidth}
    \centering
    \includegraphics[scale=0.95]{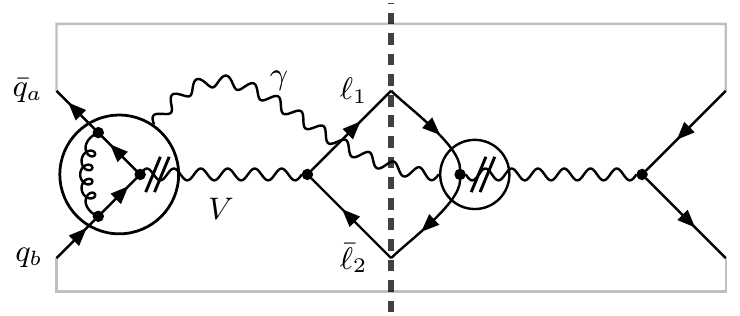}
    \qquad
    \includegraphics[scale=0.95]{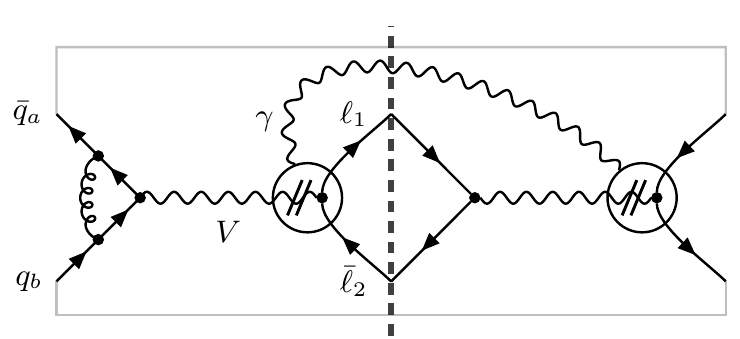}
    \subcaption{c}{Non-factorizable virtual QCD $\times$ real photonic corrections}
    %\label{fig:NNLO-nf-graphs-vr}
  \end{subfigure}
  \\[1.5em]
  \begin{subfigure}[m]{\linewidth}
    \centering
    \includegraphics[scale=0.95]{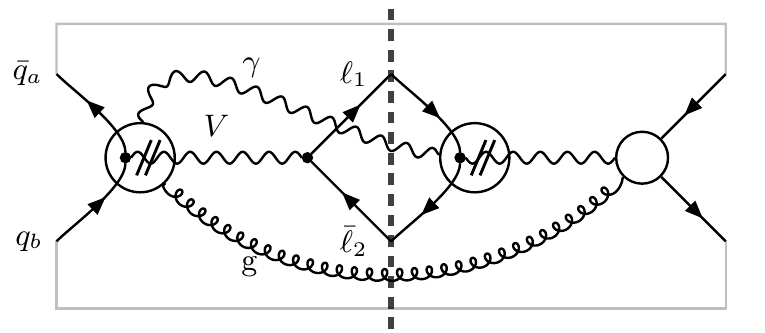}
    \qquad
    \includegraphics[scale=0.95]{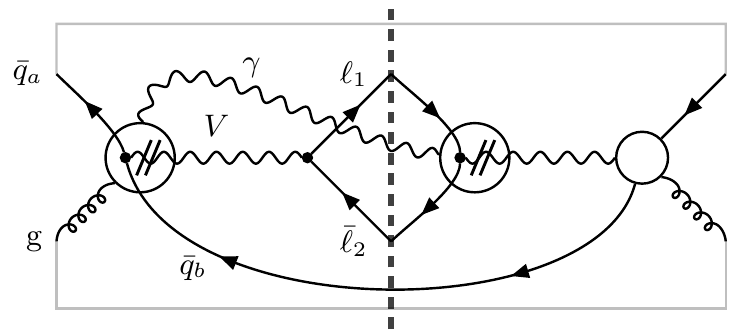}
    \subcaption{d}{Non-factorizable double-real corrections}
    %\label{fig:NNLO-nf-graphs-rr}
  \end{subfigure}
  \caption{Interference diagrams for the various contributions to the non-factorizable corrections of \order{\alphas\alpha}, with blobs representing all relevant tree structures. The encircled QCD loop diagrams with a photon attached stand for all possibilities to couple the photon to the quark line and the gauge boson $V$, see Eq.~\eqref{eq:on-shell-blob}.}
  \label{fig:NNLO-nf-graphs}
\end{figure}
%
%%%%%%%%%%
\item {\bf Real QCD $\times$ virtual EW corrections:}
  This type of corrections involves the virtual \order{\alpha} corrections to the dilepton plus jet production channels which we have considered previously for the real  \order{\alphas} QCD corrections in Eq.~\eqref{eq:rQCD}.
  The graphs for the factorizable corrections can be obtained from the diagrams of Figs.~\ref{fig:NNLOcontrib}(a),(b) upon replacing the ``$\alphas$'' in the blob by an external gluon and interfering all those diagrams with the LO amplitude for gluon emission (or its crossed variants).
  Graphs of the type shown in Fig.~\ref{fig:NNLOcontrib}(c), which consist of counterterm contributions only, do not have counterparts with real gluons.
  Finally, the non-factorizable corrections of (real QCD)$\times$(virtual EW) type stem from diagrams generically shown in Fig.~\ref{fig:NNLO-nf-graphs}(b).
  These corrections will be discussed in Sect.~\ref{sec:RV} in detail.
%%%%%%%%%%
\item {\bf Virtual QCD $\times$ real photonic corrections:}
  These corrections mainly consist of virtual \order{\alphas} corrections to dilepton plus photon production,
  \begin{equation}
    \label{eq:vQCD-rEW}
    \Paq_a(p_a) \,+\, \Pq_b(p_b) \,\to\, \Pl_1(k_1) \,+\, \Pal_2(k_2) \,+\, \Pgg(k) .
  \end{equation}
  Of course, there are also contributions from the crossed channels with photons in the initial state, but owing to their smallness already at NLO (i.e.\ without QCD corrections) their impact is certainly negligible at \order{\alphas\alpha}.

  In order to work out a PA for the photonic bremsstrahlung
  correction~\eqref{eq:vQCD-rEW}, we proceed exactly as described for
  the pure \order{\alpha} case in Sect.~\ref{sec:real-nonfact}.  Factorizable and
  non-factorizable contributions are constructed as described there,
  with the only difference that each contribution to squared matrix
  elements now contains a QCD loop of \order{\alphas}.  The
  non-factorizable parts will be calculated in Sect.~\ref{sec:VR};
  the corresponding generic diagrams are shown in
  Fig.~\ref{fig:NNLO-nf-graphs}(c).
%%%%%%%%%%
\item {\bf Double-real corrections:}
  The double-real \order{\alphas\alpha} corrections consist of the tree-level contributions to dilepton production in association with a photon and a jet, which receive contributions from the following partonic channels:
  \begin{subequations}
    \label{eq:rQCD-rEW}
    \begin{align}
      \Paq_a(p_a) \,+\, \Pq_b(p_b) &\,\to\, \Pl_1(k_1) \,+\, \Pal_2(k_2) \,+\, \Pg(k_\Pg) \,+\, \Pgg(k) , \label{eq:rr-qqb}\\
      \Pg(p_\Pg) \,+\, \Pq_b(p_b) &\,\to\, \Pl_1(k_1) \,+\, \Pal_2(k_2) \,+\, \Pq_a(k_a) \,+\, \Pgg(k) , \label{eq:rr-gq}\\
      \Pg(p_\Pg) \,+\, \Paq_a(p_a) &\,\to\, \Pl_1(k_1) \,+\, \Pal_2(k_2) \,+\, \Paq_b(k_b) \,+\, \Pgg(k) . \label{eq:rr-gqb}
    \end{align}
  \end{subequations}
  As in the (virtual QCD)$\times$(real photonic) corrections, we neglect contributions from crossed channels with photons in the initial state. 
  The PA for the double-real corrections~\eqref{eq:rQCD-rEW} is again constructed as in the pure \order{\alpha} case discussed in Sect.~\ref{sec:real-nonfact}. 
  The additional feature is the presence
  of an additional final-state parton whose kinematics will be treated
  exactly.  These double-real non-factorizable corrections will be
  discussed in Sect.~\ref{sec:RR} in detail; the corresponding generic
  graphs are shown in Fig.~\ref{fig:NNLO-nf-graphs}(d).
\end{enumerate}

%%%%%%%%%%%%%%%%%%%%%%%%%%%%%%%%%%%%%%%%%%%%%%%%%%
\subsection{Calculation of the non-factorizable \texorpdfstring{\order{\alphas\alpha}}{O(as a)} corrections}
\label{sec:calc-nf-nnlo}

In this section we calculate the various contributions to the non-factorizable corrections of \order{\alphas\alpha}, which are diagrammatically characterized in Fig.~\ref{fig:NNLO-nf-graphs}.
The calculation makes use of factorization properties of the virtual and real photonic parts of the non-factorizable \order{\alphas\alpha} corrections, which result from the soft nature of the effect. 
Extending the arguments given in the the classic paper~\cite{Yennie:1961ad} of Yennie, Frautschi and Suura (YFS), we show in Sect.~\ref{sec:yfs} that this factorization of the photonic factors even holds to any order in the strong coupling $\alphas$.
In the remainder of this section we use this insight to show that both the virtual and real-photonic corrections can be written as correction factors to squared matrix elements containing gluon loops or external gluons, i.e.\ the necessary building blocks are obtained from tree-level and one-loop calculations.
We have verified this statement diagrammatically and, for the corrections involving a gluon loop, also with EFT techniques, as discussed in Sect.~\ref{sec:VV} and Appendix~\ref{app:yfs-example} for the example of the double-virtual corrections.

%%%%%%%%%%%%%%%%%%%%%%%%%%%%%%
\subsubsection{Soft-photon radiation off a quark line---the YFS approach}
\label{sec:yfs}

Before entering the calculation of the non-factorizable corrections of \order{\alphas\alpha}, it is useful to discuss the general pattern of soft-photon radiation off a quark line with arbitrary gluon emission or exchange. 
In this investigation we closely follow the appendix of Ref.~\cite{Yennie:1961ad}, where the analogous situation is considered in pure QED.

We consider the  matrix element $T_0$ for a hard scattering process, which contains an incoming quark line with an arbitrary number of gluon attachments with momenta $q_i$ (treated as incoming by convention) and dressed by gluon and quark loops,
\begin{align}
  \label{eq:def-t0}
  T_0(p+Q) \;\equiv\;
  \raisebox{-48pt}{\includegraphics{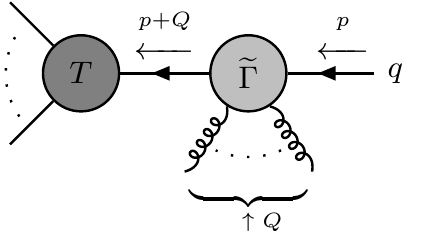}} 
  \;=\; T(p+Q)\,\frac{\ri}{\slashed p+\slashed Q}
  \widetilde\Gamma^{\Paq\Pq\Pg\dots\Pg}(-p-Q,p)u(p) .
\end{align}
Here $Q=\sum_i q_i$ is the sum of the gluon momenta incident on the
quark line, which can be hard or soft.
The $T$-blob in Eq.~\eqref{eq:def-t0} 
represents the hard scattering subprocess.
In the application to the Drell--Yan process, it is defined such that the quark
line with momentum $p+Q$ is directly connected to an irreducible
vertex function for the vector boson $V$.
The truncated Green function $\widetilde\Gamma$ is given by a chain of $N$ vertex functions for a quark--antiquark pair and incoming gluons, at arbitrary order in $\alphas$, linked by tree-like quark propagators, i.e.,
\begin{align}
  \label{eq:def-gamma-tilde}
  &\widetilde\Gamma^{\Paq\Pq \Pg\dots\Pg}(-p-Q,p) \;\equiv\;
  \raisebox{-48pt}{\includegraphics{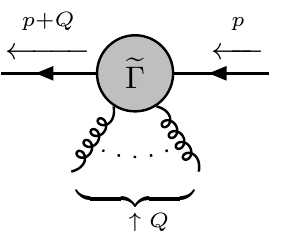}} \nonumber\\
  &\quad=\;\sum_{\mathclap{\substack{\text{gluon} \nonumber\\ \text{assignments}}}}\,
  \Gamma^{\Paq\Pq \Pg\dots\Pg}(-p-Q,p_{N-1})
  \frac{\ri}{\slashed p_{N-1}}
  \Gamma^{\Paq\Pq \Pg\dots\Pg}(-p_{N-1},p_{N-2})\dots\frac{\ri}{\slashed p_{1}}
  \Gamma^{\Paq\Pq \Pg\dots\Pg}(-p_1,p) \nonumber\\
  &\quad\equiv\;\sum_{\mathclap{\substack{\text{gluon} \\ \text{assignments}}}}\quad
  \raisebox{-48pt}{\includegraphics{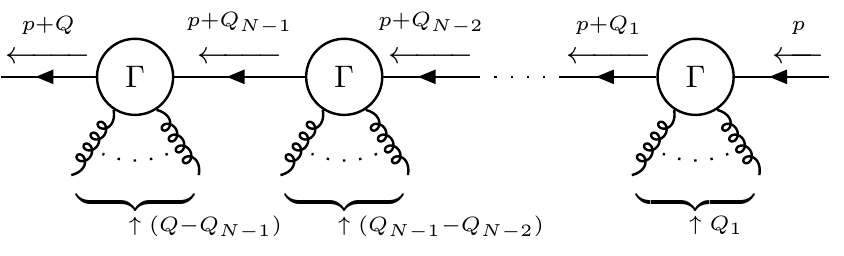}} .
\end{align}
Note that Lorentz and colour indices are suppressed in the notation.
The vertex functions $\Gamma$  are one-particle-irreducible with respect to the
quark line (called ``quark-irreducible'' in the following), but not necessarily with respect to gluon lines.
Fields and momenta in the arguments of Green functions are always taken to be incoming in this paper.
The sum over ``gluon assignments'' takes care of the fact that all distributions of gluons to quark-irreducible building blocks $\Gamma$ have to be taken into account; in particular this sum includes also summation over the number $N$.
Note, however, that by definition the
outermost $\Gamma$ insertion must involve at least one gluon attached to it, because the whole amplitude $T_0$ results from a truncated Green function (external self-energies amputated).
For simplicity of the argument we assume (without any restriction) that all external self-energy corrections are exactly cancelled by (on-shell) wave-function renormalization.
In the diagram, $p_i=p+Q_i$ is the quark momentum after the $i$th
insertion of $\Gamma^{\Paq\Pq\Pg\dots\Pg}$ and $Q_i-Q_{i-1}$ is the sum
of the gluon momenta incident at the $i$th vertex function $\Gamma^{\Paq\Pq\Pg\dots\Pg}$. 

Note that the situation sketched in Eqs.~\eqref{eq:def-t0} and \eqref{eq:def-gamma-tilde} covers all possible ways of gluon emission or virtual exchange if we interpret the gluon attached to the $\widetilde\Gamma$ vertices either as external or as virtual lines going into the $T$-blob. 

We also consider the quark-reducible truncated Green function $\widetilde\Gamma_\mu^{A\Paq\Pq \Pg\dots\Pg}(q,\bar{p},p)$ with an additional photon attachment, obtained by inserting the photon into all the quark-irreducible vertex functions and quark propagators in Eq.~\eqref{eq:def-gamma-tilde}; see Eq.~\eqref{eq:def-gamma-tilde-A} in the appendix for more details.  
Since the gluons carry no electric charge, the truncated Green functions $\widetilde\Gamma$ satisfy the QED Ward identity
\begin{equation}
  \label{eq:ward-id}
  q^\mu \, \widetilde\Gamma_{\mu}^{A\Paq\Pq \Pg\dots\Pg}(q,\bar p,p)
  = eQ_\Pq\left[ \widetilde\Gamma^{\Paq\Pq \Pg\dots\Pg}(\bar p +q,p)
    -\widetilde\Gamma^{\Paq\Pq \Pg\dots\Pg}(\bar p,p+q)\right] , 
\end{equation}
which results from contracting the photon leg with its own momentum $q$.
We prove this identity in Appendix~\ref{app:yfs}.

\begin{figure}
  \centering
  \includegraphics{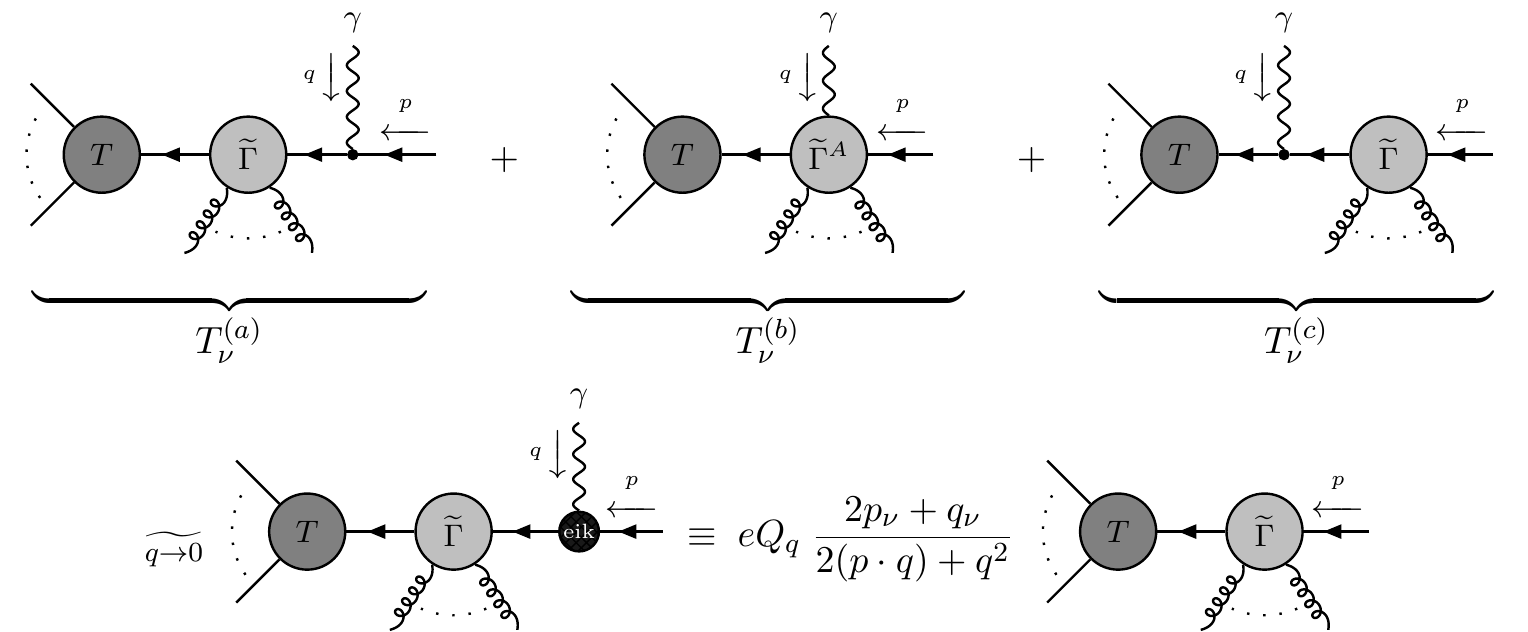}
  \caption{Diagrammatical representation of the result~\eqref{eq:YFS},
    with the vertex labelled by ``eik'' denoting the eikonal approximation given
    explicitly in the right-hand diagram on the last line.}
  \label{fig:YFS}
\end{figure}

We now consider the attachment of the photon with momentum $q$ to the quark line, which can take place either at a reducible quark line or at the truncated Green function $\widetilde\Gamma$ (see first line in Fig.~\ref{fig:YFS}),
\begin{subequations}
  \label{eq:YFS-abc}
  \begin{align}
    T_\nu^{(a)}&=T(p+Q+q)\,\frac{\ri}{\slashed p+\slashed Q+\slashed q}
    \widetilde\Gamma^{\Paq\Pq \Pg\dots\Pg}(-p-Q-q,p+q) 
    \frac{\ri}{\slashed p+\slashed q}(-\ri\re Q_\Pq)\gamma_\nu u(p) ,
    \label{eq:YFS-a}\\
    T_\nu^{(b)}&=T(p+Q+q)\,\frac{\ri}{\slashed p+\slashed Q+\slashed q}
    \widetilde\Gamma_\nu ^{A\Paq\Pq \Pg\dots\Pg}(q,-p-Q-q,p)  u(p) , 
    \label{eq:YFS-b}\\
    T_\nu^{(c)}&=T(p+Q+q)\,\frac{\ri}{\slashed p+\slashed Q+\slashed q}
    (-\ri\re Q_\Pq) \gamma_\nu
    \frac{\ri}{\slashed p+\slashed Q}
    \widetilde\Gamma^{\Paq\Pq \Pg\dots\Pg}(-p-Q,p) u(p) .
    \label{eq:YFS-c}
  \end{align}
\end{subequations}
In $T_\nu^{(a)}$ the photon is attached at the external end of the quark line, in $T_\nu^{(b)}$ it is attached to all possible points inside $\widetilde\Gamma^{\Paq\Pq \Pg\dots\Pg}$, and in $T_\nu^{(c)}$ the photon sits on the quark line between $\widetilde\Gamma^{\Paq\Pq \Pg\dots\Pg}$ and the remaining hard scattering described by $T(p+Q+q)$.
The main result that will be extensively employed in the following sections is the statement that the soft-photon limit of the sum of the three contributions in Eq.~\eqref{eq:YFS-abc} simplifies as follows:
\begin{equation}
  \label{eq:YFS}
  T_\nu^{(a)}+ T_\nu^{(b)}+ T_\nu^{(c)}\asymp{q\to 0}
  \;e Q_\Pq\; \frac{2p_\nu+q_\nu}{2(p\cdot q)+q^2}\; T_0(p+Q) .
\end{equation}
Figure~\ref{fig:YFS} diagrammatically illustrates this identity, which
holds to all orders in QCD.
The soft-photon factor multiplying the lower-order matrix element
$T_0$ is represented by the vertex labelled ``eik'', where it is
understood that the quark momentum remains on-shell after interaction
with the soft photon.
Note that $T_0$ is independent of the soft photon momentum $q$, which is a non-trivial statement on overlapping IR divergences that does not hold for individual diagrams. 
By definition of the hard subprocess, the soft photon momentum can be set to zero in the hard matrix element, $T(p+Q+q)\xrightarrow{q\to 0}T(p+Q)$, without encountering IR singularities.  
Note, however, that when this argument is applied to our calculation of the non-factorizable corrections, the photon momentum $q$ must not be neglected in the resonant $V$ propagator, which distinguishes the ESPA from the usual SPA.
The result in Eq.~\eqref{eq:YFS} holds for arbitrary gluon momenta $q_i$ in $Q=\sum_iq_i$, i.e.\ gluons can be hard or soft.  
The proof of Eq.~\eqref{eq:YFS} follows the one of the analogous result in QED~\cite{Yennie:1961ad}, which is essentially based on the Ward identity of Eq.~\eqref{eq:ward-id}, and is spelled out in some detail in Appendix~\ref{app:yfs}.%
\footnote{ Note that for a soft-gluon insertion into a one-loop QCD amplitude an extra contribution appears in addition to the eikonal factor~\cite{Catani:2000pi}. This term is a non-abelian effect proportional to the structure constants of the gauge group, i.e.\ our result for a photon insertion~\eqref{eq:YFS} is consistent with Ref.~\cite{Catani:2000pi}.}
Note that it has not been specified whether the soft photon with
momentum $q$ is a real or virtual photon, so that Eq.~\eqref{eq:YFS}
can also be applied to subdiagrams appearing inside a larger loop
diagram.
When the soft-photon line is closed to a loop by connecting to a
charged-particle line entering the $T$-blob in~\eqref{eq:def-t0}, the
identity~\eqref{eq:YFS} or its analogue in QED can be
applied  as well, so the soft photon
always couples via the eikonal vertex introduced in Fig.~\ref{fig:YFS}.

%%%%%%%%%%%%%%%%%%%%%%%%%%%%%%
\subsubsection{Double-virtual corrections}
\label{sec:VV}

The double-virtual non-factorizable corrections consist of two different contributions, which are illustrated in Fig.~\ref{fig:NNLO-nf-graphs}(a) by interference diagrams:
First, the \order{\alphas\alpha} corrections to the $\Paq_a\Pq_b\to\Pl_1\Pal_2$ amplitude, which interfere with the Born diagram (the left diagram in Fig.~\ref{fig:NNLO-nf-graphs}(a)), and second, the interference between two one-loop-corrected amplitudes with corrections of \order{\alphas} and \order{\alpha}, respectively, 
\begin{equation}
\label{eq:NNLO-vv}
  \begin{aligned}
    \bigl\lvert \M^{\Paq_a\Pq_b\to\Pl_1\Pal_2} \bigr\rvert^2 
    \;\Bigr\rvert_{\nf}^{\Vs\otimes\Vew} = 
    & \hphantom{{}+{}} 2\Re\left\{
    \delta\M_{\Vs\otimes\Vew,\nf}^{\Paq_a\Pq_b\to\Pl_1\Pal_2}
    \left(\M_{0,\PA}^{\Paq_a\Pq_b\to\Pl_1\Pal_2}\right)^* \right\} \\
    &+ 2\Re\left\{
    \delta\M_{\Vew,\nf}^{\Paq_a\Pq_b\to\Pl_1\Pal_2}
    \left(\delta\M_{\Vs,\PA}^{\Paq_a\Pq_b\to\Pl_1\Pal_2}\right)^* \right\} .
  \end{aligned}
\end{equation}

The latter correction is
shown in the right-hand diagram in Fig.~\ref{fig:NNLO-nf-graphs}(a). It can be directly obtained from the results of Sect.~\ref{sec:NLO} and reads
\begin{equation}
  \label{eq:mix_vv2}
  \Re\left\{
  \delta\M_{\Vew,\nf}^{\Paq_a\Pq_b\to\Pl_1\Pal_2}
  \left(\delta\M_{\Vs,\PA}^{\Paq_a\Pq_b\to\Pl_1\Pal_2}\right)^* \right\}  
  = \Re\left\{ \delta_{\Vew,\nf}^{\Paq_a\Pq_b\to\Pl_1\Pal_2}  \left(\delta_{\Vs}^{\Paq_a\Pq_b\to\Pl_1\Pal_2} \right)^*\right\}
  \left\lvert\M_{0,\PA}^{\Paq_a\Pq_b\to\Pl_1\Pal_2}\right\rvert^2 .
\end{equation}
The expressions for $\delta_{\Vew,\nf}^{\Paq_a\Pq_b\to\Pl_1\Pal_2}$ and $\delta_{\Vs}^{\Paq_a\Pq_b\to\Pl_1\Pal_2}$ are given in Eqs.~\eqref{eq:nf-virt} and
\eqref{eq:nlo-qcd}, respectively.  

\begin{figure}
  \centering
  \includegraphics[scale=0.75]{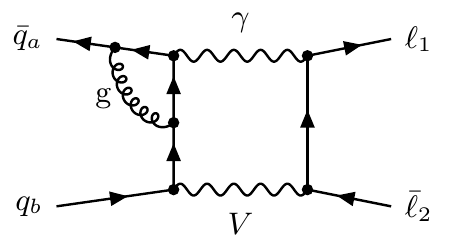} \;
  \includegraphics[scale=0.75]{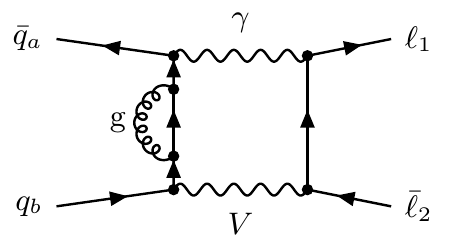} \;
  \includegraphics[scale=0.75]{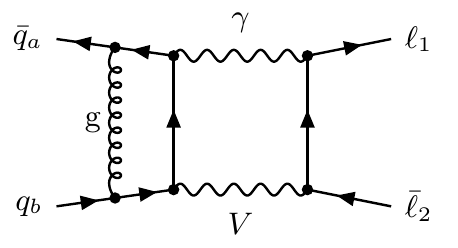} \;
  \includegraphics[scale=0.75]{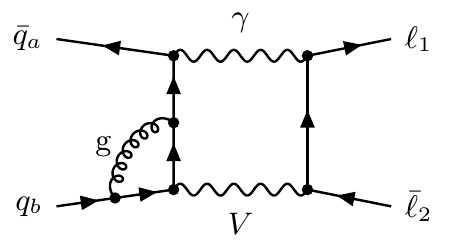}
\caption{Example diagrams for the non-factorizable double-virtual corrections.} 
\label{fig:QCD-EW-virt-if}
\end{figure}

The non-factorizable part of the two-loop correction
$\delta\M_{\Vs\otimes\Vew,\nf}^{\Paq_a\Pq_b\to\Pl_1\Pal_2}$ in Eq.~\eqref{eq:NNLO-vv} is defined in
analogy to the NLO case discussed in Sect.~\ref{sec:virt-nonfact}.
The contributing two-loop diagrams
can be obtained from the NLO diagrams of Fig.~\ref{fig:nlo-nf-virt} by attaching a virtual gluon in all possible ways to the quark line. 
For the example of the initial--final state interference diagram of Fig.~\ref{fig:nlo-nf-virt}(a), the resulting diagrams are displayed in Fig.~\ref{fig:QCD-EW-virt-if}.  
As at NLO, there are diagrams
containing only non-factorizable contributions as well as diagrams
possessing both factorizable and non-factorizable parts.  For the
latter type of diagrams the non-factorizable part is defined via the
generalization of Eq.~\eqref{eq:def-nonfact},
\begin{equation}
  \label{eq:def-vv-nonfact}
  \delta\M_{\Vs\otimes\Vew,\nf}^{\Paq_a\Pq_b\to\Pl_1\Pal_2}=
  \left\{ \delta\M_{\Vs\otimes\Vew}^{\Paq_a\Pq_b\to\Pl_1\Pal_2}
    - \delta\M_{\Vs\otimes\Vew,\fact}^{\Paq_a\Pq_b\to\Pl_1\Pal_2}
  \right\}_{p_V^2\to \MV^2} .
\end{equation}

A complication compared to the NLO calculation originates from the possibility of
overlapping QCD and soft-photon corrections, as in the first and third
diagram of Fig.~\ref{fig:QCD-EW-virt-if}.  After non-trivial
cancellations, which are discussed in detail below, the two-loop
corrections factorize from the Born amplitude,
\begin{equation}
  \label{eq:mix_vv1}
  \delta\M_{\Vs\otimes\Vew,\nf}^{\Paq_a\Pq_b\to\Pl_1\Pal_2} \,=\, 
  \delta_{\Vew,\nf}^{\Paq_a\Pq_b\to\Pl_1\Pal_2} \, 
  \delta_{\Vs}^{\Paq_a\Pq_b\to\Pl_1\Pal_2} \,
  \M_{0,\PA}^{\Paq_a\Pq_b\to\Pl_1\Pal_2} ,
\end{equation}
with identical correction factors as in Eq.~\eqref{eq:mix_vv2} above.
The final result of the double-virtual corrections to the cross section then reads
\begin{equation}
  \label{eq:mix_vv}
  \begin{aligned}
    \rd\sigma_{\Paq_a\Pq_b,\nf}^{\Vs\otimes\Vew}
    =\;& 2 \biggl[
    \Re\left\{ \delta_{\Vew,\nf}^{\Paq_a\Pq_b\to\Pl_1\Pal_2} \left(\delta_{\Vs}^{\Paq_a\Pq_b\to\Pl_1\Pal_2} \right)^* \right\}
    + \Re\left\{ \delta_{\Vew,\nf}^{\Paq_a\Pq_b\to\Pl_1\Pal_2} \; \delta_{\Vs}^{\Paq_a\Pq_b\to\Pl_1\Pal_2} \right\}
    \biggr] \rd\sigma_{\Paq_a\Pq_b,\PA}^0 \\
    =\;& 4\Re\left\{ \delta_{\Vew,\nf}^{\Paq_a\Pq_b\to\Pl_1\Pal_2} \right\}
    \Re\left\{ \delta_{\Vs}^{\Paq_a\Pq_b\to\Pl_1\Pal_2} \right\}  
    \rd\sigma_{\Paq_a\Pq_b,\PA}^0 .
  \end{aligned}
\end{equation}

We still have to fill the gap of proving Eq.~\eqref{eq:mix_vv1}, a task that we have attacked in three different ways: \ref{sec:vv-yfs} using the results from the YFS approach of Sect.~\ref{sec:yfs}, \ref{sec:vv-mb} evaluating the non-factorizable parts of the two-loop diagrams directly using the Mellin--Barnes technique, and~\ref{sec:vv-eft} using the EFT inspired expansion of the loop integrals in momentum regions.

%%%%%%%%%%
\paragraph{YFS argument}
\label{sec:vv-yfs}

In order to prove the factorization of the two-loop corrections into a
product of one-loop factors in Eq.~\eqref{eq:mix_vv1} we deduce two
results from Eq.~\eqref{eq:YFS}.  Firstly, we consider the situation
in which the $\widetilde\Gamma$ blocks in Eqs.~\eqref{eq:YFS-abc} and
\eqref{eq:YFS} (see also Fig.~\ref{fig:YFS}) do not involve external
gluons, but only one internal QCD loop, i.e.\ the $\widetilde\Gamma$'s
are proportional to $\alphas$, counting powers of $\alphas$ only.
\begin{figure}
  \centering
  \includegraphics[scale=0.9]{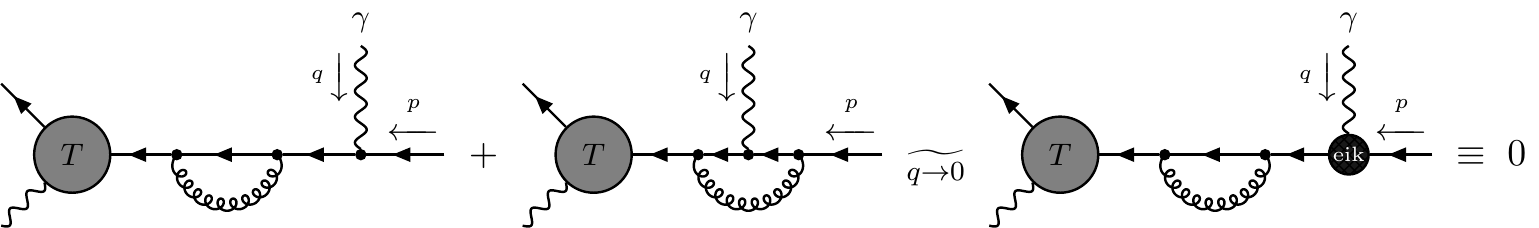}
  \caption{Application of Eq.~\eqref{eq:YFS} to self-energy diagrams.}
  \label{fig:YFS-se}
  \vspace*{2em}
  \Vcenter{\includegraphics[scale=0.9]{images/diags/DY_2loopSelf}} $+$
  \Vcenter{\includegraphics[scale=0.9]{images/diags/DY_2loop}} 
  $\asymp{\text{non-fact.}} 0$
  \caption{Cancellation of diagrams in the non-factorizable double-virtual QCD$\times$EW corrections due to the identity~\eqref{eq:YFS-se} illustrated in Fig.~\ref{fig:YFS-se}.} 
  \label{fig:vertex-box-cancel}
\end{figure}
Owing to our on-shell wave-function renormalization of the external quark field the $T^{(c)}$ part is absent, i.e.\ $T^{(c)}=0$, and likewise the whole r.h.s.\ of Eq.~\eqref{eq:YFS}.
The parts $T^{(a)}$ and $T^{(b)}$, however, are non-vanishing and, thus, cancel in the sum as shown in Fig.~\ref{fig:YFS-se}.
The respective $\widetilde\Gamma$ functions of $T^{(a)}$ and $T^{(b)}$ are given by the one-loop quark self-energy factor $\ri\Sigma^\Pq$ and by the quark--photon vertex function $\Gamma^{A\Paq\Pq}$, so that Eq.~\eqref{eq:YFS} turns into the identity
\begin{equation}
  \label{eq:YFS-se}
  T(p+Q+q)\,\frac{\ri}{\slashed p+\slashed q}\left\{
  \ri\Sigma^\Pq(p+q) \frac{\ri}{\slashed p+\slashed q}\ (-\ri\re Q_\Pq\gamma_\nu)
  +\Gamma_\nu ^{A\Paq\Pq}(q,-p-q,p)
  \right\} u(p)
  \asymp{q\to 0}0 . 
\end{equation}
Furthermore, since the identity is also valid for the case where the soft photon is virtual, the photon can as well be connected to one of the other charged-particle lines entering the hard function $T$.  
Applied to the non-factorizable double-virtual corrections, this leads to the cancellation of all diagrams involving a QCD correction to the quark--photon vertex against diagrams with a QCD self-energy correction to an internal quark line. 
In the example diagrams from Fig.~\ref{fig:QCD-EW-virt-if} this leads to the cancellation of the two diagrams shown in Fig.~\ref{fig:vertex-box-cancel}.

As the second application of Eq.~\eqref{eq:YFS}, we take the truncated Green function $\widetilde\Gamma$ to be the tree-level quark--gluon vertex.  
\begin{figure}
  \centering
  \includegraphics[scale=0.9]{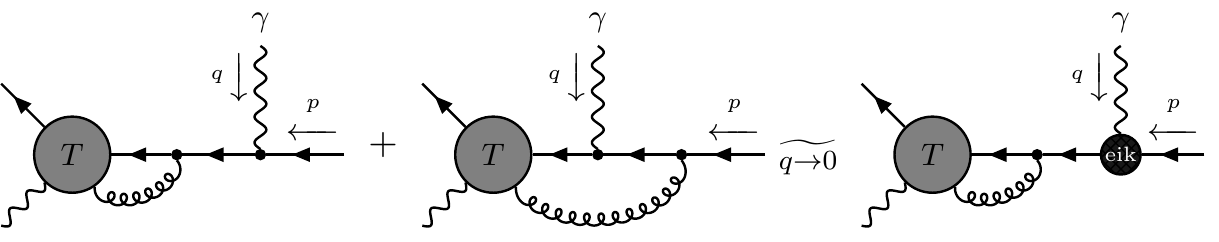}
  \caption{Application of Eq.~\eqref{eq:YFS} to vertex diagrams.}
  \label{fig:YFS-vert}
  \vspace*{2em}
  \Vcenter{\includegraphics[scale=0.9]{images/diags/DY_2loopVert}} $+$
  \Vcenter{\includegraphics[scale=0.9]{images/diags/DY_2loopBox}}
  $\asymp{\text{non-fact.}}$
  \Vcenter{\includegraphics[scale=0.9]{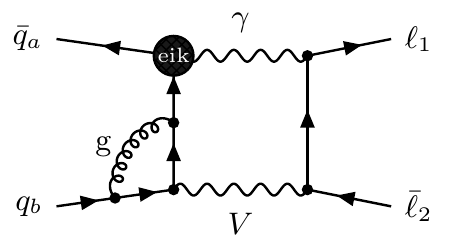}}
  \caption{Simplification of diagrams in the non-factorizable double-virtual QCD$\times$EW corrections.} 
  \label{fig:double-box-cancel}
\end{figure}
In this case the term $T^{(b)}$ in Eq.~\eqref{eq:YFS} vanishes since there is no tree-level gluon--photon--antiquark--quark vertex.
As discussed above, the emitted gluon can be connected to the hard subprocess by a loop so that we obtain the identity shown in Fig.~\ref{fig:YFS-vert}.
In the application to the double-virtual non-factorizable corrections, the resulting identity shows that the sum of the diagrams involving a QCD correction to the (off-shell) $V$-boson--antiquark--quark vertex and box-type diagrams simplifies to diagrams with a one-loop QCD vertex with on-shell quarks and antiquarks inserted into the one-loop non-factorizable corrections. 
This simplification is illustrated in Fig.~\ref{fig:double-box-cancel}.

The remaining  non-factorizable double-virtual corrections can be obtained  from the example diagrams in Fig.~\ref{fig:QCD-EW-virt-if} by moving the photon from the lepton line to the antilepton or $V$-boson lines, and by attaching it to the incoming quark instead of the antiquark line. 
In all diagrams, the two identities in Figs.~\ref{fig:YFS-se} and \ref{fig:YFS-vert} can be applied in an analogous way.
As a result, the sum of all diagrams where the $V$ boson is attached directly to the QCD loop is represented in PA by the NLO diagrams for the non-factorizable corrections without gluon exchange times the QCD correction factor~\eqref{eq:nlo-qcd} for the on-shell $V\Paq_a\Pq_b$ vertex, while all other diagrams where $V$ is not attached to the QCD loop cancel each other.
This establishes the identity in Eq.~\eqref{eq:mix_vv1}.

%%%%%%%%%%
\paragraph{Mellin--Barnes calculation}
\label{sec:vv-mb}

In this calculation of the non-factorizable contributions, the resonance-enhanced terms are extracted from the double-virtual corrections on a diagram-by-diagram basis by means of the Mellin--Barnes technique~\cite{Smirnov:2006ry}.
For manifestly non-factorizable diagrams, following an analogous power-counting argument as presented in Ref.~\cite{Denner:1997ia},
a one-to-one correspondence can be established between the
resonant non-factorizable contributions and the terms that diverge linearly for soft photon momenta in the on-shell limit ($p_V^2=\MV^2$).
In the case where the virtual photon is attached to the intermediate gauge-boson propagator, i.e.\ diagrams that contain both factorizable and non-factorizable parts, Eq.~\eqref{eq:def-vv-nonfact} defines the gauge-invariant prescription to extract the non-factorizable contributions.
Here, all terms that are not IR singular in the on-shell limit with respect to the photon momentum cancel in the difference.
In summary, the non-factorizable corrections receive contributions only from the soft-photon region of the loop integrals. 
In order to identify all terms that exhibit this leading IR-singular behaviour in the photon momentum, first, a tensor reduction into scalar integrals is performed for the internal ``QCD sub-loop''.
Then, by simple power counting in the photon momentum, terms with a sub-leading singular behaviour can be singled out and omitted before proceeding with the evaluation of the two-loop integrals.
In particular, terms involving additional factors of the photon momentum in the numerator do not lead to resonant contributions and can be already neglected at this stage.
As a consequence, only scalar two-loop integrals appear in the final result of the non-factorizable corrections.
They are evaluated using the Mellin--Barnes method, which can be employed to perform a systematic expansion in powers of $\tfrac{p_V^2-\mu_V^2}{\MV^2}$, where only the leading contribution is needed for the resonance-enhanced non-factorizable corrections.
The explicit calculation of the two diagrams shown in Fig.~\ref{fig:vertex-box-cancel} and their cancellation is presented in Appendix~\ref{app:yfs-mellin}.

%%%%%%%%%%
\paragraph{Effective-field-theory inspired calculation}
\label{sec:vv-eft}

In the application of the EFT approach to the double-virtual
\order{\alphas\alpha} corrections, the contributing two-loop
integrals are expanded in the hard, soft, and collinear momentum
regions. In the EFT language, hard loop momenta contribute to a
renormalization of Wilson coefficients of $V$-boson production and
decay operators. These contributions can be identified with the
factorizable corrections in the conventional language. In contrast,
soft and collinear loop momenta correspond to loop corrections within
the EFT. 
 Since only soft loops can connect production and decay
stages without throwing the vector boson off shell, initial--final
collinear or hard loops do not contribute at leading power
in the resonance expansion.  Furthermore, soft gluons can only be
exchanged between the initial-state quark legs which leads to
scaleless integrals. Therefore, soft QCD corrections do not contribute.

We will first show that the hard$_{\QCD}$/soft$_{\EW}$ corrections
are given by the non-factorizable \order{\alpha} corrections with an
insertion of the \order{\alphas} corrections to the on-shell
production vertex. 
For illustration, consider the diagrams in Fig.~\ref{fig:QCD-EW-virt-if}.  Note that diagrams
where the soft photon is emitted from an internal hard line (such as
the first and third diagram of Fig.~\ref{fig:QCD-EW-virt-if} in case
the virtual gluon is hard) are suppressed by one power of
$\GV/\MV$ and, therefore, do not contribute to the resonant
corrections in the pole expansion. In the EFT language these
contributions correspond to higher-dimensional production operators.
The hard contribution of the second diagram involves the on-shell
quark self-energy which vanishes for massless quarks in dimensional
regularization.  Therefore only the fourth diagram has a non-vanishing
hard$_{\QCD}$/soft$_{\EW}$ contribution.  In the expansion by
regions, the soft photon momentum is neglected compared to the hard
gluon momentum, so that the QCD vertex correction is independent of the
photon momentum.  Since the soft-photon couplings to the quarks and
the $V$ boson are given by the same standard eikonal factor and
modified $V$-boson propagators as in the ESPA, this contribution is of
the same form as the final diagram in
Fig.~\ref{fig:double-box-cancel} and lead to the factorized result
of Eq.~\eqref{eq:mix_vv1}.

In order for \emph{all} non-factorizable corrections to be of the
form~\eqref{eq:mix_vv1}, it remains to be shown that the contributions
from collinear gluon momenta cancel.  In contrast to the NLO case,
individual collinear loop diagrams can be non-vanishing at the two-loop
level, as pointed out in Ref.~\cite{Beneke:2004km}.  In fact, the
first three diagrams in Fig.~\ref{fig:QCD-EW-virt-if} only receive
non-vanishing contributions from collinear gluon momenta, while the
last diagram receives contributions both from collinear and hard gluon
momenta.  The collinear sub-integrals can be evaluated by expanding
the integrands using the scaling~\eqref{eq:momentum-scalings} or,
equivalently, applying the 
soft--collinear effective theory~(SCET) Feynman rules of
Ref.~\cite{Bauer:2000yr}, and subsequently integrating over the
collinear loop momentum using the method discussed in
Sect.~\ref{sec:nlo-eft}.  After this step, one observes a pairwise
cancellation of the diagrams in agreement with the pattern established
in Paragraph~\ref{sec:vv-yfs}.  For the example of the two diagrams in
Fig.~\ref{fig:vertex-box-cancel} this is demonstrated explicitly in
Appendix~\ref{app:yfs-eft}.  Therefore, all contributions from
collinear gluon momentum regions cancel and only the
hard$_{\QCD}$/soft$_{\EW}$ region gives a non-vanishing contribution
to the non-factorizable corrections. As discussed above, this
contribution leads to the result~\eqref{eq:mix_vv1}.

%%%%%%%%%%%%%%%%%%%%%%%%%%%%%%
\subsubsection{Real QCD \texorpdfstring{$\times$}{x} virtual EW corrections}
\label{sec:RV}

\begin{figure}
  \centering
  \includegraphics[scale=0.9]{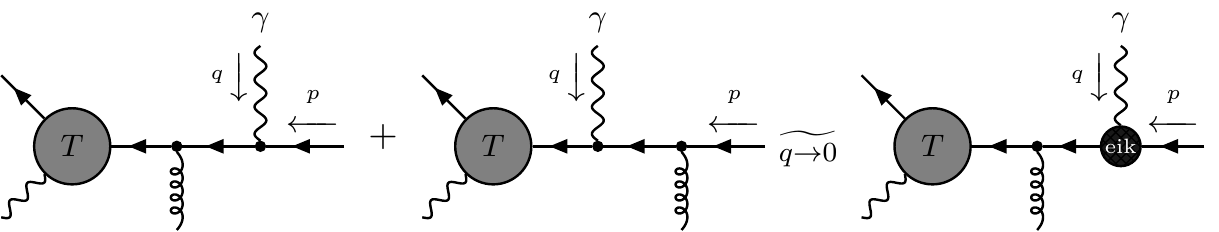}
  \caption{Application of Eq.~\eqref{eq:YFS} to gluon-emission diagrams.}
  \label{fig:YFS-RV}
  \vspace*{2em}
  \Vcenter{\includegraphics[scale=0.9]{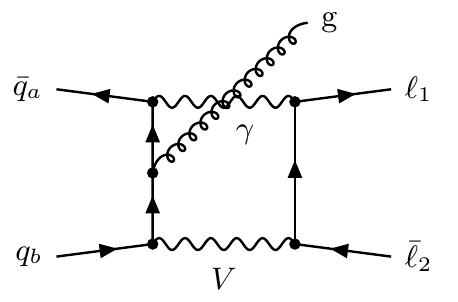}} $+$
  \Vcenter{\includegraphics[scale=0.9]{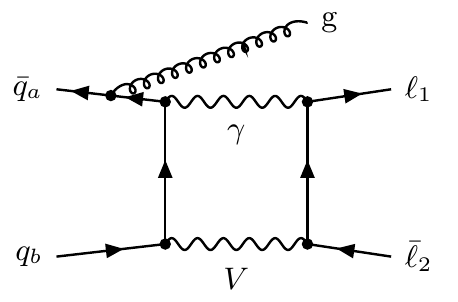}}
  $\asymp{\text{non-fact.}}$
  \Vcenter{\includegraphics[scale=0.9]{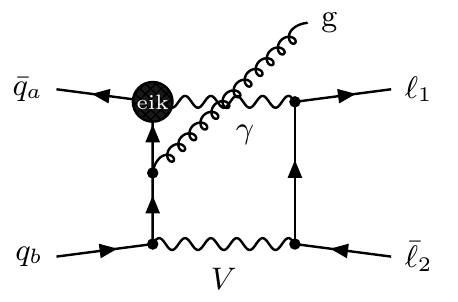}}
  \caption{Example diagrams for the (real QCD)$\times$(virtual EW) non-factorizable initial--final corrections. 
The application of the identity of Fig.~\ref{fig:YFS-RV}
 shows that the resonance expansion of the sum of the diagrams is equal to the one-loop non-factorizable initial--final soft correction to the 
 process $\Paq_a\Pq_b\to\Pl_1\Pal_2\Pg$ as indicated by the last
 diagram.}
  \label{fig:QCD-EW-real-if}
\end{figure}

The (real QCD)$\times$(virtual EW) corrections are given by the
virtual \order{\alpha} corrections to dilepton plus jet
production, with the three production channels of
Eq.~\eqref{eq:rQCD}. The diagrams contributing to the
non-factorizable part of these corrections are the same as those for
the virtual non-factorizable corrections to $V$-boson production and
decay in association with a jet.
Two generic interference diagrams are shown in Fig.~\ref{fig:NNLO-nf-graphs}(b) for the quark--antiquark induced (left-hand diagram) and gluon--quark induced (right-hand diagram) channels.
The calculation proceeds in an analogous manner as presented in Sect.~\ref{sec:virt-nonfact} for the case of $V$ production, but becomes more involved due to the additional external particle.
Furthermore, a new feature comes into play for the gluon-induced processes, where the photon exchange between two final-state particles enters the calculation.
Considering for example the $\Paq\Pq$ initial state, the non-factorizable corrections are defined in analogy to Eq.~\eqref{eq:def-nonfact} by the difference of the full diagrams and the factorizable contributions in the limit where the vector boson is on shell,
\begin{equation}
  \label{eq:NNLO-nf-rv-def}
  \delta\M_{\Vew,\nf}^{\Paq_a\Pq_b\to  \Pl_1\Pal_2\Pg}=
  \left\{ \delta\M_{\mathrm{\Vew}}^{\Paq_a\Pq_b\to \Pl_1\Pal_2\Pg}
  - \delta\M_{\Vew,\fact}^{\Paq_a\Pq_b\to \Pl_1\Pal_2\Pg}
  \right\}_{p_V^2\to \MV^2} .
\end{equation}
Note that we will make no assumption on the momentum $k_\Pg$ of the real gluon, in particular it is not assumed to be soft.  
Therefore in the (real QCD)$\times$(virtual EW) case the resonance expansion is performed for
\begin{equation}
  \label{eq:resonance-gluon}
  p_V^2=  (k_1+k_2)^2=s_{ab}-2(p_a+p_b)\cdot k_\Pg \approx\MV^2 ,
\end{equation}
with $s_{ab}=(p_a+p_b)^2$.
The definition of the non-factorizable corrections for the $\Pg\Pq$- and
$\Pg\Paq$-initiated channels is analogous.  As in the NLO case, only
soft-photon exchange delivers resonant contributions, so that the YFS
arguments of Sect.~\ref{sec:yfs} apply.  Since these arguments work
equally well for real and virtual gluons, the same reasoning made
explicit in the double-virtual case of the previous section implies
that the virtual non-factorizable photonic corrections factorize from
the real QCD amplitude.  As an example, the analogue of the identity from
Fig.~\ref{fig:YFS-vert} for real-gluon emission is shown in
Fig.~\ref{fig:YFS-RV}, and the application to the non-factorizable
corrections is illustrated in Fig.~\ref{fig:QCD-EW-real-if} for a set
of example diagrams.

We obtain the final result
\begin{subequations}
  \label{eq:nf-rv}
\begin{align}
  \rd\sigma_{\Paq_a\Pq_b,\nf}^{\Rs\otimes\Vew}
  =&\; 2\Re\left\{ \delta_{\Vew,\nf}^{\Paq_a\Pq_b\to\Pl_1\Pal_2\Pg} \right\}
  \rd\sigma_{\Paq_a\Pq_b,\PA}^{\Rs} , \\
  \rd\sigma_{\Pg\Pq_b,\nf}^{\Rs\otimes\Vew}
  =&\; 2\Re\left\{ \delta_{\Vew,\nf}^{\Pg\Pq_b\to\Pl_1\Pal_2\Pq_a} \right\}
  \rd\sigma_{\Pg\Pq_b,\PA}^{\Rs} , \\
  \rd\sigma_{\Pg\Paq_a,\nf}^{\Rs\otimes\Vew}
  =&\; 2\Re\left\{ \delta_{\Vew,\nf}^{\Pg\Paq_a\to\Pl_1\Pal_2\Paq_b} \right\}
  \rd\sigma_{\Pg\Paq_a,\PA}^{\Rs} ,
\end{align}  
\end{subequations}
where the tree-level cross sections $\rd\sigma^\Rs_\PA$ result from the
real NLO QCD corrections in Eqs.~\eqref{eq:dipole-nlo} 
and \eqref{eq:dipole-nlo-ch} upon neglecting non-resonant contributions.  
The virtual correction
factors have been computed both by directly evaluating the difference
of the one-loop integrals in Eq.~\eqref{eq:NNLO-nf-rv-def} and alternatively using
the EFT inspired method.  
The explicit results for each channel are
given by
\begin{subequations}
\label{eq:nf-rv-ew}
\begin{align}
  %%%%%%%%%%
  \delta_{\Vew,\nf}^{\Paq_a\Pq_b\to\Pl_1\Pal_2\Pg}
  =& - \frac{\alpha}{2\pi}
  \sum_{\substack{ii'=ab,ba\\f=1,2}} \eta_i Q_i\,\eta_f Q_f \Biggl\lbrace
  2 +\Li_{2}\left(1+\frac{\MV^2-t_{i'\Pg}}{t_{if}}\right) \nonumber\\
  &\quad +\left[ \frac{c_\epsilon}{\epsilon} - 2\ln\left(\frac{\mu_V^2-s_{12}}{\mu \MV}\right) \right]
  \left[ 1-\ln\left(\frac{\MV^2-t_{i'\Pg}}{-t_{if}}\right) \right] \Biggr\rbrace ,\\
  %%%%%%%%%%
  \delta_{\Vew,\nf}^{\Pg\Pq_b\to\Pl_1\Pal_2\Pq_a}
  =& - \frac{\alpha}{2\pi}
  \sum_{f=1,2}\eta_f Q_f \Biggl\lbrace
  \left[ \frac{c_\epsilon}{\epsilon} - 2\ln\left(\frac{\mu_V^2-s_{12}}{\mu \MV}\right) \right] \nonumber\\
  &\quad\times
  \left[ (Q_b-Q_a)
    -Q_b\ln\left(\frac{\MV^2-t_{\Pg a}}{-t_{bf}}\right)
    +Q_a\ln\left(\frac{s_{\Pg b}-\MV^2}{s_{af}}\right)\right] \nonumber\\
  &\quad+ 2(Q_b-Q_a) 
  + Q_b\Li_{2}\left(1+\frac{\MV^2-t_{\Pg a}}{t_{bf}}\right)
  - Q_a\Li_{2}\left(1-\frac{s_{\Pg b}-\MV^2}{s_{af}}\right) \Biggr\rbrace  ,\\
  \delta_{\Vew,\nf}^{\Pg\Paq_a\to\Pl_1\Pal_2\Paq_b}
  =& - \delta_{\Vew,\nf}^{\Pg\Pq_b\to\Pl_1\Pal_2\Pq_a}
  \biggr\rvert_{a\leftrightarrow b} . 
\end{align}
\end{subequations}
In the limits of soft and/or collinear gluons, these correction factors can be related to the one in Eq.~\eqref{eq:nf-virt} for the hard process without the additional QCD radiation,
\begin{subequations}
  \label{eq:limits-RV}
  \begin{align}
    &\delta_{\Vew,\nf}^{\Paq_a\Pq_b\to\Pl_1\Pal_2\Pg}(p_a,p_b,k_1,k_2,k_\Pg)
    &&\xrightarrow{\;k_\Pg\text{ soft}\;}&&
    \delta_{\Vew,\nf}^{\Paq_a\Pq_b\to\Pl_1\Pal_2}(p_a,p_b,k_1,k_2) ,\\
    &\delta_{\Vew,\nf}^{\Paq_a\Pq_b\to\Pl_1\Pal_2\Pg}(p_a,p_b,k_1,k_2,k_\Pg)
    &&\xrightarrow{\;k_\Pg\to(1-x)p_a\;}&&
    \delta_{\Vew,\nf}^{\Paq_a\Pq_b\to\Pl_1\Pal_2}(xp_a,p_b,k_1,k_2)   ,\\ 
    &\delta_{\Vew,\nf}^{\Paq_a\Pq_b\to\Pl_1\Pal_2\Pg}(p_a,p_b,k_1,k_2,k_\Pg)
    &&\xrightarrow{\;k_\Pg\to(1-x)p_b\;}&&
    \delta_{\Vew,\nf}^{\Paq_a\Pq_b\to\Pl_1\Pal_2}(p_a,xp_b,k_1,k_2)   ,\\ 
    &\delta_{\Vew,\nf}^{\Pg\Pq_b\to\Pl_1\Pal_2\Pq_a}(p_\Pg,p_b,k_1,k_2,k_a)
    &&\xrightarrow{\;k_a\to(1-x)p_\Pg\;}&&
    \delta_{\Vew,\nf}^{\Paq_a\Pq_b\to\Pl_1\Pal_2}(xp_\Pg,p_b,k_1,k_2)   ,\\ 
    &\delta_{\Vew,\nf}^{\Pg\Paq_a\to\Pl_1\Pal_2\Paq_b}(p_\Pg,p_a,k_1,k_2,k_b)
    &&\xrightarrow{\;k_b\to(1-x)p_\Pg\;}&&
    \delta_{\Vew,\nf}^{\Paq_a\Pq_b\to\Pl_1\Pal_2}(p_a,xp_\Pg,k_1,k_2)  .
  \end{align}
\end{subequations}
These properties are crucial in the cancellation of IR singularities, as will be discussed in more detail in Sect.~\ref{sec:master}.

%%%%%%%%%%%%%%%%%%%%%%%%%%%%%%
\subsubsection{Virtual QCD \texorpdfstring{$\times$}{x} real photonic corrections}
\label{sec:VR}

The non-factorizable (virtual QCD)$\times$(real photonic) corrections
are constructed from the one-loop QCD amplitude for dilepton plus
photon production~\eqref{eq:vQCD-rEW} by an extension of the method
used at NLO in Sect.~\ref{sec:real-nonfact}.  We split photon
emission from the $V$-boson line into initial- and final-state
radiation parts via the partial fractioning~\eqref{eq:prop-id}. 
The non-factorizable corrections arise from
the interference of diagrams with initial-state radiation with those
with final-state radiation.  The contributing diagrams are obtained by
attaching a virtual gluon in all possible ways to the quark line in
the NLO real non-factorizable corrections in
Fig.~\ref{fig:nlo-pa-real}(c).  The two possible contributions are
illustrated in Fig.~\ref{fig:NNLO-nf-graphs}(c) in terms of
interference diagrams.  Again, a resonance enhancement only occurs if
the momentum of the real photon is soft.  The contributions of the form of the
right-hand diagram are already factorized into the virtual QCD
corrections to $V$-boson production and the final--initial
real-photonic corrections. In contrast, the contributions shown in the
left-hand diagram involve the virtual QCD corrections to $V$-boson
plus photon production. The contributing diagrams are obtained from
those for the double-virtual corrections (c.f.\
Fig.~\ref{fig:QCD-EW-virt-if}) by turning the virtual photon into a
real photon by detaching it from the final-state lepton line.

\begin{figure}
  \centering
  \Vcenter{\includegraphics{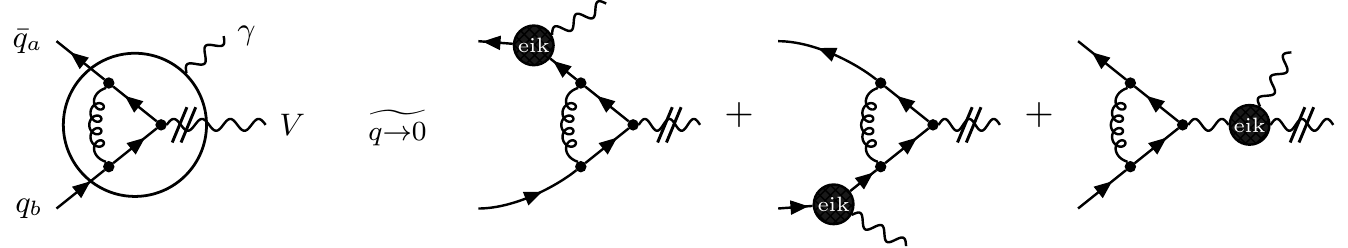}} 
  \caption{Simplifications of the (virtual QCD)$\times$(real photonic
    initial-state) corrections appearing in the left-hand diagram of
    Fig.~\ref{fig:NNLO-nf-graphs}(c) as a result of the YFS
    identities of Figs.~\ref{fig:YFS-se} and \ref{fig:YFS-vert}.}
  \label{fig:vr-is}
\end{figure}

The fastest way to calculate the latter contributions is again to make
use of the YFS arguments described in Sect.~\ref{sec:yfs}, where we
did not specify whether the soft photon with momentum $q$ is virtual or
real.  This is true, in particular, for the identities shown in
Figs.~\ref{fig:YFS-se} and \ref{fig:YFS-vert} used in the
derivation~\ref{sec:vv-yfs} of Sect.~\ref{sec:VV} for the
double-virtual case. In the same way as for the double-virtual
corrections, these simplifications imply that the (virtual
QCD)$\times$(real photonic) initial-state corrections to the matrix
element factorize into the QCD on-shell vertex corrections $
\delta_{\Vs}^{\Paq_a\Pq_b\to\Pl_1\Pal_2}$ 
of Eq.~\eqref{eq:nlo-qcd} and the eikonal current
$\mathcal{J}_{\pro}$ of Eq.~\eqref{eq:nlo-nf-real-delta}. This result
is illustrated diagrammatically in Fig.~\ref{fig:vr-is}.
The same result is obtained using the two other methods of
Sect.~\ref{sec:VV}, since the cancellations of the diagrams in
Figs.~\ref{fig:YFS-se} and \ref{fig:YFS-vert} can be observed
already after performing the integral over the gluon
loop momentum and holds both for real or virtual soft photons. 

The result for the complete (virtual QCD)$\times$(real photonic)
non-factorizable corrections, thus, is
\begin{equation}
  \rd\sigma_{\Paq_a\Pq_b,\nf}^{\Vs\otimes\Rew}
  \,=\, \delta_{\Rew,\nf}^{\Paq_a\Pq_b\to\Pl_1\Pal_2\Pgg} 
  \rd\sigma_{\Paq_a\Pq_b,\PA}^{\Vs} 
  \,=\, \delta_{\Rew,\nf}^{\Paq_a\Pq_b\to\Pl_1\Pal_2\Pgg} 
  \;2\Re\left\{ \delta_{\Vs}^{\Paq_a\Pq_b\to\Pl_1\Pal_2} \right\}
  \rd\sigma^{0}_{\Paq_a\Pq_b,\PA} ,
\end{equation}
with the real-photon correction factor~\eqref{eq:nlo-nf-real-delta}.
The soft-photon factor required for the regularization of the soft
singularities using the slicing method is identical to the one at NLO
and given in Eq.~\eqref{eq:nf-slice}.

%%%%%%%%%%%%%%%%%%%%%%%%%%%%%%
\subsubsection{Double-real corrections}
\label{sec:RR}

The non-factorizable double-real corrections are calculated by applying the construction of Sect.~\ref{sec:real-nonfact} to the three partonic channels~\eqref{eq:rQCD-rEW} for dilepton production in association with a jet and a photon. 
As for the (real QCD)$\times$(virtual EW) corrections the kinematics of the extra final-state QCD parton will be treated exactly so that the resonance condition is given by Eq.~\eqref{eq:resonance-gluon} for the case of the subprocess $\Paq_a\Pq_b\to\Pl_1\Pal_2\Pgg\Pg$, and analogous expressions for the other two subprocesses.
The non-factorizable contributions are again given by the interference of diagrams with soft-photon emission from the production and decay stages of the vector boson. 
Diagrams for the quark--antiquark and quark--gluon initial states are shown in Fig.~\ref{fig:NNLO-nf-graphs}(d). 
Note that for the channels with an initial-state gluon, photon radiation from the production subprocess is not limited to initial-state radiation because of photon emission off the final-state quark or antiquark.

The YFS arguments of Sect.~\ref{sec:yfs} prove that the non-factorizable real photonic corrections, which are induced by soft photons with $E_\Pgg\lesssim\GV$ only, factorize from the respective squared amplitudes of the processes~\eqref{eq:rQCD} with external (hard or soft) partons.
The squared matrix elements for the double-real corrections, 
\begin{subequations}
  \label{eq:nf-rr}
  \begin{align}
    \left\lvert\M_{\Rs\otimes\Rew}^{\Paq_a\Pq_b\to\Pl_1\Pal_2\Pgg\Pg}\right\rvert^2_\nf
    &{}\;=\; \delta_{\Rew,\nf}^{\Paq_a\Pq_b\to\Pl_1\Pal_2\Pgg\Pg} 
    \left\lvert\M_{\Rs,\PA}^{\Paq_a\Pq_b\to\Pl_1\Pal_2\Pg}\right\rvert^2 , \\
    \left\lvert\M_{\Rs\otimes\Rew}^{\Pg\Pq_b\to\Pl_1\Pal_2\Pgg\Pq_a}\right\rvert^2_\nf
    &{}\;=\; \delta_{\Rew,\nf}^{\Pg\Pq_b\to\Pl_1\Pal_2\Pgg\Pq_a} 
    \left\lvert\M_{\Rs,\PA}^{\Pg\Pq_b\to\Pl_1\Pal_2\Pq_a}\right\rvert^2 , \\
    \left\lvert\M_{\Rs\otimes\Rew}^{\Pg\Paq_a\to\Pl_1\Pal_2\Pgg\Paq_b}\right\rvert^2_\nf
    &{}\;=\; \delta_{\Rew,\nf}^{\Pg\Paq_a\to\Pl_1\Pal_2\Pgg\Paq_b} 
    \left\lvert\M_{\Rs,\PA}^{\Pg\Paq_a\to\Pl_1\Pal_2\Paq_b}\right\rvert^2 ,
  \end{align}
\end{subequations}
are, thus, given by the ones for the tree matrix elements with an additional final-state parton,  multiplied by the correction factors
\begin{subequations}
  \label{eq:deltanf-rr}
  \begin{align}
    \delta_{\Rew,\nf}^{\Paq_a\Pq_b\to\Pl_1\Pal_2\Pgg\Pg} 
    =&\;
    -2\Re\Bigl\{
    \mathcal{J}_\pro^\mu(p_a,p_b,p_a+p_b-k_\Pg)
    \bigl(\mathcal{J}_{\dec,\mu}(k_1,k_2)\bigr)^* \Bigr\} , \\
    \delta_{\Rew,\nf}^{\Pg\Pq_b\to\Pl_1\Pal_2\Pgg\Pq_a} 
    =&\;
    -2\Re\Bigl\{
    \mathcal{J}_\pro^\mu(-k_a,p_b,p_\Pg+p_b-k_a)
    \bigl(\mathcal{J}_{\dec,\mu}(k_1,k_2)\bigr)^* \Bigr\} , \\
    \delta_{\Rew,\nf}^{\Pg\Paq_a\to\Pl_1\Pal_2\Pgg\Paq_b} 
    =&\;
    -2\Re\Bigl\{
    \mathcal{J}_\pro^\mu(p_a,-k_b,p_\Pg+p_a-k_b)
    \bigl(\mathcal{J}_{\dec,\mu}(k_1,k_2)\bigr)^* \Bigr\} , 
  \end{align}
\end{subequations}
for the non-factorizable real photonic corrections, which are derived from the modified eikonal currents~\eqref{eq:eik}.
The difference between the three versions of $\mathcal{J}_\pro^\mu$ in Eq.~\eqref{eq:deltanf-rr} lies only in the momentum insertions; the first two arguments of $\mathcal{J}_\pro^\mu$ refer to the momenta of the photon-radiating (anti)quark legs, the third to the resonant internal $V$ boson which takes care of the momentum loss induced by parton emission before the resonance. 
Note that the gluon momentum (but not the soft photon momentum) now enters momentum conservation, i.e.\ $p_a+p_b=k_1+k_2+k_\Pg$ for process~\eqref{eq:rr-qqb}, etc., in the evaluation of $\delta_{\Rew,\nf}^{\,\cdots}$ via Eq.~\eqref{eq:deltanf-rr}.

As already done at NLO, we  regularize the soft-photon singularity by applying a slicing cut $\Delta E\ll\GV$ on the photon energy.
Soft photons obeying $\Delta E<E_\Pgg\lesssim\GV\ll \MV$ are included in the integration of $\delta_{\Rew,\nf}^{\,\cdots}$ in Eq.~\eqref{eq:deltanf-rr}, and extremely soft photons with $E_\Pgg<\Delta E$ are included via eikonal integrals based on the usual soft-photon asymptotics as given in Eq.~\eqref{eq:nf-slice} for the NLO case.
Here we have to generalize this result to take into account the hard-gluon kinematics and the possibility to have gluons in the initial state.
The soft factors explicitly read
\begin{subequations}
  \label{eq:slice-nf-rr}
\begin{align}
  %%%%%%%%%%
   \delta_{\soft}^{\Paq_a\Pq_b\to\Pl_1\Pal_2\Pg}
  =& \,\frac{\alpha}{\pi}
  \sum_{{\substack{ii'=ab,ba\\ff'=12,21}}} \eta_i Q_i\,\eta_f Q_f \Biggl\lbrace
  \left[ \frac{c_\epsilon}{\epsilon} - 2\ln\left(\frac{2\Delta E}{\mu}\right) \right]
  \left[ 1-\ln\left(\frac{\MV^2-t_{i'\Pg}}{-t_{if}}\right) \right] \nonumber\\
  &+\frac{s_{ab}+\MV^2}{s_{ab}-\MV^2}\ln\left(\frac{s_{ab}}{\MV^2}\right)
  - \frac{1}{2}\ln^2\left(\frac{s_{ab}}{\MV^2}\right)
  - \Li_{2}\left(-\frac{s_{\Pg f}}{\MV^2}\right)
  - \Li_{2}\left(\frac{s_{\Pg f^\prime}}{s_{ab}}\right) \nonumber\\
  &+\Li_{2}\left(-\frac{t_{i'f}}{t_{if}}\right)
  - \Li_{2}\left(\frac{t_{i\Pg}}{s_{ab}+t_{i\Pg}}\right)
  - \Li_{2}\left(\frac{-t_{i'\Pg}}{\MV^2-t_{i'\Pg}}\right) \Biggr\rbrace , \\ 
  %%%%%%%%%%
  \delta_{\soft}^{\Pg\Pq_b\to\Pl_1\Pal_2\Pq_a}
  =& \,\frac{\alpha}{\pi}
  \sum_{ff'=12,21}\eta_f Q_f \Biggl\lbrace
  \left[ \frac{c_\epsilon}{\epsilon} - 2\ln\left(\frac{2\Delta E}{\mu}\right) \right] \nonumber\\
  &\times
  \left[ Q_b-Q_a
    -Q_b\ln\left(\frac{\MV^2-t_{\Pg a}}{-t_{bf}}\right)
    +Q_a\ln\left(\frac{s_{\Pg b}-\MV^2}{s_{af}}\right)\right] \nonumber\\
  &+(Q_b-Q_a) \biggl[
  \frac{s_{\Pg b}+\MV^2}{s_{\Pg b}-\MV^2}\ln\left(\frac{s_{\Pg b}}{\MV^2}\right)
  -\Li_{2}\left(-\frac{s_{af}}{\MV^2}\right)
  -\Li_{2}\left(\frac{s_{af^\prime}}{s_{\Pg b}}\right)
  \biggr]\nonumber\\
  &+Q_b\biggl[
  \Li_{2}\left(-\frac{t_{\Pg f}}{t_{bf}}\right)
  -\Li_{2}\left(\frac{t_{ba}}{s_{\Pg b}+t_{ba}}\right)
  -\Li_{2}\left(\frac{-t_{\Pg a}}{\MV^2-t_{\Pg a}}\right)
  -\frac{1}{2}\ln^2\left(\frac{s_{\Pg b}}{\MV^2}\right)
  \biggr]\nonumber\\
  &-Q_a\left[
  \Li_{2}\left(-\frac{\MV^2}{s_{\Pg b}}\,\frac{s_{af^\prime}}{s_{af}}\right)
  +\Li_{2}\left(1- \frac{s_{\Pg b}}{\MV^2}\right)
  \right] \Biggr\rbrace , \\
  \delta_{\soft}^{\Pg\Paq_a\to\Pl_1\Pal_2\Paq_b}
  =& - \delta_{\soft}^{\Pg\Pq_b\to\Pl_1\Pal_2\Pq_a}
  \biggr\rvert_{a\leftrightarrow b} ,
\end{align}
\end{subequations}
where $s_{ii'}=(p_i+p_{i'})^2$ and $t_{if}=(p_i-k_f)^2$.
In the soft-gluon limit, the correction factors reduce to the respective case without the gluon emission from Sect.~\ref{sec:real-nonfact},
\begin{subequations}
\label{eq:limits-RR}
\begin{align}
  &\delta_{\soft}^{\Paq_a\Pq_b\to\Pl_1\Pal_2\Pg}(p_a,p_b,k_1,k_2,k_\Pg)
  &&\xrightarrow{\;k_\Pg\text{ soft}\;}&&
  \delta_{\soft}^{\Paq_a\Pq_b\to\Pl_1\Pal_2}(p_a,p_b,k_1,k_2).
\end{align}
In the limits of collinear quark--gluon splittings in the initial states, the frame dependence of the slicing factors becomes apparent, and we obtain
\begin{align}
  &\delta_{\soft}^{\Paq_a\Pq_b\to\Pl_1\Pal_2\Pg}(p_a,p_b,k_1,k_2,k_\Pg)
  &&\xrightarrow{\;k_\Pg\to(1-x)p_a\;}&&
  \delta_{\soft,\boost(x,1)}^{\Paq_a\Pq_b\to\Pl_1\Pal_2}(xp_a,p_b,k_1,k_2) , \\ 
  &\delta_{\soft}^{\Paq_a\Pq_b\to\Pl_1\Pal_2\Pg}(p_a,p_b,k_1,k_2,k_\Pg)
  &&\xrightarrow{\;k_\Pg\to(1-x)p_b\;}&&
  \delta_{\soft,\boost(1,x)}^{\Paq_a\Pq_b\to\Pl_1\Pal_2}(p_a,xp_b,k_1,k_2) , \\ 
  &\delta_{\soft}^{\Pg\Pq_b\to\Pl_1\Pal_2\Pq_a}(p_\Pg,p_b,k_1,k_2,k_a)
  &&\xrightarrow{\;k_a\to(1-x)p_\Pg\;}&&
  \delta_{\soft,\boost(x,1)}^{\Paq_a\Pq_b\to\Pl_1\Pal_2}(xp_\Pg,p_b,k_1,k_2) , \\ 
  &\delta_{\soft}^{\Pg\Paq_a\to\Pl_1\Pal_2\Paq_b}(p_\Pg,p_a,k_1,k_2,k_b)
  &&\xrightarrow{\;k_b\to(1-x)p_\Pg\;}&&
  \delta_{\soft,\boost(1,x)}^{\Paq_a\Pq_b\to\Pl_1\Pal_2}(p_a,xp_\Pg,k_1,k_2) .
\end{align}
\end{subequations}
The generic factor $\delta_{\soft,\boost(z_a,z_b)}^{\Paq_a\Pq_b\to\Pl_1\Pal_2}(p_a,p_b,k_1,k_2)$ is the slicing factor obtained for the process $\Paq_a\Pq_b\to\Pl_1\Pal_2$, evaluated not in the partonic centre-of-mass frame (given by the rest frame of the momentum $p_a+p_b$), but in the rest frame of the momentum $\tfrac{p_a}{z_a}+\tfrac{p_b}{z_b}$, which is Lorentz boosted along the beam axis
\begin{align}
  \label{eq:nf-soft-boost}
  \delta_{\soft,\boost(z_a,z_b)}^{\Paq_a\Pq_b\to\Pl_1\Pal_2}(p_a,p_b,k_1,k_2)
  =&\; \frac{\alpha}{\pi}
  \sum_{{\substack{ii'=ab,ba\\f=1,2}}}\eta_i Q_i\,\eta_f Q_f \Biggl\lbrace
  \left[ \frac{c_\epsilon}{\epsilon} - 2\ln\left(\frac{2\Delta E}{\mu}\right) \right] 
  \left[ 1-\ln\left(\frac{\MV^2}{-t_{if}}\right)\right] \nonumber\\
  &+\frac{z_a+z_b}{z_a-z_b}\ln\left(\frac{z_a}{z_b}\right)
  +\Li_{2}\left(-\frac{z_i}{z_{i'}}\frac{t_{i'f}}{t_{if}}\right)
  +\Li_{2}\left(1-\frac{z_{i'}}{z_i}\right) \nonumber\\
  &-\Li_{2}\left(\left(1-\frac{z_{i'}}{z_i}\right)\frac{-t_{if}}{\MV^2}\right)
  -\Li_{2}\left(\left(1-\frac{z_i}{z_{i'}}\right)\frac{-t_{i'f}}{\MV^2}\right)
  \Biggr\rbrace . 
\end{align}
For the proper cancellation of the overlapping (QCD)$\times$(photonic) IR singularities, these correction factors are employed in the context of the dipole subtraction formalism.

%%%%%%%%%%%%%%%%%%%%%%%%%%%%%%%%%%%%%%%%%%%%%%%%%%
\subsection{Complete result for the non-factorizable corrections}
\label{sec:master}

The non-factorizable corrections to the cross section are obtained by integrating the different contributions calculated in Sect.~\ref{sec:calc-nf-nnlo} over the respective phase spaces,
\begin{align}
  \label{eq:generic-master}
  \sighat_\nf^{\NNLO_{\rs\otimes\rew}} \;=\;
  & \int_{2} \rd\sigma_\nf^{\Vs\otimes\Vew}
  + \int_{3} \rd\sigma_\nf^{\Rs\otimes\Vew}
  + \int_{2} \rd\sigma_\nf^{\Cs\otimes\Vew} \nonumber\\
  &+ \iint\limits_{2+\Pgg} \rd\sigma_\nf^{\Vs\otimes\Rew}
  + \iint\limits_{3+\Pgg} \rd\sigma_\nf^{\Rs\otimes\Rew}
  + \iint\limits_{2+\Pgg} \rd\sigma_\nf^{\Cs\otimes\Rew} \nonumber\\
  \;=\;& \int_{2} \rd\sigma_\PA^{\Vs} \;2\Re\left\{\delta_{\Vew,\nf}^{2\to2}\right\}
  + \int_{3} \rd\sigma_\PA^{\Rs} \;2\Re\left\{\delta_{\Vew,\nf}^{2\to3}\right\}
  + \int_{2} \rd\sigma_\PA^{\Cs} \;2\Re\left\{\delta_{\Vew,\nf}^{2\to2}\right\} \nonumber\\
  &+ \iint\limits_{2+\Pgg} \rd\sigma_\PA^{\Vs} \; \delta_{\Rew,\nf}^{2\to2+\Pgg} 
  + \iint\limits_{3+\Pgg} \rd\sigma_\PA^{\Rs} \; \delta_{\Rew,\nf}^{2\to3+\Pgg} 
  + \iint\limits_{2+\Pgg} \rd\sigma_\PA^{\Cs} \; \delta_{\Rew,\nf}^{2\to2+\Pgg} .
\end{align}
Here $\delta_{\Vew,\nf}^{2\to2}$ is a shorthand for the virtual
NLO non-factorizable corrections~\eqref{eq:nf-virt}, while the factor
$\delta_{\Vew,\nf}^{2\to3} $ is a generic expression for the
virtual non-factorizable corrections to the various real-QCD
correction channels given in Eq.~\eqref{eq:nf-rv}. Similar abbreviations have
been introduced for the real non-factorizable corrections. In the
first line of Eq.~\eqref{eq:generic-master}  a
collinear counterterm $ \rd\sigma_\nf^{\Cs\otimes\Vew}$ has been added to 
subtract collinear
singularities remaining in the sum of the double-virtual and the (real
QCD)$\times$(virtual EW) corrections that are absorbed into the PDFs. The
term $\rd\sigma_\nf^{\Cs\otimes\Rew}$ plays the same role for
the sum of the (virtual QCD)$\times$(real EW) and the double-real
corrections in the second line. Since the non-factorizable EW
corrections contain only soft singularities, all collinear
singularities are purely of QCD origin. 
The collinear subtraction term exploits factorization properties in the collinear limits and is therefore universal in the sense that it is constructed as a convolution of the Altarelli--Parisi splitting kernels with the lower-order (in $\alphas$) cross section.
Therefore, the results of Sect.~\ref{sec:nlo-nonfact} for the non-factorizable EW corrections can be used to write these collinear counterterms schematically as a product of $\rd\sigma_\PA^{\Cs}$, which is obtained from the collinear counterterm appearing in the NLO QCD corrections in Eq.~\eqref{eq:nlo-dipole} upon neglecting non-resonant contributions, and the photonic correction factors $\delta_{\Vew,\nf}^{2\to2}$ and $\delta_{\Rew,\nf}^{2\to2+\Pgg}$, respectively.  
Each of the six
integrals in Eq.~\eqref{eq:generic-master} are separately IR
divergent, however, all divergences cancel in the sum.  
As described
for the NLO calculation in Sect.~\ref{sec:NLO}, IR singularities
from QCD emission are rearranged using the dipole subtraction
formalism, while the soft singularities from photon emission are
regularized by the slicing method.

We first apply the dipole subtraction formalism to Eq.~\eqref{eq:generic-master} in order to cancel the IR singularities associated with the QCD corrections.
For the double-real contribution, we construct the dipole terms in the following way,
\begin{equation}
  \label{eq:dipole-RR}
  \iint\limits_{3+\Pgg} \biggl\{
  \rd\sigma_\PA^{\Rs} \; \delta_{\Rew,\nf}^{2\to3+\Pgg} \bigl(\Phi_{3+\Pgg}\bigr)
  - \sum_\text{dipoles}\rd\sigma_\PA^0
  \; \delta_{\Rew,\nf}^{2\to2+\Pgg} \bigl(\widetilde\Phi_{2+\Pgg}\bigr)
  \!\otimes\!\CSV \biggr\} , 
\end{equation}
where the relative correction factors $\delta_{\Rew,\nf}^{2\to n+\Pgg}$ of the non-factorizable photonic corrections match the differential cross sections they are multiplied with.
In particular, the correction factor to the dipoles $\CSV$ is consistently evaluated on the respective dipole phase space.
In the singular regions, where the additional QCD emission tends towards a soft and/or collinear configuration, the correction factors coincide 
\begin{equation}
  \delta_{\Rew,\nf}^{2\to3+\Pgg} 
  \;\xrightarrow[\text{QCD partons}]{\text{soft and/or collinear}}\;
  \delta_{\Rew,\nf}^{2\to2+\Pgg} ,
\end{equation}
and a proper cancellation of divergences is ensured.

Similarly to the double-real case, the dipole terms for the (real QCD)$\times$(virtual EW) corrections are constructed as follows,
\begin{equation}
  \label{eq:dipole-RV}
  \int_{3} \biggl\{
    \rd\sigma_\PA^{\Rs} \;2\Re\Bigl\{\delta_{\Vew,\nf}^{2\to3}\bigl(\Phi_3\bigr)\Bigr\}
    - \sum_\text{dipoles}\rd\sigma_\PA^0
    \;2\Re\Bigl\{\delta_{\Vew,\nf}^{2\to2}\bigl(\widetilde\Phi_2\bigr)\Bigr\}  
    \!\otimes\!\CSV \biggr\} .
\end{equation}
The relation between the correction factors $\delta_{\Vew,\nf}^{2\to3}$ and $\delta_{\Vew,\nf}^{2\to2}$ in the soft and/or collinear limits given in Eq.~\eqref{eq:limits-RV} ensures that the cancellation in the singular regions remains unaffected although different correction factors are introduced for the real cross-section and subtraction terms.

As in the NLO QCD case described in Sect.~\ref{sec:QCD}, the subtraction functions are integrated over the one-parton phase space containing the soft and collinear QCD singularities, leading to the \CSI, \CSK, and \CSP terms.
This integration is essentially unchanged in the \order{\alphas\alpha} non-factorizable corrections, since the additional \order{\alpha} factors $\delta_{\Rew,\nf}^{2\to2+\Pgg}$ and $\delta_{\Vew,\nf}^{2\to2}$ multiplying the dipoles \CSV in Eqs.~\eqref{eq:dipole-RR} and \eqref{eq:dipole-RV}, respectively, do not depend on the singular parton kinematics, but merely appear as constant factors in the singular integrals.
Introducing the insertion operators of the dipole subtraction formalism, as described in Sect.~\ref{sec:QCD}, we obtain
\begin{align}
  \label{eq:dipole-master}
  \sighat_\nf^{\NNLO_{\rs\otimes\rew}} 
  =& \hphantom{{}+{}}\int_{2} \biggl\{ \rd\sigma_\PA^{\Vs} + \rd\sigma_\PA^0\!\otimes\!\CSI \biggr\}
  \;2\Re\left\{\delta_{\Vew,\nf}^{2\to2}\right\} \nonumber\\
  &+ \int_{3} \biggl\{
  \rd\sigma_\PA^{\Rs} \;2\Re\Bigl\{\delta_{\Vew,\nf}^{2\to3}\Bigr\}
  - \sum_\text{dipoles}\rd\sigma_\PA^0\;2\Re\Bigl\{\delta_{\Vew,\nf}^{2\to2}\Bigr\}  
  \!\otimes\!\CSV \biggr\} \nonumber\\
  &+ \int_0^1\rd x\int_2
  \rd\sigma_\PA^0\;2\Re\left\{\delta_{\Vew,\nf}^{2\to2}\right\} \!\otimes\!(\CSK+\CSP) \nonumber\\
  &+ \iint\limits_{2+\Pgg}  
  \biggl\{ \rd\sigma_\PA^{\Vs} + \rd\sigma_\PA^0\!\otimes\!\CSI \biggr\}
  \; \delta_{\Rew,\nf}^{2\to2+\Pgg} \nonumber\\
  &+ \iint\limits_{3+\Pgg} \biggl\{
  \rd\sigma_\PA^{\Rs} \; \delta_{\Rew,\nf}^{2\to3+\Pgg}
  - \sum_\text{dipoles}\rd\sigma_\PA^0
  \; \delta_{\Rew,\nf}^{2\to2+\Pgg}\!\otimes\!\CSV \biggr\} \nonumber\\
  &+ \int_0^1\rd x\iint\limits_{2+\Pgg}
  \rd\sigma_\PA^0\; \delta_{\Rew,\nf}^{2\to2+\Pgg}\!\otimes\!(\CSK+\CSP) ,
\end{align}
where we have suppressed the dependence on the phase-space kinematics for the sake of transparency.
It is implicitly assumed that all correction factors are evaluated with the same kinematics as the differential cross sections $\rd\sigma$ in front of them.

All lines of Eq.~\eqref{eq:dipole-master} are now individually free of
QCD singularities, however, the integrals over the photon phase space
in the last three lines still gives rise to soft singularities that
cancel against the IR singularities in the virtual photonic corrections in
the first three lines.
These soft singularities are dealt with using the slicing method as described in
Sect.~\ref{sec:real-nonfact} for the NLO EW case.  As a result we
are able to arrange the six contributions in
Eq.~\eqref{eq:dipole-master}  into a form where all IR divergences are cancelled in the integrands explicitly,
\begin{equation}
  \label{eq:final-master}
  \sighat_\nf^{\NNLO_{\rs\otimes\rew}}
  =\sigreg_\nf^{\Vs\otimes\Vew}
  +\sigreg_\nf^{\Vs\otimes\Rew}
  +\sigreg_\nf^{\Rs\otimes\Vew}
  +\sigreg_\nf^{\Rs\otimes\Rew}
  +\sigreg_\nf^{\Cs\otimes\Vew}
  +\sigreg_\nf^{\Cs\otimes\Rew} ,
\end{equation}
where  each term is an IR-finite object and
its phase-space integration can be performed numerically in four
dimensions. However, individual terms depend on an artificial energy-slicing parameter $\Delta E$ that cancels in the sum. 
Equation~\eqref{eq:final-master} will be our master formula for the
numerical evaluation discussed in Sect.~\ref{sec:numerics}.
In the follwing we describe the construction of the  individual ingredients. 
In order to discuss some subtle points in the implementation that are not obvious in  the compact notation used here, we give the explicit expressions for all contributions for  the quark--antiquark and quark--gluon induced channels in Appendix~\ref{app:dipole}.

The first two terms in Eq.~\eqref{eq:final-master} arise from the sum
of the double-virtual and the (virtual QCD)$\times$(real
photonic) corrections, including the insertion operators from the dipole
formalism.  Analogously to the NLO case discussed in the end of
Sect.~\ref{sec:real-nonfact}, the integral over the photon
momentum with $E_\Pgg<\Delta E$ in the real photonic corrections 
is computed analytically and added to the double-virtual corrections,
\begin{align} 
%%%%%%%%%%%%%%%%%%%%
  \label{eq:master-VV}
  \sigreg_\nf^{\Vs\otimes\Vew}&= \int_2 \biggl\{
  \rd\sigma_\PA^{\Vs} + \rd\sigma_\PA^0\!\otimes\!\CSI \biggr\}\;
  \Bigl[2\Re\left\{\delta_{\Vew,\nf}^{2\to2}\right\}
    +\delta_{\soft}^{2\to2}(\Delta E)\Bigr], \\
%%%%%%%%%%%%%%%%%%%%
  \label{eq:master-VR}
  \sigreg_\nf^{\Vs\otimes\Rew}&=
  \iint\limits_{\substack{2+\Pgg\\E_\Pgg>\Delta E}} \biggl\{
  \rd\sigma_\PA^{\Vs} + \rd\sigma_\PA^0\!\otimes\!\CSI \biggr\}
  \; \delta_{\Rew,\nf}^{2\to2+\Pgg} ,
\end{align}
where the integrated soft-photon factor $\delta_{\soft}^{2\to2}$ is the same as at NLO, see Eq.~\eqref{eq:nf-slice}. 

For the contributions involving real QCD corrections, special care must be taken since the procedure of applying a cut on the photon energy depends on
the frame of reference.
In order not to spoil the cancellation of the IR singularities 
treated with the subtraction method, the
slicing cut on the photon energy in $\sigreg_\nf^{\Rs\otimes\Rew}$ is
applied in the partonic centre-of-mass frame of the $2\to3+\Pgg$
process for both the double-real cross section and the dipole terms.
This results in the following expressions for the IR-finite real-gluon contributions to the cross section,
\begin{align}
%%%%%%%%%%%%%%%%%%%
  \label{eq:master-RV}
  \sigreg_\nf^{\Rs\otimes\Vew}=& \int_{3} \biggl\{
  \rd\sigma_\PA^{\Rs} \;
  \left[2\Re\Bigl\{\delta_{\Vew,\nf}^{2\to3}\Bigr\} 
    +\delta_{\soft}^{2\to3}(\Delta E)  \right] \nonumber\\
  &-
  \sum_\text{dipoles}\rd\sigma_\PA^0\;\left[2\Re\Bigl\{\delta_{\Vew,\nf}^{2\to2}\Bigr\}
    + \delta_{\soft,\boost(x_\text{dip})}^{2\to2}(\Delta E)\right]
  \!\otimes\!\CSV \biggr\} ,
\\
%%%%%%%%%%%%%%%%%%%
  \label{eq:master-RR}
  \sigreg_\nf^{\Rs\otimes\Rew}\,
  =&\iint\limits_{\substack{3+\Pgg\\E_\Pgg>\Delta E}} \biggl\{
  \rd\sigma_\PA^{\Rs} \; \delta_{\Rew,\nf}^{2\to3+\Pgg} -
  \sum_\text{dipoles}\rd\sigma_\PA^0 \;
  \delta_{\Rew,\nf}^{2\to2+\Pgg}\!\otimes\!\CSV \biggr\} ,
\end{align}
using a generic notation $\boost(x_\text{dip})$ for the boost 
in the soft-photon factor of Eq.~\eqref{eq:nf-soft-boost}.
Note that the dipole kinematics in Eq.~\eqref{eq:master-RR} is
boosted along the beam axis, c.f.\ Eq.~\eqref{eq:dipole_ii}.
Therefore, 
the analytical integration over the soft-photon phase space results in
the boosted slicing factor $\delta_{\soft,\boost}^{2\to2}$, given in Eq.~\eqref{eq:nf-soft-boost},
for the integrated dipole terms appearing in Eq.~\eqref{eq:master-RV}.
Inspecting the limiting behaviour of $\delta_{\soft}^{2\to3}$ and 
$\delta_{\soft,\boost}^{2\to2}$ in the
singular limits, as given in Eq.~\eqref{eq:limits-RR}, 
one can easily verify that expression~\eqref{eq:master-RV} is
IR finite.

Finally, we consider the convolution terms with additional virtual or
real EW corrections in the third and sixth line
of Eq.~\eqref{eq:dipole-master}. 
The cut on the photon energy $E_\Pgg>\Delta E$ in the bremsstrahlung contribution 
is applied in the rest frame of the incoming momenta \emph{before} the collinear
splittings, i.e.\ the rest frame of the momenta $p_a/x+p_b$ or $p_a+p_b/x$, where 
$p_a$ and $p_b$ denote the momenta entering the
hard scattering cross section $\sigma_\PA^0$.
This choice is consistent with the cut prescription chosen in the 
the double-real corrections in
Eq.~\eqref{eq:master-RR}. 
For the resulting IR-finite contributions to
the cross section we obtain
\begin{align}
%%%%%%%%%%%%%%%%%
 \label{eq:master-CV}
 \sigreg_\nf^{\Cs\otimes\Vew}=&
 \int_0^1\rd x\int_2
 \rd\sigma_\PA^0\;\Bigl[  2\Re\left\{\delta_{\Vew,\nf}^{2\to2}\right\}
 +\delta_{\soft,\boost(x)}^{2\to2}(\Delta E)\Bigr]\!\otimes\!(\CSK+\CSP) , \\
%%%%%%%%%%%%%%%%%
 \label{eq:master-CR}
 \sigreg_\nf^{\Cs\otimes\Rew} =& \int_0^1\rd x\iint\limits_{\substack{2+\Pgg\\E_\Pgg>\Delta E}}
 \rd\sigma_\PA^0\; \delta_{\Rew,\nf}^{2\to2+\Pgg}\!\otimes\!(\CSK+\CSP) .
\end{align}

%%%%%%%%%%%%%%%%%%%%%%%%%%%%%%%%%%%%%%%%%%%%%%%%%%%%%%%%%%%%%%%%%%%%%%
\section{Numerical results for non-factorizable \texorpdfstring{\order{\alphas\alpha}}{O(as a)} corrections}
\label{sec:numerics}

The results presented in this section are evaluated using the identical setup as in the NLO calculation, which is described in Sect.~\ref{sec:nlo-num}.

\begin{figure}
  \centering
  \includegraphics[scale=0.8]{{{images/plots/Wp.mtll.rel.Oaas}}}
  \qquad
  \includegraphics[scale=0.8]{{{images/plots/Wp.ptl.rel.Oaas}}}
  \caption{Relative non-factorizable corrections of \order{\alphas\alpha} to the
distributions in the transverse mass (left) and transverse lepton momentum (right)
for $\PWp$ production at the LHC, broken up into contributions of partonic $\Pq\Paq/\Pq\Pg$ initial states
and virtual/real photonic contributions.}
  \label{fig:Wp-Oaas-nf}
  \vspace*{2em}
  \includegraphics[scale=0.8]{{{images/plots/Z.mll.rel.Oaas}}}
  \qquad
  \includegraphics[scale=0.8]{{{images/plots/Z.ptl.rel.Oaas}}}
  \caption{Relative non-factorizable corrections of \order{\alphas\alpha} to the
distributions in the invariant mass (left) and transverse lepton momentum (right)
for \PZ production at the LHC, broken up into contributions of partonic $\Pq\Paq/\Pq\Pg$ initial states
and virtual/real photonic contributions.}
  \label{fig:Z-Oaas-nf}
\end{figure}

Figure~\ref{fig:Wp-Oaas-nf} shows our numerical results for the \order{\alphas\alpha} non-factorizable corrections to the $M_{\rT,\Pgn\Pl}$ and $p_{\rT,\Pl}$ distributions for $\PWp$ production at the LHC in terms of relative corrections factors ($\delta$) to the LO prediction.
The respective absolute distributions and the results of the NLO calculation are shown in Fig.~\ref{fig:distNLO-W}.
The contributions induced by virtual photons, indicated by $\alpha_{\text{virt}}$, are IR regularized upon adding the real soft-photon counterpart which accounts for the emission of photons of energy $E_\Pgg<\Delta E\ll\GV$.
Correspondingly, the real photonic contributions denoted by $\alpha_{\text{real}}$ only consider the emission of photons with $E_\Pgg>\Delta E$.
Therefore, both contributions, $\alpha_{\text{virt}}$ and $\alpha_{\text{real}}$, depend on the unphysical slicing parameter $\Delta E$ which cancels in their sum.
Note that the photon-energy cut $\Delta E$, which we numerically set to $\Delta E=10^{-4}\times\sqrt{\hat{s}}/2$, is much smaller than the width \GV of the resonating gauge boson.
Our results are further broken down into the quark--antiquark and the gluon-induced channels, denoted by $\Pq\Paq$ and $\Pq\Pg$, receptively.%
\footnote{The antiquark--gluon induced contributions are included in the contributions labelled $\Pq\Pg$ in the figures.}
The individual contributions from virtual and real photons roughly reach the $0.5\%$ level in case of the $M_{\rT,\Pgn\Pl}$ distribution and even grow to a several percent in the $p_{\rT,\Pl}$ distribution.
The relatively large corrections observed in the transverse-momentum distributions above the threshold are induced by the recoil of the \PW~boson against a hard jet in the real QCD corrections which was also observed for the NLO QCD corrections as shown in the middle plots of Fig.~\ref{fig:distNLO-W}.
However, in the $\Delta E$-independent sum of all non-factorizable corrections, these contributions cancel almost perfectly leading to a net correction that is way below the $0.1\%$ level and, thus, phenomenologically negligible.

The numerical results of the non-factorizable \order{\alphas\alpha} corrections to the neutral-current process are shown in Fig.~\ref{fig:Z-Oaas-nf}, which comprise the relative correction factors to the $M_{\Pl\Pl}$ and $p_{\rT,\Pl}$ distribution.
The respective absolute distributions and results of the NLO calculation are shown in Fig.~\ref{fig:distNLO-Z}.
These corrections turn out to be even smaller than in the charged-current case, the individual virtual and real photonic contributions staying below the $0.02\%$ and $0.5\%$ level for the invariant-mass and the transverse-momentum distribution, respectively.
The smallness of these corrections can be understood by inspecting the behaviour of the non-factorizable corrections under the interchange of the momenta of the two leptons.
For the neutral-current process, the correction factors $\delta_\nf$ are antisymmetric under such an interchange ($k_1\leftrightarrow k_2$), which can be directly seen from the respective formulas by using $Q_1=Q_2$ and $\eta_1=-\eta_2$.
This property of $\delta_\nf$ effectively projects out the antisymmetric part of the cross section that they multiply, which is highly suppressed compared to the symmetric part that enters the LO prediction in the normalization.
Similar suppression effects were observed in the literature in the comparison of non-factorizable corrections between $\PZ\PZ$~\cite{Denner:1998rh} and $\PW\PW$~\cite{Beenakker:1997bp,Beenakker:1997ir,Denner:1997ia} production.
Apart from the overall size, the corrections show a similar behaviour as in the case of \PW~production discussed above, leading to the same conclusion that the mixed QCD--electroweak non-factorizable corrections are phenomenologically negligible.

Of course, one could have speculated on this suppression, since the impact of non-factorizable photonic corrections is already at the level of some $0.1\%$ at NLO. 
However, the \order{\alphas\alpha} corrections mix EW and QCD effects, so that small photonic corrections might have been enhanced by the strong jet recoil effect observed in the $p_{\rT,\Pl}$ distribution (c.f.\ Sect.~\ref{sec:nlo-result}).
This enhancement is seen in the virtual and real corrections separately, but not in their sum.
Furthermore, the existence of gluon-induced ($\Pq\Pg$) channels implies a new feature in the non-factorizable corrections. 
In the $\Pq\Paq$ channels, and thus in the full NLO part of the non-factorizable corrections, the soft-photon exchange proceeds between initial- and final-state particles, whereas in the $\Pq\Pg$ channels at \order{\alphas\alpha} the photon is also exchanged between final-state particles. 
The known suppression mechanisms in non-factorizable corrections work somewhat differently in those cases~\cite{Melnikov:1995fx}.

%%%%%%%%%%%%%%%%%%%%%%%%%%%%%%%%%%%%%%%%%%%%%%%%%%%%%%%%%%%%%%%%%%%%%%
\section{Conclusions}
\label{sec:concl}

Single-\PW- and single-\PZ-boson production via the Drell--Yan mechanism 
do not only represent important reference processes for detector calibration,
parton-distribution fits, etc., at hadron colliders, such as the LHC,
their investigation can also contribute to searches for new physics and strengthen
tests of the Standard Model. 
In the high-energy tails of distributions, these processes extend
the reach in energy in the searches for new gauge bosons $\PW'$ and $\PZ'$,
and in the resonance regions they allow for precision measurements of
the \PW-boson mass and the effective weak mixing angle. 
In particular, the latter task of high-precision physics requires a further
increase in accuracy to the highest possible level.
On the QCD side, corrections are known up to the next-to-next-to-leading
order with improvements from resummations or parton showers beyond fixed order, 
and on the electroweak side
next-to-leading-order corrections as well as universal higher-order
corrections from renormalization, collinear final-state radiation, and
Sudakov-type high-energy logarithms are available.
This leaves the largest unknown component of radiative corrections in the
sector of mixed QCD--electroweak contributions of \order{\alphas\alpha}.

In this paper, we have described how the missing \order{\alphas\alpha}
corrections can be systematically
calculated in the resonance region in terms of a pole approximation, which 
simplifies the actual calculation in comparison to a full treatment 
of the $2\to2$ scattering process at the two-loop level and which poses
further advantages.
This pole approximation is based on the leading 
terms of a consistent expansion of the cross section about the resonance pole, which 
classifies the corrections in terms of factorizable and non-factorizable contributions.
The factorizable corrections 
can be attributed to the \PW/\PZ production and decay subprocesses individually, 
while the  non-factorizable contributions
link production and decay by soft-photon exchange.

In order to describe the concept and to validate the pole approximation, we
have applied it to the electroweak next-to-leading-order corrections of 
\order{\alpha} and
compared it to the full next-to-leading-order result. In the resonance region
the approximation reproduces the full result within fractions of $1\%$, i.e.\
at a phenomenologically satisfactory level. In detail, the factorizable contributions
to the \PW/\PZ production subprocess turn out to be mostly suppressed to this level, 
apart from off-shell tails in transverse-momentum distributions where recoil effects
in the QCD corrections are overwhelming. 
Non-factorizable contributions are generally suppressed to the $0.1\%$ level.
The bulk of corrections are, thus, factorizable corrections to the decay subprocesses,
with the most important contribution resulting from collinear final-state radiation.
This fact is, of course, known. Here we have quantified the size of the individual
components and find that the pole approximation nicely isolates the by far
dominant corrections in \order{\alpha}.

At \order{\alphas\alpha}, we have classified the individual contributions to
the pole approximation:
\begin{enumerate}
\item
Factorizable corrections of the type ``initial--initial'',
i.e.\ two-loop \order{\alphas\alpha} corrections to \PW/\PZ production and corresponding
real-emission parts. From the analysis of the corresponding \order{\alpha}
contributions we do not expect very significant corrections from this part.
Note also that this contribution involves some extra uncertainty, since no
fully adapted set of parton distributions including \order{\alphas\alpha}
corrections is available (and might never be). All the more it is important to 
consistently isolate this contribution---as the pole approximation does---from 
the rest of the corrections, in order to
properly assess its uncertainty.
\item
Factorizable corrections of the type ``final--final'',
i.e.\ two-loop \order{\alphas\alpha} corrections to the leptonic \PW/\PZ decays.
Such corrections are entirely furnished by virtual contributions from
two-loop counterterms. They will be small in size and will not have any visible effect 
on the shape of distributions, and thus are expected to be phenomenologically insignificant.
\item
Factorizable corrections of the type ``initial--final'',
i.e.\ \order{\alphas} corrections to W/Z production in combination with \order{\alpha}
corrections to the leptonic \PW/\PZ decays. They comprise reducible
contributions of the type (one-loop)$\times$(one-loop) on the virtual side, mixed virtual/real
contributions induced by one-loop graphs with single real photon or gluon emission, and
double-real contributions with photon and gluon emission.
The largest effect at the \order{\alphas\alpha} is to be expected from this
contribution, because these reducible contributions combine large QCD corrections
to the production with large electroweak corrections to the \PW/\PZ decay.
The leading part of collinear final-state photon radiation is known to factorize
from the QCD-corrected cross section, but only a proper evaluation of these
initial--final corrections can reveal to which accuracy simplified factorization
approaches hold.
\item
Non-factorizable corrections with QCD corrections to the production and soft-photon
exchange between production and decay.
We have worked out this contribution in this paper in very detail,
first proving that the corresponding matrix elements all show factorization
of the soft-photonic part, where ``soft'' means that the photon energy can be of the
order of the \PW/\PZ decay widths. The proof of this fact, which even holds to
any power in \alphas, is based on arguments taken over from the classic
1961 paper~\cite{Yennie:1961ad} of Yennie, Frautschi, and Suura.
For the purely virtual corrections, we have checked this statement in some
explicit diagrammatical calculations and by an analysis in the framework of an
effective field theory for unstable particles.
Our numerical results show that the sum of all non-factorizable \order{\alphas\alpha}
corrections has a negligible impact on differential cross sections.
This result is plausible in view of the strong suppression of those effects already in
\order{\alpha}, but could not be safely predicted in advance because of a
conceivable enhancement due to QCD recoil effects.

As a byproduct, our results generalize the proof~\cite{Fadin:1993dz} that
non-factorizable corrections vanish after integration over
the resonance (as e.g.\ in the total cross section) to the
case of QCD-corrected cross sections.
\end{enumerate}

In summary, we have worked out a concept for evaluating \order{\alphas\alpha}
corrections to Drell--Yan processes in a pole approximation near the resonances
and have explicitly shown that the corresponding non-factorizable contributions
are phenomenologically negligible, i.e.\ that corrections factorize in the
resonance region in this sense.
The most important \order{\alphas\alpha} contribution will be given by
initial--final QCD$\times$electroweak corrections, whose evaluation is in
progress and will be discussed in a forthcoming paper.

%%%%%%%%%%%%%%%%%%%%%%%%%%%%%%%%%%%%%%%%%%%%%%%%%%%%%%%%%%%%%%%%%%%%%%
\section*{Acknowledgements}

This project is supported by the German Research Foundation (DFG) via grant DI 784/2-1.

%%%%%%%%%%%%%%%%%%%%%%%%%%%%%%%%%%%%%%%%%%%%%%%%%%%%%%%%%%%%%%%%%%%%%%
\appendix
\section*{Appendix}

\section{Soft-photon radiation off a quark line---the proof of factorization}
\label{app:yfs}

In this appendix we prove identity~\eqref{eq:YFS} used in the calculation of the non-factorizable  
\order{\alphas\alpha} corrections in Sect.~\ref{sec:calc-nf-nnlo}.
To this end, we show 
how the arguments given in
Ref.~\cite{Yennie:1961ad} to derive the corresponding identity in pure
QED extend to the case of QCD$\otimes$QED.  In pure QED, the
cancellations leading to Eq.~\eqref{eq:ward-id} are a result of the well-known
Ward identity for the one-particle irreducible
photon--antifermion--fermion vertex function, where the photon
polarization vector is replaced by the photon momentum. Since the
gluons are electrically neutral, this Ward identity
generalizes to photon--antifermion--fermion vertex functions with an
arbitrary numbers of additional gluons,
\begin{equation}
\label{eq:ward-id-gamma}
  q^\mu \, \Gamma_{\mu}^{A\Paq\Pq \Pg\dots\Pg}(q,\bar p,p)
  = eQ_\Pq\left[ \Gamma^{\Paq\Pq \Pg\dots\Pg}(\bar p +q,p)
    -\Gamma^{\Paq\Pq \Pg\dots\Pg}(\bar p,p+q)\right] ,
\end{equation}
where we have suppressed the Lorentz and colour indices associated with the gluons.
The fact that the $\Gamma$ vertex functions used here are only one-particle
irreducible with respect to the quark line does not spoil this identity.

The function $\Gamma_{\mu}^{A\Paq\Pq\Pg\dots\Pg}$
appears in the truncated Green function for the quark line with a
soft-photon insertion that is obtained from~\eqref{eq:def-gamma-tilde}
by inserting a photon in all possible ways into one of the $N$
vertex functions or directly into one of the $(N-1)$ internal quark lines, 
\begin{align}
  \widetilde\Gamma^{A\Paq\Pq\Pg\dots\Pg}_\mu(q,\bar p,p)\; = \;
  \sum_{\mathclap{\substack{\text{gluon} \\ \text{assignments}}}}\quad \Biggl\lbrace
  & \sum_{k=1}^N \Gamma^{\Paq\Pq\Pg\dots\Pg}(\bar p,p_{N-1}+q) \frac{\ri}{\slashed p_{N-1}+\slashed q} \nonumber\\
  & \cdots\; \frac{\ri}{\slashed p_k+\slashed q}
  \Gamma_\mu^{A\Paq\Pq\Pg\dots\Pg}(q,-p_{k}-q,p_{k-1})\frac{\ri}{\slashed p_{k-1}}
  \;\cdots\;
  \frac{\ri}{\slashed p_{1}} \Gamma^{\Paq\Pq\Pg\dots\Pg}(-p_1,p) \nonumber\\
  +& \sum_{k=1}^{N-1} \Gamma^{\Paq\Pq\Pg\dots\Pg}(\bar p,p_{N-1}+q) \frac{\ri}{\slashed p_{N-1}+\slashed q} \nonumber\\
  & \cdots\; \frac{\ri}{\slashed p_k+\slashed q} (-\ri e Q_\Pq \gamma_\mu) \frac{\ri}{\slashed p_k} \;\cdots\;
  \frac{\ri}{\slashed p_{1}} \Gamma^{\Paq\Pq\Pg\dots\Pg}(-p_1,p)
  \Biggr\rbrace .
  \label{eq:def-gamma-tilde-A}
\end{align}
The truncated Green function~\eqref{eq:def-gamma-tilde-A} satisfies the
Ward identity~\eqref{eq:ward-id}, as can be
derived using Eq.~\eqref{eq:ward-id-gamma} and the
partial-fractioning identity
\begin{equation}
  \label{eq:ward-part-frac}
  \frac{\ri}{\slashed p+\slashed q}\, (-\ri e Q_\Pq\slashed q) \,\frac{\ri}{\slashed p} \;=\; 
  e Q_\Pq \left[\, \frac{\ri}{\slashed p} \,-\, \frac{\ri}{\slashed p+\slashed q} \,\right] .
\end{equation}

In order to derive our main identity~\eqref{eq:YFS}, consider the
difference of $T^{(a)}$ as defined in Eq.~\eqref{eq:YFS-a} and the
right-hand side of Eq.~\eqref{eq:YFS}. In the soft-photon limit, it can
be simplified to
\begin{align}
  & T_\nu^{(a)}
  - e Q_\Pq\; \frac{2p_\nu+q_\nu}{2(p\cdot q)+q^2}\; T_0(p+Q)
  \nonumber\\
  & \asymp{q\to 0}\,
  T(p+Q)\left\{\,\frac{\ri}{\slashed p+\slashed Q+\slashed q}
  \widetilde\Gamma^{\Paq\Pq \Pg\dots\Pg}(-p-Q-q,p+q) 
  -\frac{\ri}{\slashed p+\slashed Q}
  \widetilde\Gamma^{\Paq\Pq\Pg\dots\Pg}(-p-Q,p)
  \right\} \nonumber\\
  & \qquad\times e Q_\Pq\; \frac{2p_\nu+q_\nu}{2(p\cdot q)+q^2}
  u(p)\nonumber \\
  &\quad=    T(p+Q)\left\{\,\frac{\ri}{\slashed p+\slashed Q+\slashed q}
  (-q^\mu )\, \widetilde\Gamma_{\mu}^{A\Paq\Pq \Pg\dots\Pg}(q,-p-Q-q,p) 
  \right. \nonumber\\
  &\qquad\left.  + e Q_\Pq\;\left[\frac{\ri}{\slashed p+\slashed Q+\slashed q}
  - \frac{\ri}{\slashed p+\slashed Q} \right]
  \widetilde\Gamma^{\Paq\Pq\Pg\dots\Pg}(-p-Q,p)  \right\}
  \frac{2p_\nu+q_\nu}{2(p\cdot q)+q^2}  u(p).
\label{eq:YFS-diff-a}
\end{align}
In the first step the soft photon momentum has been
neglected in the hard part $T$, while in the second step the Ward
identity~\eqref{eq:ward-id} has been applied.  The first term in the
curly braces involves the same vertex function
$\widetilde\Gamma_{\mu}^{A\Paq\Pq\Pg\dots\Pg}$ as the expression $T^{(b)}$
defined in Eq.~\eqref{eq:YFS-b}, while the terms in the last line
of~\eqref{eq:YFS-diff-a} can be brought to a form similar to $T^{(c)}$ defined in Eq.~\eqref{eq:YFS-c}
using the partial-fractioning identity~\eqref{eq:ward-part-frac}.
Using Eq.~\eqref{eq:YFS-diff-a}, we obtain the following result for the soft limit of the l.h.s.\ of Eq.~\eqref{eq:ward-id},
\begin{equation}
  \left[\, T_\nu^{(a)}+ T_\nu^{(b)}+ T_\nu^{(c)} \,\right]
  \asymp{q\to0}
  \;e Q_\Pq\; \frac{2p_\nu+q_\nu}{2(p\cdot q)+q^2} T_0(p+Q)   
  +  T_\nu^{(\mathrm{int})}\,,
  \label{eq:YFS-proof}
\end{equation}
where we have introduced the expression
\begin{align}
  T_\nu^{(\mathrm{int})} \,=\; T(p+Q)\biggl\{\,
  &\frac{\ri}{\slashed p+\slashed Q+\slashed q}
  \widetilde\Gamma_{\mu}^{A\Paq\Pq \Pg\dots\Pg}(q,-p-Q-q,p) 
  \nonumber\\
  &+ \frac{\ri}{\slashed p+\slashed Q+\slashed q}
  (-\ri e Q_q \gamma_\mu )\frac{\ri}{\slashed p+\slashed Q} 
  \widetilde\Gamma^{\Paq\Pq\Pg\dots\Pg}(-p-Q,p)  \biggr\}\, u(p) \; G^\mu{}_\nu ,
  \label{eq:YFS-int}
\end{align}
with 
\begin{equation}
G^\mu{}_\nu=  \delta^\mu{}_\nu- q^\mu \frac{2p_\nu+q_\nu}{2(p\cdot q)+q^2}.
\end{equation}
As in the pure QED case~\cite{Yennie:1961ad} we obtain the result that soft-photon emission from external lines is described by an eikonal factor 
times the lower-order matrix element, while soft-photon emission from
internal lines is described by the replacement $\gamma_\nu\to
\gamma_\mu\,G^\mu{}_\nu$ in the photon--antifermion--fermion
vertices.

The result~\eqref{eq:YFS-proof} leads to the desired
identity~\eqref{eq:YFS} if it can be shown that the term
$T^{(\mathrm{int})}$ is subleading compared to the eikonal term 
(describing emission from external lines) in the
soft-photon limit $q\to 0$.
This has to be true independently of the gluon momenta $q_i$
flowing into the quark line 
which can be hard or soft.  In order to
show this, first consider all diagrams where the photon with momentum
$q$ couples directly to the quark line, i.e.\ the term in the last line
of Eq.~\eqref{eq:YFS-int} and those diagrams in the truncated Green
function $\widetilde\Gamma_{\mu}^{A\Paq\Pq\Pg\dots\Pg}$ where the
photon is not attached to an internal closed quark loop.  
The attachment of the photon to the quark chain
after the momentum flow $Q_k$ is described by the replacement
\begin{equation}
  \label{eq:prop-repl}
  \frac{1}{\slashed p+\slashed{Q}_k} \,\longrightarrow\,
  \frac{1}{\slashed p+\slashed{Q}_k+\slashed q}
  \,\gamma_\mu\,G^\mu{}_\nu\,\frac{1}{\slashed p+\slashed{Q}_k} ,
\end{equation}
in the lower-order matrix element $T_0$ and the appropriate momentum
shift $Q_l\to Q_l+q$ in the propagators with $l>k$.  We observe that
in the case where $q$ and $Q_k$ become \emph{simultaneously} soft,
this substitution amounts to the replacement of one small denominator
by two, potentially worsening the IR divergence behaviour compared to
that of $T_0$.  Hard momenta $Q_k$, on the other hand, do not lead to
singularities and constitute subleading contributions that need not be
considered further.  The case of a soft gluon momentum and a hard
photon momentum is not relevant for us since only soft photons lead to
a resonance enhancement and contribute to the non-factorizable
corrections.

In order to study the potentially problematic case where $q$ and $Q_k$ in 
Eq.~\eqref{eq:prop-repl} become simultaneously soft, we first rewrite the 
r.h.s.\ of Eq.~\eqref{eq:prop-repl} as follows,
\begin{equation}
  \label{eq:insert-photon}
  \begin{aligned}
  &\frac{\slashed p+\slashed{Q}_k+\slashed q}{q^2+2q\cdot(p+Q_k)+2p\cdot{Q}_k+Q_k^2}
  \,\gamma_\mu\,G^\mu{}_\nu\,\frac{\slashed p+\slashed{Q}_k} {2 p\cdot{Q}_k+Q_k^2} \\
  &= \frac{1}{q^2+2q\cdot(p+Q_k)+2p\cdot{Q}_k+Q_k^2}
  \biggl[ \left\{\slashed p\,,\,\gamma_\mu\,G^\mu{}_\nu \right\} (\slashed p+\slashed{Q}_k)
  - \gamma_\mu\,G^\mu{}_\nu\; \slashed p\;\slashed{Q}_k \biggr]
  \frac{1}{2 p\cdot{Q}_k+Q_k^2} +\dots  ,
  \end{aligned}
\end{equation}
where the on-shell condition $p^2=0$ for massless quarks was used.
The anticommutator can be seen to vanish in the soft-photon limit,
\begin{equation}
  \left\{\slashed p\,,\,\gamma_\mu\,G^\mu{}_\nu \right\} 
  = 2\,p_\mu\,G^\mu{}_\nu 
  = -q_\nu + q^2\,\frac{2p_\nu+q_\nu}{2(p\cdot q) + q^2}
  = -q_\mu\,G^\mu{}_\nu 
  \,\asymp{q\to0}\,0.
\end{equation}
Inspecting the double-soft limit $q\to \lambda q$, $Q_k\to \lambda
Q_k$ of Eq.~\eqref{eq:insert-photon}, we observe that the scaling
behaviour of the right-hand side of Eq.~\eqref{eq:prop-repl} is not
$\lambda^{-2}$ as naively estimated, but rather only $\lambda^{-1}$,
which is of the same order as the $Q_k\sim 0$ limit of the
matrix element $T_0$ without soft-photon insertion.
Therefore, the term $T^{(\mathrm{int})}$ in~\eqref{eq:YFS-proof} is
suppressed by one power of $\lambda$ compared to the first term on
the right-hand side which establishes our main
formula~\eqref{eq:ward-id}.

The arguments presented up to this point are sufficient to prove the
factorization \eqref{eq:YFS} at \order{\alphas}.
In order to extend the reasoning to any order in $\alphas$, however, 
we still have to lift two assumptions made in our line of arguments:
\begin{enumerate}
\item
Above we ignored the possibility that the soft photon may be attached to
a closed quark line implicitly contained in one of the $\Gamma$ vertex
functions. We, thus, have to show
that the attachment of a soft photon to a closed quark loop in the Green function  $\widetilde\Gamma_{\mu}^{A\Paq\Pq\Pg\dots\Pg}$ does not lead to an IR enhancement. 
To this end, we consider a diagram with a closed quark loop and an arbitrary number of external or internal gluon lines attached to it.
Following the pure QED case discussed in  Ref.~\cite{Yennie:1961ad}, we observe that taking the derivative with respect to the loop momentum running along the fermion loop yields the set of diagrams obtained by attaching a photon with zero momentum to the original quark loop in all possible ways. Therefore, the attachment of a soft photon to a quark loop vanishes since it leads to an integral of a total derivative.
\item
We finally have to admit the case where the $\widetilde\Gamma$ and $\Gamma$
vertex functions involve additional external quark lines. By definition of
$\widetilde\Gamma$ and $\Gamma$, an external quark field $\Pq'$ will always appear in
$\Paq'\Pq'$ pairs in the field arguments, i.e.\ as 
$\widetilde\Gamma_{\mu}^{A\Paq\Pq\Paq'\Pq'\Pg\dots\Pg}$,
$\Gamma_{\mu}^{A\Paq\Pq\Paq'\Pq'\Pg\dots\Pg}$, etc. In the following we analyze the 
situation of one additional $\Pq'$ line flowing through $\widetilde\Gamma$, etc.;
the generalization to more external and/or identical quarks is obvious.
The first step in the argument is to establish Ward identities
for $\widetilde\Gamma_{\mu}^{A\Paq\Pq\Paq'\Pq'\Pg\dots\Pg}$ and
$\Gamma_{\mu}^{A\Paq\Pq\Paq'\Pq'\Pg\dots\Pg}$ analogous to 
Eqs.~\eqref{eq:ward-id-gamma} and \eqref{eq:ward-id}.
Obviously the analogue to Eq.~\eqref{eq:ward-id-gamma} reads
\begin{align}
  &q^\mu \, \Gamma_{\mu}^{A\Paq\Pq\Paq'\Pq'\Pg\dots\Pg}(q,\bar p,p,\bar p',p') \nonumber\\
  &\quad = \,eQ_\Pq\left[ \Gamma^{\Paq\Pq\Paq'\Pq' \Pg\dots\Pg}(\bar p +q,p,\bar p',p')
  -\Gamma^{\Paq\Pq\Paq'\Pq' \Pg\dots\Pg}(\bar p,p+q,\bar p',p')\right] \nonumber\\
  & \qquad{}+ eQ_{\Pq'}\left[ \Gamma^{\Paq\Pq\Paq'\Pq' \Pg\dots\Pg}(\bar p,p,\bar p' +q,p')
  -\Gamma^{\Paq\Pq\Paq'\Pq' \Pg\dots\Pg}(\bar p,p,\bar p',p'+q)\right],
\end{align}
with $Q_{\Pq'}$ denoting the relative electric charge of quark $\Pq'$.
This identity can be split into two separate equations by making use of the fact that the quark charges $Q_\Pq$ and $Q_{\Pq'}$ are independent.
The identity concerning the $Q_q$ part reads
\begin{equation}
  q^\mu \, \Gamma_{\mu}^{A_\Pq\Paq\Pq\Paq'\Pq' \Pg\dots\Pg}(q,\bar p,p,\bar p',p')
  = eQ_\Pq\left[ \Gamma^{\Paq\Pq\Paq'\Pq' \Pg\dots\Pg}(\bar p +q,p,\bar p',p')
    -\Gamma^{\Paq\Pq\Paq'\Pq' \Pg\dots\Pg}(\bar p,p+q,\bar p',p')\right],
\end{equation}
where the $A_\Pq$ field argument of $\Gamma_{\mu}^{A_\Pq\Paq\Pq \Pg\dots\Pg}$
indicates that the photon is attached to the $\Pq$-line in all possible ways, but nowhere 
else. The identity for the $\Pq'$-line looks analogous.
Ward identity \eqref{eq:ward-id} generalizes in a similar fashion to
\begin{equation}
  q^\mu \, \widetilde\Gamma_{\mu}^{A_\Pq\Paq\Pq\Paq'\Pq' \Pg\dots\Pg}(q,\bar p,p,\bar p',p')
  = eQ_\Pq\left[ \widetilde\Gamma^{\Paq\Pq\Paq'\Pq' \Pg\dots\Pg}(\bar p +q,p,\bar p',p')
    -\widetilde\Gamma^{\Paq\Pq\Paq'\Pq' \Pg\dots\Pg}(\bar p,p+q,\bar p',p')\right]
\end{equation}
and analogously for the $\Pq'$-line.

For photons attached to the $\Pq$-line obviously the proof of Eq.~\eqref{eq:YFS}
now proceeds as before, but photons attached to the $\Pq'$-line require further inspection.
If the external $\Pq'$ or $\Paq'$ carries a hard momentum, then external
$\Pq'/\Paq'$-lines can be treated
like the $\Pq$-line in the above reasoning, leading to an analogue of
Eq.~\eqref{eq:YFS} with $\Pq$ replaced by $\Pq'/\Paq'$.
The remaining case that $\Pq'$ is soft can be obtained from the case of
hard $\Pq'$ as limiting case $p'\to0$ (analogously for soft $\Paq'$), 
since the limit $p'\to0$ does not increase the singular behaviour
of the eikonal factor $\propto(2p'_\nu+q_\nu)/(2(p'\cdot q)+q^2)$
in the soft-photon region, neither
for virtual ($q^2\ne0$) nor for real ($q^2=0$) photons of momentum $q$.
A breakdown of the factorization \eqref{eq:YFS} would require an increase in
the degree of the soft-photon singularity in the phase-space region of soft $p'$,
because the soft-$p'$ region itself implies a phase-space suppression 
which takes the form of a factor of \order{\GV} in our case.

Suppose the prediction (including only QCD corrections)
for some observable involves a phase-space integration
over the $\Pq'$-momentum $p'$, leading to a finite, well-defined result
(possibly after summing different IR-divergent contributions).
In the region of hard quark momenta $p'$,
the \order{\alpha} correction involves the integration over the soft photon
momentum $q$ which produces the usual logarithmic soft singularities
for virtual and real photons, with a finite result for their sum.
As argued above, the phase-space integration over the $\Pq'$-momentum $p'$
(based on the eikonal approximation for the photon momentum $q$)
can be extended to the soft-$p'$ region without problems, since the
eikonal factor does not imply new singularities for $p'\to0$.
Note also that we silently already used this kind of reasoning 
in the integration of the non-factorizable \order{\alphas\alpha} corrections 
over the phase space of the outgoing quark or antiquark in the
$\Pq\Pg$ and $\Paq\Pg$ channels, where the outgoing quark or antiquark can become soft.
\end{enumerate}

%%%%%%%%%%%%%%%%%%%%%%%%%%%%%%%%%%%%%%%%%%%%%%%%%%%%%%%%%%%%%%%%%%%%%%
\section{Explicit calculation of two-loop non-factorizable diagrams}
\label{app:yfs-example}

In Sect.~\ref{sec:calc-nf-nnlo} we have used arguments based on gauge invariance to show that the non-factorizable two-loop \order{\alphas\alpha} corrections can be written as a simple product of the QCD vertex correction and the \order{\alpha} non-factorizable corrections. 
This property, for instance, relies on the cancellation of diagrams with overlapping QCD and EW singularities as displayed in Figs.~\ref{fig:YFS-se} and \ref{fig:vertex-box-cancel}. 
In this appendix we demonstrate this cancellation explicitly for the example of the two diagrams in Fig.~\ref{fig:vertex-box-cancel} using the two methods sketched in Sects.~\ref{sec:VV}~\ref{sec:vv-mb} and \ref{sec:vv-eft}, i.e.\ a direct calculation using the Mellin--Barnes method and the EFT approach.
We denote the contributions of the two diagrams to the two-loop matrix element by
\begin{equation}
  \label{eq:nnlo-example}
  \M^{(a)} \,\equiv\, \Vcenter{\includegraphics{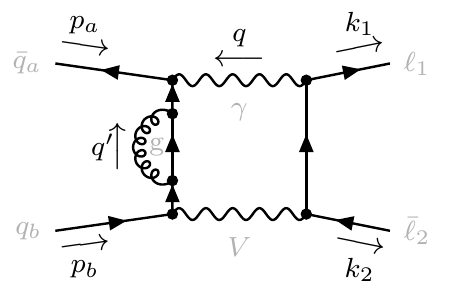}} , \qquad
  \M^{(b)} \,\equiv\, \Vcenter{\includegraphics{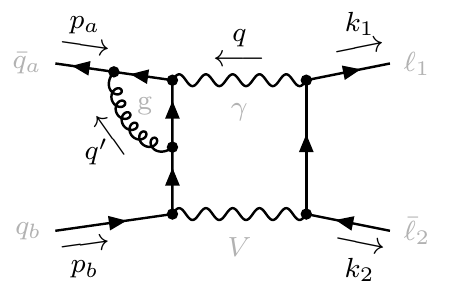}} .
\end{equation}
The resonant contributions of the two-loop diagrams arise from a soft photon momentum $q$. 
As discussed below in the two approaches, in the soft-photon limit the diagrams factorize into the LO matrix element $\M_0$ and a scalar two-loop integral according to
\begin{equation}
  \label{eq:example-def-I}
  \M^{(i)} \;\sim\; 
  \frac{C_\mathrm{F} \alphas}{4\pi}\, \frac{Q_aQ_1\alpha}{2\pi}
  \;\M_0\; (1-\epsilon)\,(-t_{a1})\,(\mu_V^2-s_{12}^2)\;
  I^{(i)}(s_{12},t_{a1}) .
\end{equation}
Using the two methods we will show by explicit calculation that the diagrams cancel, i.e.\
\begin{equation}
  I^{(a)}(s_{12},t_{a1})+I^{(b)}(s_{12},t_{a1})=0 .
\end{equation}
These cancellations can already be seen after performing the integral over the gluon momentum, 
i.e.\ before integrating over the photon momentum,
so that the result directly extends to diagrams where the soft photon is attached to the intermediate $V$ boson or the $\Pal_2$ final-state lepton, and to real soft-photon emission. 
We have also explicitly verified the identity shown in Fig.~\ref{fig:double-box-cancel} using the two methods which involves the same soft-photon loop integral after integrating out the gluon sub-loop. 
Therefore we will not give the details of this calculation here.

%%%%%%%%%%%%%%%%%%%%%%%%%%%%%%%%%%%%%%%%%%%%%%%%%%
\subsection{Mellin--Barnes method}
\label{app:yfs-mellin}

In this section we discuss the cancellation of the two diagrams defined in Eq.~\eqref{eq:nnlo-example} by explicitly computing each diagram separately and employing the Mellin--Barnes method to extract the contributions enhanced by resonant factors.

In Sect.~\ref{sec:VV}~\ref{sec:vv-mb} we have established the correspondence between the non-fac\-to\-ri\-zable corrections and the leading linear IR singularity in the photon momentum at the on-shell point ($p_V^2=\MV^2$).
This allows for the extraction of the relevant resonant contribution before evaluating the two-loop integrals.
To this end, we perform the standard tensor reduction of the internal ``QCD sub-loop'' and by a simple power-counting argument identify all terms that do not contribute to the linear IR singularity and, therefore, can be omitted in the calculation of the non-factorizable corrections.
Applying this procedure leads to \emph{scalar} master integrals only, since all terms involving additional factors of the photon momentum in the numerator are less singular and can be neglected. 
Only retaining these leading singular terms, i.e.\ applying the ESPA, the second two-loop diagram in Eq.~\eqref{eq:nnlo-example} reduces to a correction factor multiplying the Born matrix element, as anticipated in Eq.~\eqref{eq:example-def-I},
where the scalar two-loop integral is given by the expression
\begin{align}
  \label{eq:2loop-master}
  & I^{(b)}(s_{12},t_{a1}) 
  \;=\quad  
  \raisebox{-3.8em}{\includegraphics{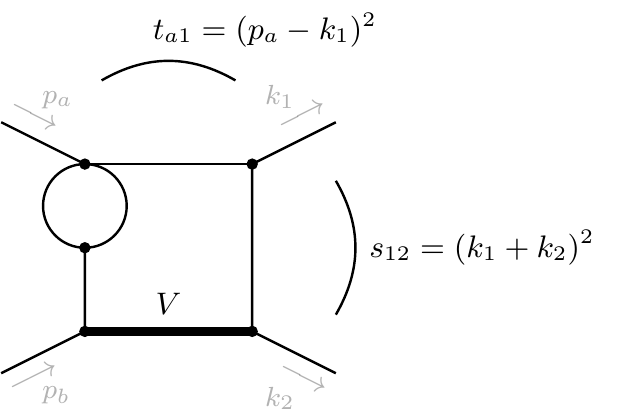}}  
  \\[.5em]
  &\quad= \left[\frac{(2\pi\mu)^{2\epsilon}}{\ri \pi^2}\right]^2
  \int \frac{\rd^D q}{q^2\,(q+p_a)^2\,\left[(q+p_a+p_b)^2-\mu_V^2\right]\,(q+k_1)^2}
  \int\frac{\rd^D q'}{q'^2\,(q'+q+p_a)^2} \nonumber\\
  &\quad= (4\pi\mu^2)^{2\epsilon}
  \frac{\Gamma(\epsilon)\,\Gamma(1-\epsilon)^2\,\Gamma(2+2\epsilon)}
  {\Gamma(1+\epsilon)\,\Gamma(2-2\epsilon)} \nonumber\\ 
  &\qquad \times\int_0^1 \rd^4x\; \delta\Bigl(1-\sum_{i=1}^4 x_i\Bigr) \,x_1^{\epsilon}\,
  \left[ x_1x_3\,(-t_{a1}) +  x_2x_4\,(\mu_V^2-s_{12}) 
  + (1-x_2)x_4\,\MV^2 \right]^{-2-2\epsilon} , \nonumber
\end{align}
and all external momenta $p_i$, $k_i$ correspond to on-shell massless particles $p_i^2=k_i^2=0$. 
In the last step the integration over the two-loop momenta $q$ and $q'$ was performed after introducing the Feynman parameters $x_i$ and the term $(\mu_V^2-s_{12})$ was extracted in the anticipation of the expansion we apply in the next step.

The integration over the four Feynman parameters $x_i$ can be performed at the expense of introducing two Mellin--Barnes integrals,
\begin{align}
  I^{(b)}(s_{12},t_{a1}) 
  =& \frac{1}{\MV^4} \left(\frac{4\pi\mu^2}{\MV^2}\right)^{2\epsilon}
  \frac{\Gamma(\epsilon)\,\Gamma(1-\epsilon)^2}
  {\Gamma(1+\epsilon)\,\Gamma(2-2\epsilon)\,\Gamma(-3\epsilon)} \nonumber\\
  &\frac{1}{(2\pi\ri)^2} \int_{-\ri\infty}^{+\ri\infty} \rd^2 z
  \left(\frac{\mu_V^2-s_{12}}{\MV^2}\right)^{z_1}
  \left(\frac{-t_{a1}}{\MV^2}\right)^{z_2} 
  \Gamma(2+2\epsilon+z_1+z_2)\,\Gamma(1+z_1) \nonumber\\
  & \Gamma(-1-3\epsilon-z_1)\,\Gamma(-z_1) 
  \frac{\Gamma(1+z_2)\,\Gamma(1+\epsilon+z_2)\,
  \Gamma(-1-2\epsilon-z_2)\,\Gamma(-z_2) }{\Gamma(1-\epsilon+z_2)} ,
\end{align}
where the integration contour of the complex variable $z_i$ is taken to the right of the ``left poles'' $\Gamma(\ldots+z_i)$ and left of the ``right poles'' $\Gamma(\ldots-z_i)$.
Closing the integration contour of $z_1$ to the right and collecting the residues of the poles in $z_1$ allows us to perform a systematic expansion in powers of $\tfrac{\mu_V^2-s_{12}}{\MV^2}$.
The leading term emerges from the residue at $z_1=(-1-3\epsilon)$, whereas all further contributions are not enhanced by resonant propagator factors and can therefore be neglected in the PA,
\begin{align}
  I^{(b)}(s_{12},t_{a1}) 
  &\sim \frac{1}{\MV^4} \left(\frac{4\pi\mu^2}{\MV^2}\right)^{2\epsilon}
  \frac{\Gamma(\epsilon)\,\Gamma(1-\epsilon)^2\,\Gamma(1+3\epsilon)}
  {\Gamma(1+\epsilon)\,\Gamma(2-2\epsilon)}
  \left(\frac{\mu_V^2-s_{12}}{\MV^2}\right)^{-1-3\epsilon} \nonumber\\
  &\quad \frac{1}{(2\pi\ri)} \int_{-\ri\infty}^{+\ri\infty} \rd z
  \left(\frac{-t_{a1}}{\MV^2}\right)^{z}
  \Gamma(1+\epsilon+z)\,\Gamma(1+z)\,\Gamma(-1-2\epsilon-z)\,\Gamma(-z) .
\end{align}
The second Mellin--Barnes integral can be performed in the usual manner by first extracting the divergent terms in $\epsilon$, which correspond to the residues of the poles that pinch the integration contour in the limit $\epsilon\to0$, and then performing an expansion in $\epsilon$ for the finite remainder. 
The final result for the scalar two-loop integral reads
\begin{align}
  \label{eq:two-loop-mellin}
  I^{(b)}(s_{12},t_{a1})\;=\;
  &\frac{\Gamma(1+\epsilon)^2}{(\mu_V^2-s_{12})(-t_{a1})}
  \left(\frac{\mu_V^2-s_{12}}{\MV^2}\right)^{-3\epsilon}
  \left(\frac{4\pi\mu^2}{-t_{a1}}\right)^{2\epsilon} 
  \Biggl\lbrace  
  \frac{1}{2\epsilon^3} + \frac{1}{\epsilon^2} \nonumber\\
  & +\frac{1}{\epsilon} \left[2+\frac{5\pi^2}{12}+ \Li_{2}\biggl(1+\frac{t_{a1}}{\MV^2}\biggr)\right] 
  +2\Li_{3}\biggl(\frac{-t_{a1}}{\MV^2}\biggr)+\Li_{3}\biggl(1+\frac{t_{a1}}{\MV^2}\biggr) \nonumber\\
  &-2\ln\biggl(\frac{-t_{a1}}{\MV^2}\biggr)
  \left[\frac{\pi^2}{6}-\Li_{2}\biggl(1+\frac{t_{a1}}{\MV^2}\biggr)\right]
  +\ln^2\biggl(\frac{-t_{a1}}{\MV^2}\biggr)\ln\biggl(1+\frac{t_{a1}}{\MV^2}\biggr) \nonumber\\
  & -6\zeta(3)+\frac{5\pi^2}{6}+2\Li_{2}\biggl(1+\frac{t_{a1}}{\MV^2}\biggr)+4+\order{\epsilon}+\order{s_{12}-\mu_V^2}
  \Biggr\rbrace .
\end{align}

The calculation of the other two-loop amplitude $\M^{(a)}$ in Eq.~\eqref{eq:nnlo-example} containing the quark self-energy insertion proceeds along the same lines as discussed above.
Applying the tensor reduction to the internal QCD sub-loop generates a term $(\slashed{q}+\slashed{p}_a)$ in the numerator which cancels one of the internal quark propagators.
The resulting loop structure of the denominators corresponds to the two-loop master integral~\eqref{eq:2loop-master}.
Keeping only the leading singular parts in the photon momentum, i.e.\ neglecting all photon momenta $q$ in the numerators, we again obtain an expression of the form~\eqref{eq:example-def-I}, 
where the correction factor multiplying the Born amplitude involves the same scalar master integral as in the case $\M^{(b)}$ and, furthermore, only differs by a global sign.
Thus the non-factorizable contributions of the two diagrams of Eq.~\eqref{eq:nnlo-example} cancel exactly, which was to be shown.

%%%%%%%%%%%%%%%%%%%%%%%%%%%%%%%%%%%%%%%%%%%%%%%%%%
\subsection{Effective-field-theory inspired method} 
\label{app:yfs-eft}

We now calculate the two diagrams of
Eq.~\eqref{eq:nnlo-example} using the
expansion in momentum regions discussed in Sect.~\ref{sec:nlo-eft}.
As discussed in Sect.~\ref{sec:VV}~\ref{sec:vv-eft}, only the soft photon momenta and collinear gluon momenta can contribute at leading power in the resonance expansion.

Consider first the vertex--box diagram $\M^{(b)}$ of Eq.~\eqref{eq:nnlo-example}.%
\footnote{As in the one-loop example in Eq.~\eqref{eq:nbar-coll} the $\bar n$-collinear region leads to a scaleless integral over the gluon loop momentum and does not contribute.}
For a soft photon momentum $q$ and a $n$-collinear-gluon momentum $q'$ it is evaluated by expanding the propagator denominators according to
\begin{equation}
  \begin{aligned}
    (p_a+q')^2&= (q'_-+p_{a-})q'_++q^{'2}_\perp \;\sim\; \GV\MV ,\\
     (p_a+q'+q)^2&\approx (q'_-+p_{a-})(q_+'+q_+)+  q^{'2}_\perp \;\sim\; \GV\MV ,
  \end{aligned}
\end{equation}
and simplifying the numerator Dirac structure using the Dirac equation
for the collinear momentum, $\bar v(\frac{p_{a+}}{2} n)\slashed n=0$,
and the identity $\{\slashed n,\slashed q'_\perp\}=0$,
\begin{align}
  &\bar v(p_a) \gamma^\rho(\slashed p_a+\slashed q')\gamma^\mu 
  (\slashed p_a+\slashed q'+\slashed q)\gamma_\rho 
  (\slashed p_a+\slashed q')\cdots  \nonumber\\ 
  &\to   \bar v(p_a) \gamma^\rho
  \left[(p_{a-}+q'_-)\frac{\slashed n}{2}+\slashed q'_\perp\right]\gamma^\mu
   \left[(p_{a-}+q'_-)\frac{\slashed n}{2}+\slashed q'_\perp\right]\gamma_\rho 
   \frac{\slashed n}{2} p_{a-} \cdots \nonumber\\
&=(D-2)\, (2p_{a}^\mu) \, \bar v(p_a) \, q^{'2}_\perp\cdots .
\end{align}
The same result is obtained by applying the SCET Feynman rules given
in Ref.~\cite{Bauer:2000yr}, see also Ref.~\cite{Stewart:2010qs}.
Using also the standard eikonal approximation for the soft photon, the two-loop integral factorizes from the LO diagram according to Eq.~\eqref{eq:example-def-I}, with the two-loop integral given by
\begin{multline}
  I_{\mathrm{EFT}}^{(b)}= \left\lbrack\frac{(2\pi\mu)^{2\epsilon}}{\ri \pi^2}
  \right\rbrack^2
  \int \rd^D q\,\frac{1}{ q^2 \left[2(p_V\cdot q)+p_V^2-\mu_V^2\right]
  (2 k_1\cdot q)(2 p_a\cdot q) }  \\
  \times   \int \rd^Dq'\,
  \frac{{q'_\perp}^2 }{(q'_-q'_++q^{'2}_\perp+\ri0)
  \left[(q'_-+p_{a-})q'_++q^{'2}_\perp +\ri0\right]
  \left[ (q'_-+p_{a-})(q'_++q_+)+ q^{'2}_\perp+\ri0\right]} ,
\end{multline}
where $p_V=k_1+k_2$.
As in the
one-loop example in Sect.~\ref{sec:nlo-eft}, the integral over
$q'_+$ can be performed with the theorem of residues, noting that
the integral vanishes for $q'_- > 0$ ($q'_-<-p_-)$, since all poles
in $q'_+$ lie in the lower (upper) half-plane. For $-p_-<q'_-<0$
the integral can be performed by picking up the pole
$q'_+=-\frac{q^{\prime2}_\perp+\ri\epsilon}{q'_-}$ in the upper
half-plane.
 The integrals over the remaining components of the
collinear loop momentum are then standard and lead to the result
\begin{align}
  I_{\mathrm{EFT}}^{(b)}\,=\,  
  -\frac{\left(-4\pi\mu^2 \right)^\epsilon\Gamma(\epsilon)\,\Gamma(1-\epsilon)^2}{
  2\,\Gamma(2-2\epsilon)}
  \frac{(2\pi\mu)^{2\epsilon}}{\ri \pi^2}
  \int \frac{\rd^D q}{ q^2 \lbrack 2(p_V\cdot q)+p_V^2-\mu_V^2\rbrack
  (2k_1\cdot q)(2 p_a\cdot q)^{1+\epsilon} } .
\end{align}
The integral over the soft photon momentum is identical to that arising in the initial--final-state NLO soft corrections up to the shift of the power of the propagator $(p_a\cdot q)\to(p_a\cdot q)^{1+\epsilon} $.  Using the methods outlined in Sect.~\ref{sec:nlo-eft}, the integral can be performed to all orders in the dimensional-regularization parameter $\epsilon$,
\begin{align}
  \label{eq:two-loop-eft}
  I_{\mathrm{EFT}}^{(b)}(s_{12},t_{a1})=&
 \frac{\Gamma(\epsilon)\,\Gamma(-2\epsilon)\,\Gamma(1-\epsilon)^2\,\Gamma(1+3\epsilon)}{
  \Gamma(2-2\epsilon)}\frac{1}{t_{a1}(\mu_V^2-s_{12})}
 \left(\frac{\mu_V^2-s_{12}}{\MV^2}\right)^{-3\epsilon}
 \left(\frac{4\pi\mu^2}{\MV^2}\right)^{2\epsilon } \nonumber\\
  &\Biggl[\frac{\Gamma^2(-\epsilon)}{\Gamma(-2\epsilon)}
  \left(1+\frac{\MV^2}{t_{a1}}\right)^\epsilon
  \left(-\frac{\MV^2}{t_{a1}}\right)^{2\epsilon}
  +\frac{1}{\epsilon}\,
  {}_2F_1\left(1,-2\epsilon,1-\epsilon, 
  -\frac{\MV^2}{t_{a1}}\right)
  \Biggr] .
\end{align}
The expansion of the hypergeometric function in $\epsilon$ can be obtained with the help of Refs.~\cite{Huber:2005yg,Huber:2007dx},
\begin{equation}
\label{eq:hyp-exp}
\begin{aligned}
  {}_2F_1(1,-2\epsilon,1-\epsilon,x)=
  &\;1+2\epsilon \ln(1-x)
  +\epsilon^2\left[-2\Li_2(x)+\ln^2(1-x)\right]\\
  &+\epsilon^3\Bigl[2\Li_3(1-x)-2\Li_3(x)
  +\ln(1-x)\Bigl(\frac{1}{3}\ln^2(1-x)\\
  &+\ln(1-x)\ln(x) 
  -\frac{\pi^2}{3}\Bigr)-2\zeta(3)
  \Bigr]
  +\order{\epsilon^4} .
\end{aligned}
\end{equation}
Using this result to expand Eq.~\eqref{eq:two-loop-eft} in $\epsilon$, this expression can be seen to be equivalent to the  result of the Mellin--Barnes calculation~\eqref{eq:two-loop-mellin} after the use of some functional identities of the dilogarithm and trilogarithm functions.

Next, consider  the left-hand diagram $\M^{(a)}$ of Eq.~\eqref{eq:nnlo-example}, which is given by the one-loop initial--final diagram with  an insertion of the quark self-energy\footnote{The quark self-energy in SCET is given entirely by the collinear contribution which is identical to the full self-energy in QCD~\cite{Bauer:2000yr}.}
\begin{equation}
  \ri\Sigma^{q}(p)=
  \frac{\alphas}{4\pi}C_\mathrm{F} (1-\epsilon)\, \slashed p \,
  \frac{\Gamma(\epsilon)\,\Gamma(1-\epsilon)^2}{\Gamma(2-2\epsilon)}
  \left(-\frac{4\pi\mu^2}{p^2+\ri0}\right)^\epsilon .
\end{equation}
Therefore the contribution of this diagram to the two-loop matrix
element factors from the LO diagram according to
Eq.~\eqref{eq:example-def-I} with the loop integral given by
\begin{align}
  I_{\mathrm{EFT}}^{(a)}= 
  \frac{\left(4\pi\mu^2 \right)^\epsilon\Gamma(\epsilon)\,\Gamma(1-\epsilon)^2}{2\,\Gamma(2-2\epsilon)} 
  \,\frac{(2\pi\mu)^{2\epsilon}}{\ri \pi^2}
  \int & \frac{\rd^D q}{ q^2 \lbrack 2(p_V\cdot q)+p_V^2-\mu_V^2\rbrack
  (2 k_1\cdot q)(2 p_a\cdot q) } 
  \nonumber\\
  &\times \frac{(\slashed p_a+\slashed q)}{(-2p_a\cdot q)^{-\epsilon}}
  \frac{1}{\slashed p_a+\slashed q}.
\end{align}
Obviously $ I_{\mathrm{EFT}}^{(a)}+ I_{\mathrm{EFT}}^{(b)}=0$, i.e.\ the two diagrams
in Eq.~\eqref{eq:nnlo-example} cancel, as was to be shown.

%%%%%%%%%%%%%%%%%%%%%%%%%%%%%%%%%%%%%%%%%%%%%%%%%%%%%%%%%%%%%%%%%%%%%%
\section{Explicit form of the IR-safe contributions to the non-factorizable corrections}
\label{app:dipole}

In this appendix we provide the explicit expressions for each of the
contributions to the non-factorizable corrections in our master
formula~\eqref{eq:final-master} for the quark--antiquark and the
quark--gluon channels, similarly to the NLO QCD corrections in
Eq.~\eqref{eq:dipole-nlo-ch}. 
The respective expressions for the
gluon--antiquark channel follow analogously.

The double-virtual corrections~\eqref{eq:master-VV} and the (virtual
QCD)$\times$(real photonic) corrections~\eqref{eq:master-VR}
are obtained by dressing the virtual part of the NLO QCD corrections in
Eq.~\eqref{eq:dipole-nlo-qq} with the appropriate non-factorizable EW
correction factors, i.e.\ with the sum of the virtual corrections~\eqref{eq:nf-virt} and the integrated soft-slicing
factor~\eqref{eq:nf-slice} and with the real-correction factor~\eqref{eq:nlo-nf-real-delta}, respectively,
\begin{subequations}
\label{eq:master-VX-qq}
\begin{align}
  %%%%%%%%%%%%%%%%%%%%%%%%%%%
  \label{eq:master-VV-qq}
  \sigreg_{\Paq_a\Pq_b,\nf}^{\Vs\otimes\Vew}&=
  \int_2
  \biggl\{ \rd\sigma_{\Paq_a\Pq_b,\PA}^{\Vs}
  + \rd\sigma_{\Paq_a\Pq_b,\PA}^0\!\otimes\!\CSI  \biggr\}
  \biggl[2\Re\Bigl\{\delta_{\Vew,\nf}^{\Paq_a\Pq_b\to\Pl_1\Pal_2}\Bigr\} 
  + \delta_{\soft}^{\Paq_a\Pq_b\to\Pl_1\Pal_2}(\Delta E) \biggr] , \\
  %%%%%%%%%%%%%%%%%%%%%%%%%%%
  \label{eq:master-VR-qq}
  \sigreg_{\Paq_a\Pq_b,\nf}^{\Vs\otimes \Rew}&=
  \iint\limits_{\substack{2+\Pgg\\E_\Pgg>\Delta E}}  
  \biggl\{ \rd\sigma_{\Paq_a\Pq_b,\PA}^{\Vs}
  + \rd\sigma_{\Paq_a\Pq_b,\PA}^0
  \!\otimes\!\CSI \biggr\}
  \; \delta_{\Rew,\nf}^{\Paq_a\Pq_b\to\Pl_1\Pal_2\Pgg} .
\end{align}
\end{subequations}
As in the NLO QCD corrections, only the quark--antiquark induced
channel yields a non-vanishing contribution.

The IR-regularized (real QCD)$\times$(virtual EW)
corrections~\eqref{eq:master-RV} and the double-real
corrections~\eqref{eq:master-RR} are obtained from the real-emission
part of the NLO QCD cross sections~\eqref{eq:dipole-nlo-ch} by
multiplying with the appropriate EW correction factors, i.e.\
the sum of the virtual corrections and the integrated slicing factors
given in Eqs.~\eqref{eq:slice-nf-rr} and \eqref{eq:nf-rv-ew}, and the
real corrections~\eqref{eq:deltanf-rr}, respectively. For the
quark--antiquark channel, the explicit expressions are
\begin{subequations}
\label{eq:master-RX-qq}
\begin{align} 
  %%%%%%%%%%%%%%%%%%%%%%%%%%%%
  \label{eq:master-RV-qq}
  \sigreg_{\Paq_a\Pq_b,\nf}^{\Rs\otimes\Vew}
  =& \int_3\biggl\{
  \rd\sigma_{\Paq_a\Pq_b,\PA}^{\Rs}
  \biggl[ 2\Re\Bigl\{\delta_{\Vew,\nf}^{\Paq_a\Pq_b\to\Pl_1\Pal_2\Pg}\Bigr\} 
  + \delta_{\soft}^{\Paq_a\Pq_b\to\Pl_1\Pal_2\Pg}(\Delta E) \biggr] \nonumber\\
  &- \rd\sigma_{\Paq_a\Pq_b,\PA}^0\bigl(\widetilde\Phi_{2,(\Paq_a\Pg)\Pq_b}\bigr) \biggl[
  2\Re\Bigl\{\delta_{\Vew,\nf}^{\Paq_a\Pq_b\to\Pl_1\Pal_2}\Bigr\} 
  + \delta_{\soft,\boost(x_{\Pg,\Paq_a\Pq_b},1)}^{\Paq_a\Pq_b\to\Pl_1\Pal_2}(\Delta E) \biggr]
  \!\otimes\!\CSV^{\Paq_a,\Paq_a} \nonumber\\
  &- \rd\sigma_{\Paq_a\Pq_b,\PA}^0\bigl(\widetilde\Phi_{2,(\Pq_b\Pg)\Paq_a}\bigr) \biggl[
  2\Re\Bigl\{\delta_{\Vew,\nf}^{\Paq_a\Pq_b\to\Pl_1\Pal_2}\Bigr\} 
  + \delta_{\soft,\boost(1, x_{\Pg,\Pq_b\Paq_a})}^{\Paq_a\Pq_b\to\Pl_1\Pal_2}(\Delta E) \biggr]
  \!\otimes\!\CSV^{\Pq_b,\Pq_b} 
  \biggr\} , 
\end{align}
\begin{align}
  %%%%%%%%%%%%%%%%%%%%%%%%%%%% 
  \label{eq:master-RR-qq} 
  \sigreg_{\Paq_a\Pq_b,\nf}^{\Rs\otimes \Rew}
  =\iint\limits_{\substack{3+\Pgg\\E_\Pgg>\Delta E}}\biggl\{
  \rd\sigma_{\Paq_a\Pq_b,\PA}^{\Rs}
  \;\delta_{\Rew,\nf}^{\Paq_a\Pq_b\to\Pl_1\Pal_2\Pgg\Pg}
  &- \rd\sigma_{\Paq_a\Pq_b,\PA}^0\bigl(\widetilde\Phi_{2+\Pgg,(\Paq_a\Pg)\Pq_b}\bigr)
  \;\delta_{\Rew,\nf}^{\Paq_a\Pq_b\to\Pl_1\Pal_2\Pgg}
  \!\otimes\!\CSV^{\Paq_a,\Paq_a} \nonumber\\[-2em]
  &- \rd\sigma_{\Paq_a\Pq_b,\PA}^0\bigl(\widetilde\Phi_{2+\Pgg,(\Pq_b\Pg)\Paq_a}\bigr)
  \;\delta_{\Rew,\nf}^{\Paq_a\Pq_b\to\Pl_1\Pal_2\Pgg}
  \!\otimes\!\CSV^{\Pq_b,\Pq_b} 
  \biggr\} .
\end{align}  
\end{subequations}
Note that it is implied that the correction factors are evaluated on
the same phase space as the differential cross section they multiply.
In particular, the two correction factors that appear as the second
and third terms inside the curly braces in Eq.~\eqref{eq:master-RV-qq}
are not identical, as they are evaluated for different kinematics
that correspond to the two possible initial-state splittings
of the incoming quarks.
The integrated slicing factors of the subtraction terms correspond to the boosted variants given in Eq.~\eqref{eq:nf-soft-boost},
where the boost along the beam axis is determined by the dipole-variable $x_{i,ab}$ defined in Eq.~\eqref{eq:dipole_ii_a}.
The corresponding expressions for the quark--gluon initiated subprocesses read
\begin{subequations}
\label{eq:master-RX-gq}
\begin{align}
  %%%%%%%%%%%%%%%%%%%%%%%%%%%%
  \label{eq:master-RV-gq}
  \sigreg_{\Pg\Pq_b,\nf}^{\Rs\otimes \Vew}
  =& \int_3 \biggl\{
  \rd\sigma_{\Pg\Pq_b,\PA}^{\Rs}
  \biggl[ 2\Re\Bigl\{\delta_{\Vew,\nf}^{\Pg\Pq_b\to\Pl_1\Pal_2\Pq_a}\Bigr\} 
  + \delta_{\soft}^{\Pg\Pq_b\to\Pl_1\Pal_2\Pq_a}(\Delta E) \biggr] \nonumber\\
  &- \rd\sigma_{\Paq_a\Pq_b,\PA}^0\bigl(\widetilde\Phi_{2,(\Pg\Pq_a)\Pq_b}\bigr) \biggl[
  2\Re\Bigl\{\delta_{\Vew,\nf}^{\Paq_a\Pq_b\to\Pl_1\Pal_2}\Bigr\} 
  + \delta_{\soft,\boost(x_{\Pq_a,\Pg\Pq_b},1)}^{\Paq_a\Pq_b\to\Pl_1\Pal_2}(\Delta E) \biggr]
  \!\otimes\!\CSV^{\Pg,\Paq_a}
  \biggr\},
\end{align}
\begin{align}
  %%%%%%%%%%%%%%%%%%%%%%%%%%%%
  \label{eq:master-RR-gq} 
  \sigreg_{\Pg\Pq_b,\nf}^{\Rs\otimes \Rew}
  =\iint\limits_{\substack{3+\Pgg\\E_\Pgg>\Delta E}}\biggl\{
  \rd\sigma_{\Pg\Pq_b,\PA}^{\Rs}
  \;\delta_{\Rew,\nf}^{\Pg\Pq_b\to\Pl_1\Pal_2\Pgg\Pq_a}
  - \rd\sigma_{\Paq_a\Pq_b,\PA}^0\bigl(\widetilde\Phi_{2+\Pgg,(\Pg\Pq_a)\Pq_b}\bigr)
  \;\delta_{\Rew,\nf}^{\Paq_a\Pq_b\to\Pl_1\Pal_2\Pgg}
  \otimes\CSV^{\Pg,\Paq_a}
  \biggr\} .
\end{align}
\end{subequations}

The collinear counterterms with additional virtual EW~\eqref{eq:master-CV} 
and real photonic~\eqref{eq:master-CR} corrections 
are constructed from the corresponding term in
Eq.~\eqref{eq:dipole-nlo-ch} by dressing it with the respective non-factorizable correction factors,
\begin{subequations}
\label{eq:master-CX-qq}
\begin{align}
  %%%%%%%%%%%%%%%%%%%%%%%%%%%%
  \label{eq:master-CV-qq}
  \sigreg_{\Paq_a\Pq_b,\nf}^{\Cs\otimes \Vew}
  =&\quad \!\int_0^1\!\rd x\int_2
  \rd\sigma_{\Paq_a\Pq_b,\PA}^0 (xp_a,p_b)
  \biggl[2\Re\Bigl\{\delta_{\Vew,\nf}^{\Paq_a\Pq_b\to\Pl_1\Pal_2}\Bigr\} 
  + \delta_{\soft,\boost(x,1)}^{\Paq_a\Pq_b\to\Pl_1\Pal_2}(\Delta E) \biggr]
  \!\otimes\!(\CSK+\CSP)^{\Paq_a,\Paq_a} \nonumber\\
  & + \!\int_0^1\!\rd x\int_2
  \rd\sigma_{\Paq_a\Pq_b,\PA}^0 (p_a,xp_b)
  \biggl[2\Re\Bigl\{\delta_{\Vew,\nf}^{\Paq_a\Pq_b\to\Pl_1\Pal_2}\Bigr\} 
  + \delta_{\soft,\boost(1,x)}^{\Paq_a\Pq_b\to\Pl_1\Pal_2}(\Delta E) \biggr]
  \!\otimes\!(\CSK+\CSP)^{\Pq_b,\Pq_b} , \\
  %%%%%%%%%%%%%%%%%%%%%%%%%%%%
  \label{eq:master-CR-qq}
  \sigreg_{\Paq_a\Pq_b,\nf}^{\Cs\otimes \Rew}
  =&\quad \int_0^1\rd x\iint\limits_{\substack{2+\Pgg\\E_\Pgg>\Delta E}}
  \rd\sigma_{\Paq_a\Pq_b,\PA}^0 (xp_a,p_b)
  \;\delta_{\Rew,\nf}^{\Paq_a\Pq_b\to\Pl_1\Pal_2\Pgg}
  \!\otimes\!(\CSK+\CSP)^{\Paq_a,\Paq_a} \nonumber\\
  & + \int_0^1\rd x\iint\limits_{\substack{2+\Pgg\\E_\Pgg>\Delta E}}
  \rd\sigma_{\Paq_a\Pq_b,\PA}^0 (p_a,xp_b)
  \;\delta_{\Rew,\nf}^{\Paq_a\Pq_b\to\Pl_1\Pal_2\Pgg}
  \!\otimes\!(\CSK+\CSP)^{\Pq_b,\Pq_b} .
\end{align}
\end{subequations}
As discussed in Sect.~\ref{sec:master}, the boosted variants of the soft-slicing factor in Eq.~\eqref{eq:master-CV-qq}
arise from the choice of the frame of reference, in which the 
photon-energy cut $E_\Pgg>\Delta E$ in Eq.~\eqref{eq:master-CR-qq} 
is applied.
The corresponding formulas for the gluon--quark channel read
\begin{subequations}
\label{eq:master-CX-gq}
\begin{align}
  %%%%%%%%%%%%%%%%%%%%%%%%%%%%
  \label{eq:master-CV-gq}
  \sigreg_{\Pg\Pq_b,\nf}^{\Cs\otimes \Vew}
  & = \!\int_0^1\!\rd x\int_2
  \rd\sigma_{\Paq_a\Pq_b,\PA}^0 (xp_\Pg,p_b)
  \biggl[2\Re\Bigl\{\delta_{\Vew,\nf}^{\Paq_a\Pq_b\to\Pl_1\Pal_2}\Bigr\} 
  + \delta_{\soft,\boost(x,1)}^{\Paq_a\Pq_b\to\Pl_1\Pal_2}(\Delta E) \biggr]
  \!\otimes\!(\CSK+\CSP)^{\Pg,\Paq_a} , \\
  %%%%%%%%%%%%%%%%%%%%%%%%%%%%
  \label{eq:master-CR-gq}
  \sigreg_{\Pg\Pq_b,\nf}^{\Cs\otimes \Rew}
  & = \int_0^1\rd x\iint\limits_{\substack{2+\Pgg\\E_\Pgg>\Delta E}}
  \rd\sigma_{\Paq_a\Pq_b,\PA}^0 (xp_\Pg,p_b)
  \;\delta_{\Rew,\nf}^{\Paq_a\Pq_b\to\Pl_1\Pal_2\Pgg}
  \!\otimes\!(\CSK+\CSP)^{\Pg,\Paq_a} .
\end{align}
\end{subequations}

%%%%%%%%%%%%%%%%%%%%%%%%%%%%%%%%%%%%%%%%%%%%%%%%%%%%%%%%%%%%%%%%%%%%%%
\bibliographystyle{tep}
\bibliography{dynonfact}

\providecommand{\href}[2]{#2}\begingroup\raggedright\begin{thebibliography}{100}

\bibitem{Gerber:2007xk}
TeV4LHC-Top and Electroweak Working Group, C.~Gerber {\em et al.},
\href{http://arxiv.org/abs/0705.3251}{{\tt arXiv:0705.3251 [hep-ph]}}.
%%CITATION = ARXIV:0705.3251;%%.

\bibitem{Abdullin:2006aa}
TeV4LHC Working Group, S.~Abdullin {\em et al.},
\href{http://arxiv.org/abs/hep-ph/0608322}{{\tt arXiv:hep-ph/0608322
  [hep-ph]}}.
%%CITATION = HEP-PH/0608322;%%.

\bibitem{Haywood:1999qg}
S.~Haywood, {\em et al.},
\href{http://arxiv.org/abs/hep-ph/0003275}{{\tt arXiv:hep-ph/0003275
  [hep-ph]}}.
%%CITATION = HEP-PH/0003275;%%.

\bibitem{Aaltonen:2013iut}
CDF Collaboration, D0 Collaboration, T.~A. Aaltonen {\em et al.},
  \href{http://dx.doi.org/10.1103/PhysRevD.88.052018}{Phys.Rev. {\bf D88}
  (2013)  052018},
\href{http://arxiv.org/abs/1307.7627}{{\tt arXiv:1307.7627 [hep-ex]}}.
%%CITATION = ARXIV:1307.7627;%%.

\bibitem{Besson:2008zs}
ATLAS Collaboration, N.~Besson, {\em et al.},
  \href{http://dx.doi.org/10.1140/epjc/s10052-008-0774-4}{Eur.Phys.J. {\bf C57}
  (2008)  627},
\href{http://arxiv.org/abs/0805.2093}{{\tt arXiv:0805.2093 [hep-ex]}}.
%%CITATION = ARXIV:0805.2093;%%.

\bibitem{Hamberg:1990np}
R.~Hamberg, W.~van Neerven, and T.~Matsuura,
  \href{http://dx.doi.org/10.1016/0550-3213(91)90064-5}{Nucl.Phys. {\bf B359}
  (1991)  343}.
[Erratum-ibid.\ B {\bf 644} (2002) 403].
%%CITATION = NUPHA,B359,343;%%.

\bibitem{Harlander:2002wh}
R.~V. Harlander and W.~B. Kilgore,
  \href{http://dx.doi.org/10.1103/PhysRevLett.88.201801}{Phys.Rev.Lett. {\bf
  88} (2002)  201801},
\href{http://arxiv.org/abs/hep-ph/0201206}{{\tt arXiv:hep-ph/0201206
  [hep-ph]}}.
%%CITATION = HEP-PH/0201206;%%.

\bibitem{Anastasiou:2003ds}
C.~Anastasiou, L.~J. Dixon, K.~Melnikov, and F.~Petriello,
  \href{http://dx.doi.org/10.1103/PhysRevD.69.094008}{Phys.Rev. {\bf D69}
  (2004)  094008},
\href{http://arxiv.org/abs/hep-ph/0312266}{{\tt arXiv:hep-ph/0312266
  [hep-ph]}}.
%%CITATION = HEP-PH/0312266;%%.

\bibitem{Melnikov:2006di}
K.~Melnikov and F.~Petriello,
  \href{http://dx.doi.org/10.1103/PhysRevLett.96.231803}{Phys.Rev.Lett. {\bf
  96} (2006)  231803},
\href{http://arxiv.org/abs/hep-ph/0603182}{{\tt arXiv:hep-ph/0603182
  [hep-ph]}}.
%%CITATION = HEP-PH/0603182;%%.

\bibitem{Melnikov:2006kv}
K.~Melnikov and F.~Petriello,
  \href{http://dx.doi.org/10.1103/PhysRevD.74.114017}{Phys.Rev. {\bf D74}
  (2006)  114017},
\href{http://arxiv.org/abs/hep-ph/0609070}{{\tt arXiv:hep-ph/0609070
  [hep-ph]}}.
%%CITATION = HEP-PH/0609070;%%.

\bibitem{Catani:2009sm}
S.~Catani, {\em et al.},
  \href{http://dx.doi.org/10.1103/PhysRevLett.103.082001}{Phys.Rev.Lett. {\bf
  103} (2009)  082001},
\href{http://arxiv.org/abs/0903.2120}{{\tt arXiv:0903.2120 [hep-ph]}}.
%%CITATION = ARXIV:0903.2120;%%.

\bibitem{Gavin:2010az}
R.~Gavin, Y.~Li, F.~Petriello, and S.~Quackenbush,
  \href{http://dx.doi.org/10.1016/j.cpc.2011.06.008}{Comput.Phys.Commun. {\bf
  182} (2011)  2388},
\href{http://arxiv.org/abs/1011.3540}{{\tt arXiv:1011.3540 [hep-ph]}}.
%%CITATION = ARXIV:1011.3540;%%.

\bibitem{Gavin:2012sy}
R.~Gavin, Y.~Li, F.~Petriello, and S.~Quackenbush,
  \href{http://dx.doi.org/10.1016/j.cpc.2012.09.005}{Comput.Phys.Commun. {\bf
  184} (2013)  208},
\href{http://arxiv.org/abs/1201.5896}{{\tt arXiv:1201.5896 [hep-ph]}}.
%%CITATION = ARXIV:1201.5896;%%.

\bibitem{Baur:1997wa}
U.~Baur, S.~Keller, and W.~Sakumoto,
  \href{http://dx.doi.org/10.1103/PhysRevD.57.199}{Phys.Rev. {\bf D57} (1998)
  199},
\href{http://arxiv.org/abs/hep-ph/9707301}{{\tt arXiv:hep-ph/9707301
  [hep-ph]}}.
%%CITATION = HEP-PH/9707301;%%.

\bibitem{Zykunov:2001mn}
V.~Zykunov, Eur.Phys.J.direct {\bf C3} (2001)  9,
\href{http://arxiv.org/abs/hep-ph/0107059}{{\tt arXiv:hep-ph/0107059
  [hep-ph]}}.
%%CITATION = HEP-PH/0107059;%%.

\bibitem{Baur:2001ze}
U.~Baur, {\em et al.},
  \href{http://dx.doi.org/10.1103/PhysRevD.65.033007}{Phys.Rev. {\bf D65}
  (2002)  033007},
\href{http://arxiv.org/abs/hep-ph/0108274}{{\tt arXiv:hep-ph/0108274
  [hep-ph]}}.
%%CITATION = HEP-PH/0108274;%%.

\bibitem{Dittmaier:2001ay}
S.~Dittmaier and M.~Kr{\"a}mer, Phys. Rev. {\bf D65} (2002)  073007,
\href{http://arxiv.org/abs/hep-ph/0109062}{{\tt hep-ph/0109062}}.
%%CITATION = HEP-PH 0109062;%%.

\bibitem{Baur:2004ig}
U.~Baur and D.~Wackeroth,
  \href{http://dx.doi.org/10.1103/PhysRevD.70.073015}{Phys.Rev. {\bf D70}
  (2004)  073015},
\href{http://arxiv.org/abs/hep-ph/0405191}{{\tt arXiv:hep-ph/0405191
  [hep-ph]}}.
%%CITATION = HEP-PH/0405191;%%.

\bibitem{Arbuzov:2005dd}
A.~Arbuzov, {\em et al.}, \href{http://dx.doi.org/10.1140/epjc/s2006-02505-y,
  10.1140/epjc/s10052-007-0225-7}{Eur.Phys.J. {\bf C46} (2006)  407},
\href{http://arxiv.org/abs/hep-ph/0506110}{{\tt arXiv:hep-ph/0506110
  [hep-ph]}}.
%%CITATION = HEP-PH/0506110;%%.

\bibitem{CarloniCalame:2006zq}
C.~Carloni~Calame, G.~Montagna, O.~Nicrosini, and A.~Vicini,
  \href{http://dx.doi.org/10.1088/1126-6708/2006/12/016}{JHEP {\bf 0612} (2006)
   016},
\href{http://arxiv.org/abs/hep-ph/0609170}{{\tt arXiv:hep-ph/0609170
  [hep-ph]}}.
%%CITATION = HEP-PH/0609170;%%.

\bibitem{Zykunov:2005tc}
V.~Zykunov, \href{http://dx.doi.org/10.1103/PhysRevD.75.073019}{Phys.Rev. {\bf
  D75} (2007)  073019},
\href{http://arxiv.org/abs/hep-ph/0509315}{{\tt arXiv:hep-ph/0509315
  [hep-ph]}}.
%%CITATION = HEP-PH/0509315;%%.

\bibitem{CarloniCalame:2007cd}
C.~Carloni~Calame, G.~Montagna, O.~Nicrosini, and A.~Vicini,
  \href{http://dx.doi.org/10.1088/1126-6708/2007/10/109}{JHEP {\bf 0710} (2007)
   109},
\href{http://arxiv.org/abs/0710.1722}{{\tt arXiv:0710.1722 [hep-ph]}}.
%%CITATION = ARXIV:0710.1722;%%.

\bibitem{Arbuzov:2007db}
A.~Arbuzov, {\em et al.},
  \href{http://dx.doi.org/10.1140/epjc/s10052-008-0531-8}{Eur.Phys.J. {\bf C54}
  (2008)  451},
\href{http://arxiv.org/abs/0711.0625}{{\tt arXiv:0711.0625 [hep-ph]}}.
%%CITATION = ARXIV:0711.0625;%%.

\bibitem{Brensing:2007qm}
S.~Brensing, S.~Dittmaier, M.~Kr{\"a}mer, and A.~M{\"u}ck,
  \href{http://dx.doi.org/10.1103/PhysRevD.77.073006}{Phys.Rev. {\bf D77}
  (2008)  073006},
\href{http://arxiv.org/abs/0710.3309}{{\tt arXiv:0710.3309 [hep-ph]}}.
%%CITATION = ARXIV:0710.3309;%%.

\bibitem{Dittmaier:2009cr}
S.~Dittmaier and M.~Huber,
  \href{http://dx.doi.org/10.1007/JHEP01(2010)060}{JHEP {\bf 1001} (2010)
  060},
\href{http://arxiv.org/abs/0911.2329}{{\tt arXiv:0911.2329 [hep-ph]}}.
%%CITATION = ARXIV:0911.2329;%%.

\bibitem{Boughezal:2013cwa}
R.~Boughezal, Y.~Li, and F.~Petriello,
\href{http://arxiv.org/abs/1312.3972}{{\tt arXiv:1312.3972 [hep-ph]}}.
%%CITATION = ARXIV:1312.3972;%%.

\bibitem{Baur:1999hm}
U.~Baur and T.~Stelzer,
  \href{http://dx.doi.org/10.1103/PhysRevD.61.073007}{Phys.Rev. {\bf D61}
  (2000)  073007},
\href{http://arxiv.org/abs/hep-ph/9910206}{{\tt arXiv:hep-ph/9910206
  [hep-ph]}}.
%%CITATION = HEP-PH/9910206;%%.

\bibitem{CarloniCalame:2003ux}
C.~Carloni~Calame, G.~Montagna, O.~Nicrosini, and M.~Treccani,
  \href{http://dx.doi.org/10.1103/PhysRevD.69.037301}{Phys.Rev. {\bf D69}
  (2004)  037301},
\href{http://arxiv.org/abs/hep-ph/0303102}{{\tt arXiv:hep-ph/0303102
  [hep-ph]}}.
%%CITATION = HEP-PH/0303102;%%.

\bibitem{Placzek:2003zg}
W.~Placzek and S.~Jadach,
  \href{http://dx.doi.org/10.1140/epjc/s2003-01223-4}{Eur.Phys.J. {\bf C29}
  (2003)  325},
\href{http://arxiv.org/abs/hep-ph/0302065}{{\tt arXiv:hep-ph/0302065
  [hep-ph]}}.
%%CITATION = HEP-PH/0302065;%%.

\bibitem{Moch:2005ba}
S.~Moch, J.~Vermaseren, and A.~Vogt,
  \href{http://dx.doi.org/10.1016/j.nuclphysb.2005.08.005}{Nucl.Phys. {\bf
  B726} (2005)  317},
\href{http://arxiv.org/abs/hep-ph/0506288}{{\tt arXiv:hep-ph/0506288
  [hep-ph]}}.
%%CITATION = HEP-PH/0506288;%%.

\bibitem{Laenen:2005uz}
E.~Laenen and L.~Magnea,
  \href{http://dx.doi.org/10.1016/j.physletb.2005.10.038}{Phys.Lett. {\bf B632}
  (2006)  270},
\href{http://arxiv.org/abs/hep-ph/0508284}{{\tt arXiv:hep-ph/0508284
  [hep-ph]}}.
%%CITATION = HEP-PH/0508284;%%.

\bibitem{Idilbi:2005ky}
A.~Idilbi and X.-d. Ji,
  \href{http://dx.doi.org/10.1103/PhysRevD.72.054016}{Phys.Rev. {\bf D72}
  (2005)  054016},
\href{http://arxiv.org/abs/hep-ph/0501006}{{\tt arXiv:hep-ph/0501006
  [hep-ph]}}.
%%CITATION = HEP-PH/0501006;%%.

\bibitem{Ravindran:2007sv}
V.~Ravindran and J.~Smith,
  \href{http://dx.doi.org/10.1103/PhysRevD.76.114004}{Phys.Rev. {\bf D76}
  (2007)  114004},
\href{http://arxiv.org/abs/0708.1689}{{\tt arXiv:0708.1689 [hep-ph]}}.
%%CITATION = ARXIV:0708.1689;%%.

\bibitem{Frixione:2006gn}
S.~Frixione and B.~R. Webber,
\href{http://arxiv.org/abs/hep-ph/0612272}{{\tt arXiv:hep-ph/0612272
  [hep-ph]}}.
%%CITATION = HEP-PH/0612272;%%.

\bibitem{Alioli:2008gx}
S.~Alioli, P.~Nason, C.~Oleari, and E.~Re,
  \href{http://dx.doi.org/10.1088/1126-6708/2008/07/060}{JHEP {\bf 0807} (2008)
   060},
\href{http://arxiv.org/abs/0805.4802}{{\tt arXiv:0805.4802 [hep-ph]}}.
%%CITATION = ARXIV:0805.4802;%%.

\bibitem{Hamilton:2008pd}
K.~Hamilton, P.~Richardson, and J.~Tully,
  \href{http://dx.doi.org/10.1088/1126-6708/2008/10/015}{JHEP {\bf 0810} (2008)
   015},
\href{http://arxiv.org/abs/0806.0290}{{\tt arXiv:0806.0290 [hep-ph]}}.
%%CITATION = ARXIV:0806.0290;%%.

\bibitem{Balazs:1997xd}
C.~Balazs and C.~Yuan,
  \href{http://dx.doi.org/10.1103/PhysRevD.56.5558}{Phys.Rev. {\bf D56} (1997)
  5558},
\href{http://arxiv.org/abs/hep-ph/9704258}{{\tt arXiv:hep-ph/9704258
  [hep-ph]}}.
%%CITATION = HEP-PH/9704258;%%.

\bibitem{Landry:2002ix}
F.~Landry, R.~Brock, P.~M. Nadolsky, and C.~Yuan,
  \href{http://dx.doi.org/10.1103/PhysRevD.67.073016}{Phys.Rev. {\bf D67}
  (2003)  073016},
\href{http://arxiv.org/abs/hep-ph/0212159}{{\tt arXiv:hep-ph/0212159
  [hep-ph]}}.
%%CITATION = HEP-PH/0212159;%%.

\bibitem{Bozzi:2010xn}
G.~Bozzi, {\em et al.},
  \href{http://dx.doi.org/10.1016/j.physletb.2010.12.024}{Phys.Lett. {\bf B696}
  (2011)  207},
\href{http://arxiv.org/abs/1007.2351}{{\tt arXiv:1007.2351 [hep-ph]}}.
%%CITATION = ARXIV:1007.2351;%%.

\bibitem{Mantry:2010mk}
S.~Mantry and F.~Petriello,
  \href{http://dx.doi.org/10.1103/PhysRevD.83.053007}{Phys.Rev. {\bf D83}
  (2011)  053007},
\href{http://arxiv.org/abs/1007.3773}{{\tt arXiv:1007.3773 [hep-ph]}}.
%%CITATION = ARXIV:1007.3773;%%.

\bibitem{Becher:2011xn}
T.~Becher, M.~Neubert, and D.~Wilhelm,
  \href{http://dx.doi.org/10.1007/JHEP02(2012)124}{JHEP {\bf 1202} (2012)
  124},
\href{http://arxiv.org/abs/1109.6027}{{\tt arXiv:1109.6027 [hep-ph]}}.
%%CITATION = ARXIV:1109.6027;%%.

\bibitem{Cao:2004yy}
Q.-H. Cao and C.~Yuan,
  \href{http://dx.doi.org/10.1103/PhysRevLett.93.042001}{Phys.Rev.Lett. {\bf
  93} (2004)  042001},
\href{http://arxiv.org/abs/hep-ph/0401026}{{\tt arXiv:hep-ph/0401026
  [hep-ph]}}.
%%CITATION = HEP-PH/0401026;%%.

\bibitem{Richardson:2010gz}
P.~Richardson, {\em et al.},
  \href{http://dx.doi.org/10.1007/JHEP06(2012)090}{JHEP {\bf 1206} (2012)
  090},
\href{http://arxiv.org/abs/1011.5444}{{\tt arXiv:1011.5444 [hep-ph]}}.
%%CITATION = ARXIV:1011.5444;%%.

\bibitem{Bernaciak:2012hj}
C.~Bernaciak and D.~Wackeroth, Phys.Rev. {\bf D85} (2012)  093003,
\href{http://arxiv.org/abs/1201.4804}{{\tt arXiv:1201.4804 [hep-ph]}}.
%%CITATION = ARXIV:1201.4804;%%.

\bibitem{Barze:2012tt}
L.~Barz\`e, {\em et al.}, JHEP {\bf 1204} (2012)  037,
\href{http://arxiv.org/abs/1202.0465}{{\tt arXiv:1202.0465 [hep-ph]}}.
%%CITATION = ARXIV:1202.0465;%%.

\bibitem{Li:2012wna}
Y.~Li and F.~Petriello,
  \href{http://dx.doi.org/10.1103/PhysRevD.86.094034}{Phys.Rev. {\bf D86}
  (2012)  094034},
\href{http://arxiv.org/abs/1208.5967}{{\tt arXiv:1208.5967 [hep-ph]}}.
%%CITATION = ARXIV:1208.5967;%%.

\bibitem{Barze:2013yca}
L.~Barz\`e, {\em et al.},
  \href{http://dx.doi.org/10.1140/epjc/s10052-013-2474-y}{Eur.Phys.J. {\bf C73}
  (2013)  2474},
\href{http://arxiv.org/abs/1302.4606}{{\tt arXiv:1302.4606 [hep-ph]}}.
%%CITATION = ARXIV:1302.4606;%%.

\bibitem{Balossini:2009sa}
G.~Balossini, {\em et al.},
  \href{http://dx.doi.org/10.1007/JHEP01(2010)013}{JHEP {\bf 1001} (2010)
  013},
\href{http://arxiv.org/abs/0907.0276}{{\tt arXiv:0907.0276 [hep-ph]}}.
%%CITATION = ARXIV:0907.0276;%%.

\bibitem{Kotikov:2007vr}
A.~Kotikov, J.~H. K{\"u}hn, and O.~Veretin,
  \href{http://dx.doi.org/10.1016/j.nuclphysb.2007.07.018}{Nucl.Phys. {\bf
  B788} (2008)  47},
\href{http://arxiv.org/abs/hep-ph/0703013}{{\tt arXiv:hep-ph/0703013
  [hep-ph]}}.
%%CITATION = HEP-PH/0703013;%%.

\bibitem{Kilgore:2011pa}
W.~B. Kilgore and C.~Sturm,
  \href{http://dx.doi.org/10.1103/PhysRevD.85.033005}{Phys.Rev. {\bf D85}
  (2012)  033005},
\href{http://arxiv.org/abs/1107.4798}{{\tt arXiv:1107.4798 [hep-ph]}}.
%%CITATION = ARXIV:1107.4798;%%.

\bibitem{Bonciani:2011zz}
R.~Bonciani,
PoS {\bf EPS-HEP2011} (2011)  365.
%%CITATION = POSCI,EPS-HEP2011,365;%%.

\bibitem{Czarnecki:1996ei}
A.~Czarnecki and J.~H. K{\"u}hn,
  \href{http://dx.doi.org/10.1103/PhysRevLett.77.3955}{Phys.Rev.Lett. {\bf 77}
  (1996)  3955},
\href{http://arxiv.org/abs/hep-ph/9608366}{{\tt arXiv:hep-ph/9608366
  [hep-ph]}}.
%%CITATION = HEP-PH/9608366;%%.

\bibitem{Kara:2013dua}
D.~Kara, \href{http://dx.doi.org/10.1016/j.nuclphysb.2013.10.024}{Nucl.Phys.
  {\bf B877} (2013)  683},
\href{http://arxiv.org/abs/1307.7190}{{\tt arXiv:1307.7190}}.
%%CITATION = ARXIV:1307.7190;%%.

\bibitem{Kuhn:2005az}
J.~H. K{\"u}hn, A.~Kulesza, S.~Pozzorini, and M.~Schulze,
  \href{http://dx.doi.org/10.1016/j.nuclphysb.2005.08.019}{Nucl.Phys. {\bf
  B727} (2005)  368},
\href{http://arxiv.org/abs/hep-ph/0507178}{{\tt arXiv:hep-ph/0507178
  [hep-ph]}}.
%%CITATION = HEP-PH/0507178;%%.

\bibitem{Kuhn:2007cv}
J.~H. K{\"u}hn, A.~Kulesza, S.~Pozzorini, and M.~Schulze,
  \href{http://dx.doi.org/10.1016/j.nuclphysb.2007.12.029}{Nucl.Phys. {\bf
  B797} (2008)  27},
\href{http://arxiv.org/abs/0708.0476}{{\tt arXiv:0708.0476 [hep-ph]}}.
%%CITATION = ARXIV:0708.0476;%%.

\bibitem{Hollik:2007sq}
W.~Hollik, T.~Kasprzik, and B.~Kniehl,
  \href{http://dx.doi.org/10.1016/j.nuclphysb.2007.09.013}{Nucl.Phys. {\bf
  B790} (2008)  138},
\href{http://arxiv.org/abs/0707.2553}{{\tt arXiv:0707.2553 [hep-ph]}}.
%%CITATION = ARXIV:0707.2553;%%.

\bibitem{Denner:2009gj}
A.~Denner, S.~Dittmaier, T.~Kasprzik, and A.~M{\"u}ck,
  \href{http://dx.doi.org/10.1088/1126-6708/2009/08/075}{JHEP {\bf 0908} (2009)
   075},
\href{http://arxiv.org/abs/0906.1656}{{\tt arXiv:0906.1656 [hep-ph]}}.
%%CITATION = ARXIV:0906.1656;%%.

\bibitem{Denner:2011vu}
A.~Denner, S.~Dittmaier, T.~Kasprzik, and A.~M{\"u}ck,
  \href{http://dx.doi.org/10.1007/JHEP06(2011)069}{JHEP {\bf 1106} (2011)
  069},
\href{http://arxiv.org/abs/1103.0914}{{\tt arXiv:1103.0914 [hep-ph]}}.
%%CITATION = ARXIV:1103.0914;%%.

\bibitem{Denner:2012ts}
A.~Denner, S.~Dittmaier, T.~Kasprzik, and A.~Mück,
  \href{http://dx.doi.org/10.1140/epjc/s10052-013-2297-x}{Eur.Phys.J. {\bf C73}
  (2013)  2297},
\href{http://arxiv.org/abs/1211.5078}{{\tt arXiv:1211.5078 [hep-ph]}}.
%%CITATION = ARXIV:1211.5078;%%.

\bibitem{Smith:1989xz}
J.~Smith, D.~Thomas, and W.~van Neerven,
\href{http://dx.doi.org/10.1007/BF01557332}{Z.Phys. {\bf C44} (1989)  267}.
%%CITATION = ZEPYA,C44,267;%%.

\bibitem{Ohnemus:1992jn}
J.~Ohnemus,
\href{http://dx.doi.org/10.1103/PhysRevD.47.940}{Phys.Rev. {\bf D47} (1993)
  940}.
%%CITATION = PHRVA,D47,940;%%.

\bibitem{Ohnemus:1994qp}
J.~Ohnemus, \href{http://dx.doi.org/10.1103/PhysRevD.51.1068}{Phys.Rev. {\bf
  D51} (1995)  1068},
\href{http://arxiv.org/abs/hep-ph/9407370}{{\tt arXiv:hep-ph/9407370
  [hep-ph]}}.
%%CITATION = HEP-PH/9407370;%%.

\bibitem{Dixon:1998py}
L.~J. Dixon, Z.~Kunszt, and A.~Signer,
  \href{http://dx.doi.org/10.1016/S0550-3213(98)00421-0}{Nucl.Phys. {\bf B531}
  (1998)  3},
\href{http://arxiv.org/abs/hep-ph/9803250}{{\tt arXiv:hep-ph/9803250
  [hep-ph]}}.
%%CITATION = HEP-PH/9803250;%%.

\bibitem{Campbell:1999ah}
J.~M. Campbell and R.~K. Ellis,
  \href{http://dx.doi.org/10.1103/PhysRevD.60.113006}{Phys.Rev. {\bf D60}
  (1999)  113006},
\href{http://arxiv.org/abs/hep-ph/9905386}{{\tt arXiv:hep-ph/9905386
  [hep-ph]}}.
%%CITATION = HEP-PH/9905386;%%.

\bibitem{DeFlorian:2000sg}
D.~De~Florian and A.~Signer,
  \href{http://dx.doi.org/10.1007/s100520050007}{Eur.Phys.J. {\bf C16} (2000)
  105},
\href{http://arxiv.org/abs/hep-ph/0002138}{{\tt arXiv:hep-ph/0002138
  [hep-ph]}}.
%%CITATION = HEP-PH/0002138;%%.

\bibitem{Campbell:2011bn}
J.~M. Campbell, R.~K. Ellis, and C.~Williams,
  \href{http://dx.doi.org/10.1007/JHEP07(2011)018}{JHEP {\bf 1107} (2011)
  018},
\href{http://arxiv.org/abs/1105.0020}{{\tt arXiv:1105.0020 [hep-ph]}}.
%%CITATION = ARXIV:1105.0020;%%.

\bibitem{Fadin:1993dz}
V.~S. Fadin, V.~A. Khoze, and A.~D. Martin,
Phys. Rev. {\bf D49} (1994)  2247.
%%CITATION = PHRVA,D49,2247;%%.

\bibitem{Melnikov:1995fx}
K.~Melnikov and O.~I. Yakovlev, Nucl. Phys. {\bf B471} (1996)  90,
\href{http://arxiv.org/abs/hep-ph/9501358}{{\tt hep-ph/9501358}}.
%%CITATION = HEP-PH 9501358;%%.

\bibitem{Beenakker:1997ir}
W.~Beenakker, A.~P. Chapovsky, and F.~A. Berends, Nucl. Phys. {\bf B508} (1997)
   17,
\href{http://arxiv.org/abs/hep-ph/9707326}{{\tt hep-ph/9707326}}.
%%CITATION = HEP-PH/9707326;%%.

\bibitem{Denner:1997ia}
A.~Denner, S.~Dittmaier, and M.~Roth, Nucl. Phys. {\bf B519} (1998)  39,
\href{http://arxiv.org/abs/hep-ph/9710521}{{\tt hep-ph/9710521}}.
%%CITATION = HEP-PH/9710521;%%.

\bibitem{Beneke:2003xh}
M.~Beneke, A.~P. Chapovsky, A.~Signer, and G.~Zanderighi, Phys. Rev. Lett. {\bf
  93} (2004)  011602,
\href{http://arxiv.org/abs/hep-ph/0312331}{{\tt hep-ph/0312331}}.
%%CITATION = HEP-PH 0312331;%%.

\bibitem{Beneke:2004km}
M.~Beneke, A.~P. Chapovsky, A.~Signer, and G.~Zanderighi, Nucl. Phys. {\bf
  B686} (2004)  205,
\href{http://arxiv.org/abs/hep-ph/0401002}{{\tt hep-ph/0401002}}.
%%CITATION = HEP-PH 0401002;%%.

\bibitem{Stuart:1991xk}
R.~G. Stuart,
Phys. Lett. {\bf B262} (1991)  113.
%%CITATION = PHLTA,B262,113;%%.

\bibitem{Aeppli:1993rs}
A.~Aeppli, G.~J. van Oldenborgh, and D.~Wyler, Nucl. Phys. {\bf B428} (1994)
  126,
\href{http://arxiv.org/abs/hep-ph/9312212}{{\tt hep-ph/9312212}}.
%%CITATION = HEP-PH 9312212;%%.

\bibitem{Wackeroth:1996hz}
D.~Wackeroth and W.~Hollik, Phys. Rev. {\bf D55} (1997)  6788,
\href{http://arxiv.org/abs/hep-ph/9606398}{{\tt hep-ph/9606398}}.
%%CITATION = HEP-PH 9606398;%%.

\bibitem{Baur:1998kt}
U.~Baur, S.~Keller, and D.~Wackeroth,
  \href{http://dx.doi.org/10.1103/PhysRevD.59.013002}{Phys.Rev. {\bf D59}
  (1999)  013002},
\href{http://arxiv.org/abs/hep-ph/9807417}{{\tt arXiv:hep-ph/9807417
  [hep-ph]}}.
%%CITATION = HEP-PH/9807417;%%.

\bibitem{Sirlin:1991fd}
A.~Sirlin,
\href{http://dx.doi.org/10.1103/PhysRevLett.67.2127}{Phys.Rev.Lett. {\bf 67}
  (1991)  2127}.
%%CITATION = PRLTA,67,2127;%%.

\bibitem{Gambino:1999ai}
P.~Gambino and P.~A. Grassi,
  \href{http://dx.doi.org/10.1103/PhysRevD.62.076002}{Phys.Rev. {\bf D62}
  (2000)  076002},
\href{http://arxiv.org/abs/hep-ph/9907254}{{\tt arXiv:hep-ph/9907254
  [hep-ph]}}.
%%CITATION = HEP-PH/9907254;%%.

\bibitem{Grassi:2001bz}
P.~A. Grassi, B.~A. Kniehl, and A.~Sirlin,
  \href{http://dx.doi.org/10.1103/PhysRevD.65.085001}{Phys.Rev. {\bf D65}
  (2002)  085001},
\href{http://arxiv.org/abs/hep-ph/0109228}{{\tt arXiv:hep-ph/0109228
  [hep-ph]}}.
%%CITATION = HEP-PH/0109228;%%.

\bibitem{Denner:1991kt}
A.~Denner, \href{http://dx.doi.org/10.1002/prop.2190410402}{Fortsch.Phys. {\bf
  41} (1993)  307},
\href{http://arxiv.org/abs/0709.1075}{{\tt arXiv:0709.1075 [hep-ph]}}.
%%CITATION = ARXIV:0709.1075;%%.

\bibitem{Dittmaier:1999mb}
S.~Dittmaier, Nucl. Phys. {\bf B565} (2000)  69,
\href{http://arxiv.org/abs/hep-ph/9904440}{{\tt hep-ph/9904440}}.
%%CITATION = HEP-PH 9904440;%%.

\bibitem{Dittmaier:2008md}
S.~Dittmaier, A.~Kabelschacht, and T.~Kasprzik,
  \href{http://dx.doi.org/10.1016/j.nuclphysb.2008.03.010}{Nucl.Phys. {\bf
  B800} (2008)  146},
\href{http://arxiv.org/abs/0802.1405}{{\tt arXiv:0802.1405 [hep-ph]}}.
%%CITATION = ARXIV:0802.1405;%%.

\bibitem{Beenakker:1988bq}
W.~Beenakker, H.~Kuijf, W.~L. van Neerven, and J.~Smith,
Phys. Rev. {\bf D40} (1989)  54.
%%CITATION = PHRVA,D40,54;%%.

\bibitem{Harris:2001sx}
B.~Harris and J.~Owens,
  \href{http://dx.doi.org/10.1103/PhysRevD.65.094032}{Phys.Rev. {\bf D65}
  (2002)  094032},
\href{http://arxiv.org/abs/hep-ph/0102128}{{\tt arXiv:hep-ph/0102128
  [hep-ph]}}.
%%CITATION = HEP-PH/0102128;%%.

\bibitem{Beneke:1997zp}
M.~Beneke and V.~A. Smirnov,
  \href{http://dx.doi.org/10.1016/S0550-3213(98)00138-2}{Nucl. Phys. {\bf B522}
  (1998)  321},
\href{http://arxiv.org/abs/hep-ph/9711391}{{\tt arXiv:hep-ph/9711391}}.
%%CITATION = HEP-PH/9711391;%%.

\bibitem{Falgari:2013gwa}
P.~Falgari, A.~Papanastasiou, and A.~Signer,
  \href{http://dx.doi.org/10.1007/JHEP05(2013)156}{JHEP {\bf 1305} (2013)
  156},
\href{http://arxiv.org/abs/1303.5299}{{\tt arXiv:1303.5299 [hep-ph]}}.
%%CITATION = ARXIV:1303.5299;%%.

\bibitem{Falgari:2009zz}
P.~Falgari, {\em W-pair production near threshold in unstable-particle
  effective theory}.
\newblock PhD thesis, RWTH Aachen University, 2008.
\newblock \url{http://darwin.bth.rwth-aachen.de/opus3/volltexte/2009/2682/}.

\bibitem{Beneke:2007zg}
M.~Beneke, {\em et al.},
  \href{http://dx.doi.org/10.1016/j.nuclphysb.2007.09.030}{Nucl.Phys. {\bf
  B792} (2008)  89}, \href{http://arxiv.org/abs/0707.0773}{{\tt arXiv:0707.0773
  [hep-ph]}}.

\bibitem{Catani:1996vz}
S.~Catani and M.~Seymour,
  \href{http://dx.doi.org/10.1016/S0550-3213(96)00589-5}{Nucl.Phys. {\bf B485}
  (1997)  291},
\href{http://arxiv.org/abs/hep-ph/9605323}{{\tt arXiv:hep-ph/9605323
  [hep-ph]}}.
%%CITATION = HEP-PH/9605323;%%.

\bibitem{Beringer:1900zz}
Particle Data Group, J.~Beringer {\em et al.},
\href{http://dx.doi.org/10.1103/PhysRevD.86.010001}{Phys.Rev. {\bf D86} (2012)
  010001}.
%%CITATION = PHRVA,D86,010001;%%.

\bibitem{Bardin:1988xt}
D.~Y. Bardin, A.~Leike, T.~Riemann, and M.~Sachwitz,
\href{http://dx.doi.org/10.1016/0370-2693(88)91627-9}{Phys.Lett. {\bf B206}
  (1988)  546–550}.
%%CITATION = PHLTA,B206,539;%%.

\bibitem{Beenakker:1996kn}
W.~Beenakker, {\em et al.},
  \href{http://dx.doi.org/10.1016/S0550-3213(97)00316-7}{Nucl.Phys. {\bf B500}
  (1997)  255},
\href{http://arxiv.org/abs/hep-ph/9612260}{{\tt arXiv:hep-ph/9612260
  [hep-ph]}}.
%%CITATION = HEP-PH/9612260;%%.

\bibitem{Ball:2012cx}
R.~D. Ball, {\em et al.},
  \href{http://dx.doi.org/10.1016/j.nuclphysb.2012.10.003}{Nucl.Phys. {\bf
  B867} (2013)  244},
\href{http://arxiv.org/abs/1207.1303}{{\tt arXiv:1207.1303 [hep-ph]}}.
%%CITATION = ARXIV:1207.1303;%%.

\bibitem{Ball:2013hta}
NNPDF, R.~D. Ball {\em et al.},
  \href{http://dx.doi.org/10.1016/j.nuclphysb.2013.10.010}{Nucl.Phys. {\bf
  B877} (2013) 2, 290},
\href{http://arxiv.org/abs/1308.0598}{{\tt arXiv:1308.0598 [hep-ph]}}.
%%CITATION = ARXIV:1308.0598;%%.

\bibitem{Yennie:1961ad}
D.~Yennie, S.~C. Frautschi, and H.~Suura,
\href{http://dx.doi.org/10.1016/0003-4916(61)90151-8}{Annals Phys. {\bf 13}
  (1961)  379}.
%%CITATION = APNYA,13,379;%%.

\bibitem{Catani:2000pi}
S.~Catani and M.~Grazzini,
  \href{http://dx.doi.org/10.1016/S0550-3213(00)00572-1}{Nucl.Phys. {\bf B591}
  (2000)  435},
\href{http://arxiv.org/abs/hep-ph/0007142}{{\tt arXiv:hep-ph/0007142
  [hep-ph]}}.
%%CITATION = HEP-PH/0007142;%%.

\bibitem{Smirnov:2006ry}
V.~Smirnov, {\em {Feynman integral calculus}}.
\newblock Springer,
2006.
\newblock
%%CITATION = INSPIRE-728109;%%.

\bibitem{Bauer:2000yr}
C.~W. Bauer, S.~Fleming, D.~Pirjol, and I.~W. Stewart, Phys. Rev. {\bf D63}
  (2001)  114020,
\href{http://arxiv.org/abs/hep-ph/0011336}{{\tt hep-ph/0011336}}.
%%CITATION = HEP-PH/0011336;%%.

\bibitem{Denner:1998rh}
A.~Denner, S.~Dittmaier, and M.~Roth,
  \href{http://dx.doi.org/10.1016/S0370-2693(98)00455-9}{Phys.Lett. {\bf B429}
  (1998)  145},
\href{http://arxiv.org/abs/hep-ph/9803306}{{\tt arXiv:hep-ph/9803306
  [hep-ph]}}.
%%CITATION = HEP-PH/9803306;%%.

\bibitem{Beenakker:1997bp}
W.~Beenakker, A.~P. Chapovsky, and F.~A. Berends, Phys. Lett. {\bf B411} (1997)
   203,
\href{http://arxiv.org/abs/hep-ph/9706339}{{\tt hep-ph/9706339}}.
%%CITATION = HEP-PH/9706339;%%.

\bibitem{Stewart:2010qs}
I.~W. Stewart, F.~J. Tackmann, and W.~J. Waalewijn,
  \href{http://dx.doi.org/10.1007/JHEP09(2010)005}{JHEP {\bf 09} (2010)  005},
\href{http://arxiv.org/abs/1002.2213}{{\tt arXiv:1002.2213 [hep-ph]}}.
%%CITATION = 1002.2213;%%.

\bibitem{Huber:2005yg}
T.~Huber and D.~Maitre,
  \href{http://dx.doi.org/10.1016/j.cpc.2006.01.007}{Comput. Phys. Commun. {\bf
  175} (2006)  122},
\href{http://arxiv.org/abs/hep-ph/0507094}{{\tt arXiv:hep-ph/0507094}}.
%%CITATION = HEP-PH/0507094;%%.

\bibitem{Huber:2007dx}
T.~Huber and D.~Maitre,
  \href{http://dx.doi.org/10.1016/j.cpc.2007.12.008}{Comput. Phys. Commun. {\bf
  178} (2008)  755},
\href{http://arxiv.org/abs/0708.2443}{{\tt arXiv:0708.2443 [hep-ph]}}.
%%CITATION = 0708.2443;%%.

\end{thebibliography}\endgroup
%%%%%%%%%%%%%%%%%%%%%%%%%%%%%%%%%%%%%%%%%%%%%%%%%%%%%%%%%%%%%%%%%%%%%%

\end{document}